\documentclass[12pt]{iopart}

\usepackage{hyperref, multirow, color, bbold}
\usepackage[dvipsnames]{xcolor}
\usepackage{orcidlink}

\newcommand{\dual}{\,{}^*\!}

\usepackage{tikz, varwidth}
\usetikzlibrary{calc,positioning,shapes,shapes.geometric,shapes.misc,
  shadows,arrows,arrows.meta}

\newcommand{\blue}[1]{
  #1
}

\usepackage[normalem]{ulem} 

\usepackage{enumitem}
\newcounter{requirements}[section]
\renewcommand*{\therequirements}{\thesection.\arabic{requirements}}

\newlist{reqlist}{enumerate}{3}
\setlist[reqlist]{label=(\roman*), ref=\therequirements(\roman*),
  leftmargin=*, before=\raggedright}

\newenvironment{requirements}{%
  \refstepcounter{requirements}%
  \vspace{-0.5em}
  \paragraph{Requirements~\therequirements}%
  \reqlist
}{%
  \endreqlist
}

\newcommand{\Cornell}{Cornell Center for Astrophysics and Planetary
    Science, Cornell University, Ithaca, New York 14853, USA}
\newcommand{\Caltech}{Theoretical Astrophysics 350-17, California
  Institute of Technology, Pasadena, CA 91125, USA}

\begin{document}

\title[A discontinuous Galerkin-finite difference hybrid method for GRMHD]{A
  high-order shock capturing discontinuous Galerkin-finite difference hybrid
  method for GRMHD}

\author{Nils Deppe$^1$\orcidlink{0000-0003-4557-4115},
  Fran\c{c}ois H\'{e}bert$^1$\orcidlink{0000-0001-9009-6955},
  Lawrence E.~Kidder$^2$\orcidlink{0000-0001-5392-7342},
  and Saul A.~Teukolsky$^{2,1}$\orcidlink{0000-0001-9765-4526}}
\address{$^1$\Caltech}
\address{$^2$\Cornell}
\ead{ndeppe@caltech.edu}

  \begin{abstract}

    We present a discontinuous Galerkin-finite difference hybrid scheme that
    allows high-order shock capturing with the discontinuous Galerkin method for
    general relativistic magnetohydrodynamics. The hybrid method is conceptually
    quite simple. An unlimited discontinuous Galerkin candidate solution is
    computed for the next time step. If the candidate solution is inadmissible,
    the time step is retaken using robust finite-difference methods. Because of
    its \textit{a posteriori} nature, the hybrid scheme inherits the best
    properties of both methods. It is high-order with exponential convergence in
    smooth regions, while robustly handling discontinuities. We give a detailed
    description of how we transfer the solution between the discontinuous
    Galerkin and finite-difference solvers, and the troubled-cell indicators
    necessary to robustly handle slow-moving discontinuities and simulate
    magnetized neutron stars. We demonstrate the efficacy of the proposed method
    using a suite of standard and very challenging 1d, 2d, and 3d relativistic
    magnetohydrodynamics test problems. The hybrid scheme is designed from the
    ground up to efficiently simulate astrophysical problems such as the
    inspiral, coalescence, and merger of two neutron stars.

  \end{abstract}


\submitto{\CQG}

\section{Introduction\label{sec:dgscl introduction}}

The discontinuous Galerkin (DG) method was first presented by Reed and
Hill~\cite{reed1973triangular} to solve the neutron transport equation. Later,
in a series of seminal papers, Cockburn and Shu applied the DG method to
nonlinear hyperbolic conservation laws~\cite{Cockburn1989tvb, COCKBURN198990,
  1990MaCom..54..545C}. A very important property of the DG method is that it
guarantees linear stability in the $L_2$ norm for arbitrary high order, which
was proven for the scalar case in~\cite{jiang1994cell} and for systems
in~\cite{barth2001energy,hou2007solutions}. While this means the DG method is
very robust, DG alone is still subject to Godunov's theorem~\cite{Godunov1959}:
at high order it produces oscillatory solutions. Accordingly, it requires some
nonlinear supplemental method for stability in the presence of discontinuities
and large gradients. A large number of different methods for limiting the DG
solution to achieve such stability have been proposed.  The basic idea shared by
all the limiters is to detect troubled cells or elements (i.e., those whose
solution is too oscillatory or has some other undesirable property), then apply
some nonlinear reconstruction using the solution from neighboring elements. This
idea is largely an extension of what has worked well for finite-volume (FV) and
finite-difference (FD) shock-capturing methods.

In this paper we follow a different avenue that, to the best of our knowledge,
was first proposed in~\cite{COSTA2007970}. The idea is to supplement a
high-order spectral-type method---such as pseudospectral collocation or, in our
case, DG---with robust FV or FD shock-capturing methods. If the solution in an
element is troubled or inadmissible, the solution is projected to a FV or FD
grid and evolved with existing robust shock-capturing methods. This approach has
been applied to DG supplemented with FV in~\cite{doi:10.1002/fld.2654,
  10.1007/978-3-319-05591-6_96, Dumbser2014a, BOSCHERI2017449, Zanotti:2015mia,
  Fambri:2018udk, NUNEZDELAROSA2018113}. The major breakthrough
in~\cite{Dumbser2014a} was applying the shock detection and physical
realizability checks on the solution \emph{after} the time step is taken and
redoing the step if the solution is found to be inadmissible. We follow this
\textit{a posteriori} approach because it allows us to guarantee a physically
realizable solution (e.g., positive density and pressure), as well as allowing
us to prevent unphysical oscillations from entering the numerical solution. This
procedure is in strong contrast to classical limiting strategies, where
effectively a filter is applied to the DG solution in an attempt to remove
spurious oscillations.

High-order pseudospectral methods have proven extremely useful in producing a
large number of long and accurate gravitational waveforms from binary black hole
merger simulations~\cite{Boyle:2019kee, SpECwebsite, Scheel:2008rj,
  Szilagyi:2009qz, Lovelace:2010ne, Buchman:2012dw, Hemberger:2012jz,
  Scheel:2014ina, Szilagyi:2014fna} \blue{as well as other applications in
relativistic astrophysics~\cite{Bonazzola:1998ge, Meringolo:2021yjh,
  Hilditch:2015aba, Rashti:2021ihv, Meringolo:2020jsu, Tichy:2009zr}}. Since
binary inspirals emit gravitational radiation, the numerical solution in
\blue{most} of the computational
domain is smooth but non-constant, and so high-order methods are
preferable. During the inspiral portion of a binary neutron star merger, the
only discontinuities present are at the stellar surfaces. This suggests that
high-order methods can be used in \blue{most} of the computational domain.
\blue{Specifically, the hydro solution inside the star is smooth,
  and while outside the star the hydro evolution is not necessary, the Einstein
  equations still need to be solved and have a smooth solution. The use of
  high-order methods allows for a significant reduction in computational
  cost of the simulation, which} is
especially important for reducing the computational cost of producing a large
gravitational waveform catalog for binary neutron star mergers.

We present a detailed derivation and description of our DG-FD hybrid scheme and
how we use it to solve the equations of general relativistic
magnetohydrodynamics (GRMHD). To the best of our knowledge, the algorithm is the
first to successfully evolve a 3d magnetized TOV star using DG methods. In
\S\ref{sec:equations} we briefly review the equations of GRMHD. In \S\ref{sec:DG
  and FD methods} we give a brief overview of DG and conservative FD methods,
provide a new simple form of the moving mesh evolution equations, and discuss
the time step size restrictions of the DG and FD methods. In \S\ref{sec:limiting
  within DG} we state our requirements from a DG limiter or DG hybrid scheme,
and then give an overview of common limiters currently used, including which of
our requirements they meet. The new DG-FD hybrid scheme is described in
\S\ref{sec:dgscl DG-FD hybrid}. Specifically, we discuss how to handle the
intercell fluxes between elements using DG and FD, the idea of applying the
troubled-cell indicators \textit{a posteriori}, the troubled-cell indicators we
use, and a new perspective on how DG-FD hybrid schemes should be interpreted. In
\S\ref{sec:dgscl numerical results} we present numerical results from the
open-source code \texttt{SpECTRE}~\cite{Kidder:2016hev, deppe_nils_2021_5501002}
using our scheme and conclude in \S\ref{sec:dgscl conclusions}.

\section{Equations of GRMHD}
\label{sec:equations}

We adopt the standard 3+1 form of the spacetime metric,~(see,
e.g.,~\cite{Baumgarte:2010ndz, 2013rehy.book.....R}),
\begin{eqnarray}
  \label{eq:spacetime metric}
  ds^2 &= g_{ab}dx^a dx^b =-\alpha^2 dt^2 + \gamma_{ij}
         \left(dx^i+\beta^i dt\right) \left(dx^j +\beta^j dt\right),
\end{eqnarray}
where $\alpha$ is the lapse, $\beta^i$ the shift vector, and $\gamma_{ij}$ is
the spatial metric.  We use the Einstein summation convention, summing over
repeated indices.  Latin indices from the first part of the alphabet
$a,b,c,\ldots$ denote spacetime indices ranging from $0$ to $3$, while Latin
indices $i,j,\ldots$ are purely spatial, ranging from $1$ to $3$. We work in
units where $c = G = M_{\odot} = 1$.

\texttt{SpECTRE} currently solves equations in flux-balanced and first-order
hyperbolic form. The general form of a flux-balanced conservation law in a
curved spacetime is
\begin{eqnarray}
  \label{eq:conservation law}
  \partial_t u + \partial_iF^i = S,
\end{eqnarray}
where $u$ is the state vector, $F^i$ are the components of the flux vector, and
$S$ is the source vector.

We refer the reader to the literature~\cite{2006ApJ...637..296A, Font:2008fka,
  Baumgarte:2010ndz} for a detailed description of the equations of general
relativistic magnetohydrodynamics~(GRMHD).  If we ignore self-gravity, the GRMHD
equations constitute a closed system that may be solved on a given background
metric.  We denote the rest-mass density of the fluid by $\rho$ and its
4-velocity by $u^a$, where $u^a u_a=-1$. The dual of the Faraday tensor $F^{ab}$
is
\begin{eqnarray}
  \label{eq:Faraday dual}
  \dual F^{ab} = \frac{1}{2}\epsilon^{abcd}F_{cd},
\end{eqnarray}
where $\epsilon^{abcd}$ is the Levi-Civita tensor. Note that the
Levi-Civita tensor is defined here with the
convention~\cite{GravitationMTW}
that in flat spacetime $\epsilon_{0123}=+1$. The equations governing the
evolution of the GRMHD system are:
\begin{eqnarray}
  \nabla_a(\rho u^a) &= 0\quad (\textrm{rest-mass conservation}), \\
  \nabla_a T^{ab} &= 0\quad (\textrm{energy-momentum conservation}), \\
  \nabla_a  \dual F^{ab} &= 0\quad (\textrm{homogeneous Maxwell equation}).
\end{eqnarray}
In the ideal MHD limit the stress tensor takes the form
\begin{equation}
  T^{ab} = (\rho h)^*u^a u^b + p^* g^{ab} - b^a b^b
\end{equation}
where
\begin{equation}
  b^a = -\dual F^{ab}u_b
\end{equation}
is the magnetic field measured in the comoving frame of the fluid, and
$(\rho h)^* = \rho h + b^2$ and $p^* = p + b^2/2$ are the enthalpy density and
fluid pressure augmented by contributions of magnetic pressure
$p_{\mathrm{mag}} = b^2/2$, respectively.

We denote the unit normal vector to the spatial hypersurfaces as $n^a$, which is
given by
\begin{eqnarray}
  \label{eq:spacetime unit normal vector}
  n^a =& \left(1/\alpha, -\beta^i/\alpha\right)^T, \\
  \label{eq:spacetime unit normal covector}
  n_a =& (-\alpha, 0, 0, 0).
\end{eqnarray}
The spatial velocity of the fluid as measured by an observer at rest in the
spatial hypersurfaces (``Eulerian observer'') is
\begin{equation}
  \label{eq:spatial velocity}
  v^i = \frac{1}{\alpha}\left(\frac{u^i}{u^0} + \beta^i\right),
\end{equation}
with a corresponding Lorentz factor $W$ given by
\begin{eqnarray}
  \label{eq:Lorentz factor}
  W &= - u^a n_a = \alpha u^0 = \frac{1}{\sqrt{1 - \gamma_{ij}v^i v^j}} \\
  \label{eq:Lorentz factor four velocity}
  &=\sqrt{1+\gamma^{ij}u_iu_j}=\sqrt{1+\gamma^{ij}W^2v_iv_j}.
\end{eqnarray}
The electric and magnetic fields as measured by an Eulerian observer are given
by
\begin{eqnarray}
  \label{eq:Eulerian electric field}
  E^i &= F^{ia}n_a = \alpha F^{0i},\\
  \label{eq:Eulerian magnetic field}
  B^i &= -\dual F^{ia}n_a = -\alpha \dual F^{0i}.
\end{eqnarray}
Finally, the comoving magnetic field $b^a$ in terms of $B^i$ is
\begin{eqnarray}
  \label{eq:b^0}
  b^0 =& \frac{W}{\alpha}B^i v_i, \\
  \label{eq:b^i}
  b^i =& \frac{B^i + \alpha b^0 u^i}{W},
\end{eqnarray}
while $b^2=b^a b_a$ is given by
\begin{equation}
  \label{eq:b^i b_i}
  b^2 = \frac{B^2}{W^2} + (B^i v_i)^2.
\end{equation}

We now recast the GRMHD equations in a 3+1 split by projecting them along and
perpendicular to $n^a$~\cite{2006ApJ...637..296A}.  One of the main
complications when solving the GRMHD equations numerically is preserving the
constraint
\begin{eqnarray}
  \label{eq:monopole constraint}
  \partial_i (\sqrt{\gamma} B^i)=0,
\end{eqnarray}
where $\gamma=\det(\gamma_{ij})$ is the determinant of the spatial metric.
Analytically, initial data evolved using the dynamical Maxwell equations are
guaranteed to preserve the constraint. However, numerical errors generate
constraint violations that need to be controlled. We opt to use the Generalized
Lagrange Multiplier (GLM) or divergence cleaning
method~\cite{2002JCoPh.175..645D} where an additional field $\Phi$ is evolved in
order to propagate constraint violations out of the domain.  Our version is very
close to the one in~\cite{Mosta:2013gwu}.  The augmented system can still be
written in flux-balanced form, where the conserved variables are
\begin{eqnarray}
  u
  &=\sqrt{\gamma}\left(\begin{array}{c}
    D \\
    S_i \\
    \tau \\
    B^i \\
    \Phi
  \end{array}\right)
  =\left(\begin{array}{c}
    \tilde{D} \\
    \tilde{S}_i \\
    \tilde{\tau} \\
    \tilde{B}^i \\
    \tilde{\Phi}
  \end{array}\right) \nonumber \\
  &= \sqrt{\gamma}
    \left(\begin{array}{c}
      \rho W \\
      (\rho h)^* W^2 v_i - \alpha b^0 b_i \\
      (\rho h)^* W^2 - p^* - \left(\alpha b^0\right)^2 - \rho W \\
      B^i \\
      \Phi
    \end{array}\right),
  \label{eq:conserved variables}
\end{eqnarray}
with corresponding fluxes
\begin{eqnarray}
  \label{eq:conserved fluxes}
  F^i =
  \left(\begin{array}{c}
    \tilde{D} v^i_\mathrm{tr} \\
    \tilde{S}_j v^i_\mathrm{tr} + \alpha\sqrt{\gamma} p^* \delta^i_j - \alpha
    b_j \tilde{B}^i/W\\
    \tilde{\tau} v^i_\mathrm{tr} + \alpha\sqrt\gamma p^* v^i - \alpha^2 b^0
    \tilde{B}^i / W\\
    \tilde{B}^j v^i_\mathrm{tr} - \alpha v^j\tilde{B}^i + \alpha
    \gamma^{ij}\tilde{\Phi}\\
    \alpha \tilde{B}^i - \tilde{\Phi} \beta^i
  \end{array}\right),
\end{eqnarray}
and corresponding sources
\begin{eqnarray}
  \label{eq:conserved sources}
  S =
  \left(\begin{array}{c}
    0\\
    (\alpha/2)\tilde{S}^{kl} \partial_i\gamma_{kl}
    + \tilde{S}_k \partial_i \beta^k - \tilde{E}\partial_i \alpha\\
    \alpha \tilde{S}^{kl}K_{kl} - \tilde{S}^k\partial_k \alpha\\
    -\tilde{B}^j \partial_j \beta^i +
    \Phi\partial_k(\alpha\sqrt\gamma\gamma^{ik})\\
    \alpha\tilde{B}^k\partial_k\ln \alpha - \alpha K\tilde{\Phi} -
    \alpha\kappa\tilde{\Phi}
  \end{array}\right).
\end{eqnarray}
The transport velocity is defined as $v_\mathrm{tr}^i = \alpha v^i - \beta^i$
and the generalized energy $\tilde{E}$ and source $\tilde{S}^{ij}$ are given by
\begin{eqnarray}
  \label{eq:generalized energy}
  \tilde{E} &= \tilde{\tau} + \tilde{D},\\
  \label{eq:generalized source}
  \tilde{S}^{ij} &= \sqrt\gamma\left[(\rho h)^*W^2 v^i v^j +
                   p^*\gamma^{ij} -
                   \gamma^{ik}\gamma^{jl}b_kb_l\right].
\end{eqnarray}

\section{The discontinuous Galerkin and conservative finite difference
  methods\label{sec:DG and FD methods}}

We are interested in solving nonlinear hyperbolic conservation laws of the form
\begin{equation}
  \label{eq:dgscl conservation law}
  \partial_a F^a=\partial_t u+\partial_i F^i = S,
\end{equation}
where $u$ are the evolved/conserved variables, $F^i$ are the fluxes, and $S$ are
the source terms.

\subsection{Discontinuous Galerkin method\label{sec:dgscl dg methods}}

In the DG method the computational domain is divided up into non-overlapping
elements or cells, which we denote by $\Omega_k$. This allows us to write the
conservation law~\eref{eq:dgscl conservation law} as a semi-discrete
system, where time remains continuous. In the DG method one integrates the
evolution equations~\eref{eq:dgscl conservation law} against spatial basis
functions of degree $N$, which we denote by $\phi_{\breve{\imath}}$. We index
the basis functions and collocation points of the DG scheme with breve Latin
indices, e.g.~$\breve{\imath}, \breve{\jmath}, \breve{k}$. The basis functions
are defined in the reference coordinates of each element, which we denote by
$\xi^{\hat{\imath}}$.  We use hatted indices to denote tensor components in the
reference frame. The reference coordinates are mapped to the physical
coordinates using the general function
\begin{eqnarray}
  \label{eq:dgscl time independent coordinate map}
  x^i=x^i(\xi^{\hat{\imath}}).
\end{eqnarray}
We will discuss making the mapping time-dependent in \S\ref{sec:moving mesh}
below.

In the DG method we integrate the basis functions against~\eref{eq:dgscl
  conservation law},
\begin{eqnarray}
  \label{eq:dgscl integrated evolution}
  \int_{\Omega_k} d^3x\,\phi_{\breve{\imath}} \left[
  \partial_t u + \partial_i F^i
  - S\right] = 0,
\end{eqnarray}
where repeated indices are implicitly summed over.  Note that we are integrating
over the physical coordinates, not the reference coordinates
$\xi^{\hat{\imath}}$. Following the standard prescription where we integrate by
parts and replace the flux on the boundary $n_i F^i$ with a boundary term $G$ (a
numerical flux dotted into the normal to the surface), we obtain the weak form
\begin{eqnarray}
  \label{eq:dgscl weak form a}
  \int_{\Omega_k} d^3x\,\phi_{\breve{\imath}} \left[
  \partial_t u - S\right]
  - \int_{\Omega_k} d^3x\, F^i \partial_i \phi_{\breve{\imath}}
  + \oint_{\partial\Omega_k}d^2\Sigma\, \phi_{\breve{\imath}} G = 0,
\end{eqnarray}
where $\partial\Omega_k$ is the boundary of the element and $d^2\Sigma$ is the
surface element. Undoing the integration by parts gives us the equivalent strong
form
\begin{eqnarray}
  \label{eq:dgscl strong form a}
  \int_{\Omega_k} d^3x\,\phi_{\breve{\imath}} \left[
  \partial_t u  + \partial_i F^i - S\right]
  + \oint_{\partial\Omega_k}d^2\Sigma\, \phi_{\breve{\imath}} \left(G
  -n_i F^i\right) = 0,
\end{eqnarray}
where $n_i$ is the outward-pointing unit normal covector in the physical
frame. Next, we use a nodal DG method and expand the various terms using the
basis $\phi_{\breve{\imath}}$ as
\begin{eqnarray}
  \label{eq:dgscl basis expansion}
  u=\sum_{\breve{\imath}=0}^N
  u_{\breve{\imath}}\phi_{\breve{\imath}}.
\end{eqnarray}
The weak form can be written as
\begin{eqnarray}
  \label{eq:dgscl weak form}
  \int_{\Omega_k} d^3x\,\phi_{\breve{\imath}}\phi_{\breve{k}} \left[
  \partial_t u_{\breve{k}} - S_{\breve{k}}\right]
  - \int_{\Omega_k} d^3x\, F^i_{\breve{k}}
  \phi_{\breve{k}}\partial_i \phi_{\breve{\imath}}
  + \oint_{\partial\Omega_k}d^2\Sigma\, \phi_{\breve{\imath}} \phi_{\breve{k}}
  G_{\breve{k}} = 0.
\end{eqnarray}
The equivalent strong form is
\begin{eqnarray}
  \label{eq:dgscl strong form}
  \int_{\Omega_k} d^3x\,\phi_{\breve{\imath}}\phi_{\breve{k}} \left[
  \partial_t u_{\breve{k}}  + (\partial_i F^i)_{\breve{k}} -
  S_{\breve{k}}\right] + \oint_{\partial\Omega_k}d^2\Sigma\,
  \phi_{\breve{\imath}} \phi_{\breve{k}}
  \left(G_{} -n_i F^i\right)_{\breve{k}} = 0.
\end{eqnarray}
In the strong form we have expanded $\partial_i F^i$ in the basis, which
might lead to aliasing~\cite{Teukolsky:2015ega}. In practice, we have not
encountered any aliasing-driven instabilities that require filtering.

In order to simplify the scheme, we use a tensor-product basis of 1d Lagrange
interpolating polynomials with Legendre-Gauss-Lobatto collocation points.
We denote this DG scheme with 1d basis functions of degree $N$ by $P_N$. A $P_N$
scheme is expected to converge at order $\mathcal{O}(\Delta x^{N+1})$ for smooth
solutions~\cite{Hesthaven2008}, where $\Delta x$ is the 1d size of the element.
The reference elements are intervals in 1d, squares in 2d, and cubes in 3d,
where each component of the reference coordinates
$\xi^{\hat{\imath}} \in [-1,1]$. We use the map $x^i(\xi^{\hat{\imath}})$ to
deform the squares and cubes into different shapes needed to produce an
efficient covering of the domain. For example, if spherical geometries are
present, we use $x^i(\xi^{\hat{\imath}})$ to create a cubed-sphere domain.

\subsection{Conservative finite-difference methods\label{sec:dgscl fd methods}}

Conservative FD methods evolve the cell-center values, but the cell-face values
(the midpoints along each axis) are necessary for solving the Riemann problem
and computing the FD derivatives of the fluxes. Denoting the numerical flux by
$\hat{F}^i$ and the $k^{\mathrm{th}}$-order FD derivative operator by
$D^{(k)}_{\hat{\imath}}$, we can write the semi-discrete evolution equations as
\begin{eqnarray}
  \partial_t u_{\underline{i}} +
  \left(\frac{\partial \xi^{\hat{\imath}}}{\partial x^i}\right)_{\underline{i}}
  \left(D^{(k)}_{\hat{\imath}} \hat{F}^i\right)_{\underline{i}}
  =S_{\underline{i}},
\label{eq:fdevol}
\end{eqnarray}
where we use underlined indices to label FD cells/grid points.
\Eref{eq:fdevol} can be rewritten to more closely resemble the DG form
since we actually use $G$ as the numerical flux $\hat{F}^i$ on the cell
boundary.  Specifically,
\begin{eqnarray}
  \label{eq:dgscl fd semi-discrete}
  \partial_t u_{\underline{i}} +
  \frac{1}{J_{\underline{i}}}\sum_{\hat{\imath}}
  \left[D_{\hat{\imath}}^{(k)}\left(J
  \sqrt{\frac{\partial\xi^{\hat{\imath}}}{\partial x^i}\gamma^{ij}
  \frac{\partial\xi^{\hat{\imath}}}{\partial x^j}}
  G^{(\hat{\imath})}\right)\right]_{\underline{i}}
  =S_{\underline{i}},
\end{eqnarray}
where $J$ is the determinant of the Jacobian matrix
$\partial x^i/\partial \xi^{\hat{\imath}}$. This form allows our implementation
to reuse as much of the DG Riemann solvers as possible, and also makes
interfacing between the DG and FD methods easier. Ultimately, we use a
flux-difference-splitting scheme, where we reconstruct the primitive variables
to the interfaces between cells. Which reconstruction method we use is stated
for each test problem below.

\subsection{Moving mesh formulation\label{sec:moving mesh}}
Moving the mesh to follow interesting features of the solution can greatly
reduce computational cost. A moving mesh is also essential for evolutions of
binary black holes, one of our target applications, where the interior of the
black holes needs to be excised to avoid the singularities~\cite{Scheel:2006gg,
  Hemberger:2012jz}. Here we present a new form of the moving mesh evolution
equations that is extremely simple to implement and derive. We assume that the
velocity of the mesh is some spatially smooth function, though this assumption
can be removed if one uses the path-conservative methods described
in~\cite{DUMBSER20091731} based on Dal Maso-LeFloch-Murat
theory~\cite{DALMASO1995}. We write the map from the reference coordinates to
the physical coordinates as
\begin{eqnarray}
  \label{eq:dgscl time dependent transformation}
  t=\hat{t}, \;\;\; x^i=x^i(\xi^{\hat{\imath}},\hat{t}).
\end{eqnarray}
The spacetime Jacobian matrix is given by
\begin{eqnarray}
  \label{eq:dgscl time dependent Jacobian matrix}
  \frac{\partial x^a}{\partial \xi^{\hat{a}}}=
  \left(\begin{array}{cc}
    \frac{\partial t}{\partial \hat{t}} %
    & \frac{\partial t}{\partial \xi^{\hat{\imath}}}
    \\
    \frac{\partial x^i}{\partial \hat{t}} %
    & \frac{\partial x^i}{\partial \xi^{\hat{\imath}}}
  \end{array}\right) =
      \left(\begin{array}{cc}
        1 & 0 \\
        v^i_g & \frac{\partial x^i}{\partial \xi^{\hat{\imath}}}
      \end{array}\right),
\end{eqnarray}
where the mesh velocity of the physical frame is defined as
\begin{eqnarray}
  \label{eq:dgscl inertial grid velocity}
  v^i_g=\frac{\partial x^i}{\partial \hat{t}}.
\end{eqnarray}
The inverse spacetime Jacobian matrix is given by
\begin{eqnarray}
  \label{eq:dgscl time dependent inverse Jacobian matrix}
  \frac{\partial \xi^{\hat{a}}}{\partial x^{a}}=
  \left(\begin{array}{cc}
    \frac{\partial \hat{t}}{\partial t} %
    & \frac{\partial \hat{t}}{\partial x^{i}}
    \\
    \frac{\partial \xi^{\hat{\imath}}}{\partial t} %
    & \frac{\partial \xi^{\hat{\imath}}}{\partial x^{i}}
  \end{array}\right)
      =
      \left(\begin{array}{cc}
        1 & 0 \\
        v^{\hat{\imath}}_g %
        & \left(\frac{\partial x^{i}}{\partial \xi^{\hat{\imath}}}\right)^{-1}
      \end{array}\right),
\end{eqnarray}
where the mesh velocity in the reference frame is given by
\begin{eqnarray}
  \label{eq:dgscl reference grid velocity}
  v^{\hat{\imath}}_g \equiv \frac{\partial \xi^{\hat{\imath}}}{\partial t}
  =-\frac{\partial \xi^{\hat{\imath}}}{\partial x^i}v^i_g.
\end{eqnarray}
When composing coordinate maps the velocities combine as:
\begin{eqnarray}
  \label{eq:dgscl moving mesh velocity addition}
  v^i_g=
  \frac{\partial x^i}{\partial \hat{t}}=\frac{\partial x^i}{\partial \tilde{t}}
  +\frac{\partial x^i}{\partial X^{\tilde{\imath}}}
  \frac{\partial X^{\tilde{\imath}}}{\partial \hat{t}},
\end{eqnarray}
where a new intermediate frame with coordinates $\{\tilde{t},
X^{\tilde{\imath}}\}$ is defined and
$X^{\tilde{\imath}}=X^{\tilde{\imath}}\left(\xi^{\hat{\imath}},\hat{t}\right)$.

To obtain the moving mesh evolution equations, we need to transform the time
derivative in~\eref{eq:dgscl conservation law} from being with respect to
$t$ to being with respect to $\hat{t}$. Starting with the chain rule
for $\partial u/\partial \hat t$, we get
\begin{eqnarray}
  \label{eq:dgscl time derivative transform}
  \frac{\partial u}{\partial t}=\frac{\partial u}{\partial
  \hat{t}} -\frac{\partial x^i}{\partial \hat{t}}\partial_i u
  =\partial_{\hat{t}} u
  -\partial_i\left(v^i_g u\right)
  +u \partial_i v^i_g.
\end{eqnarray}
Substituting~\eref{eq:dgscl time derivative transform} into
\eref{eq:dgscl conservation law} we get
\begin{eqnarray}
  \label{eq:dgscl conservation law moving mesh}
  \partial_{\hat{t}} u + \partial_i \left(F^i - v^i_g u\right)
  = S - u \partial_i v^i_g.
\end{eqnarray}
This formulation of the moving mesh equations is simpler than the common ALE
(Arbitrary Lagrangian-Eulerian) formulation~\cite{MINOLI20111876}.

The same DG or FD scheme used to discretize~\eref{eq:dgscl conservation
  law} can be used to discretize~\eref{eq:dgscl conservation law moving
  mesh}. In the case that $v^i_g$ is an evolved variable, the
additional term should be treated as a nonconservative product using the
path-conservative formalism~\cite{DUMBSER20091731}. Finally, we note that the
characteristic fields are unchanged by the mesh movement, but the characteristic
speeds $\lambda$ are changed to
$\lambda\to\lambda - n_i v^i_g$.

\subsection{Time discretization\label{sec:dgscl time discretization}}
We evolve the semi-discrete system (be it the DG or FD discretized system) in
time using a method of lines. We use either a third-order strong-stability
preserving Runge-Kutta method~\cite{Shu1988439} or a forward self-starting
Adams-Bashforth time stepper~\cite{doi:10.1137/19M1292692, mersman1965self}.
Which method is used will be noted for each test case.

The DG method has a rather restrictive Courant-Friedrichs-Lewy (CFL) condition
that decreases as the polynomial degree $N$ of the basis is increased. The CFL
number scales roughly as $1/(2N + 1)$~\cite{cockburn2000development,
  cockburn2001runge}, which can be understood as a growth in the spectrum of the
spatial discretization operator~\cite{KRIVODONOVA20131}. For a DG discretization
in $d$ spatial dimensions, the time step $\Delta t$ must satisfy
\begin{eqnarray}
  \label{eq:dgscl time step size}
  \Delta t \le \frac{1}{d(2N + 1)}\frac{h}{|\lambda_{\max}|},
\end{eqnarray}
where $h$ is the characteristic size of the element and $\lambda_{\max}$ is the
maximum characteristic speed of the system being evolved. For comparison, FV and
FD schemes have a time step restriction of
\begin{eqnarray}
  \label{eq:dgscl time step size FD}
  \Delta t \le \frac{1}{d}\frac{h}{|\lambda_{\max}|},
\end{eqnarray}
where $h$ is the characteristic size of the FV or FD cell. \blue{However, a DG
  element has $N+1$ grid points per dimension, while FV or FD cells only have
  one, and so the CFL condition for DG is partly offset by the increase in order
  that the algorithm provides.}

\section{Limiting in the DG method\label{sec:limiting within DG}}
In this section we give an overview of what we require from a DG limiter,
followed by a brief discussion of existing limiters in the literature and which
of our requirements they meet.

\subsection{Requirements\label{sec:dgscl requirements}}
We have several requirements that, when combined, are very stringent. However,
we view these as necessary for DG to live up to the promise of a high-order
shock-capturing method. In no particular order, we require that
\begin{requirements}
  \label{req:dgscl all}
\item smooth solutions are resolved, i.e., smooth extrema are not
  flattened,\label{req:dgscl smooth}
\item unphysical oscillations are removed,\label{req:dgscl remove oscillations}
\item physical realizability of the solution is guaranteed,\label{req:dgscl
    physical}
\item sub-cell or sub-element resolution is possible, i.e., discontinuities are
  resolved inside the element, not just at boundaries,\label{req:dgscl sub-cell}
\item curved hexahedral elements are supported,\label{req:dgscl curved elements}
\item slow-moving shocks are resolved,\label{req:dgscl slow shocks}
\item moving meshes are supported,\label{req:dgscl moving mesh}
\item higher than fourth-order DG can be used.\label{req:dgscl high order}
\end{requirements}
Requirement~\ref{req:dgscl sub-cell} is necessary to justify the restrictive
time step size,~\eref{eq:dgscl time step size}. That is, if discontinuities
are only resolved at the boundaries of elements, the DG scheme results in
excessive smearing. In such a scenario it becomes difficult to argue for using
DG over FV or FD methods. While in principle it is possible to use adaptive mesh
refinement or $hp$-adaptivity to switch to low-order DG at discontinuities,
effectively switching to a low-order FV method, we are unaware of
implementations that are capable of doing so for high-order DG.

We note that achieving higher-than-fourth order is especially challenging with
many of the existing limiters. Since FV and FD methods of fourth or higher order
are becoming more common, we view high order as being crucial for DG to be
competitive with existing FV and FD methods, especially given the restrictive
time step size.

\subsection{Overview of existing DG limiters\label{sec:dgscl existing limiters}}
Aside from the FV subcell limiters~\cite{doi:10.1002/fld.2654,
  10.1007/978-3-319-05591-6_96, Dumbser2014a}, DG limiters operate on the
solution after a time step or substep is taken so as
to remove spurious oscillations and
sometimes also to correct unphysical values. This is generally achieved by some
nonlinear reconstruction using the solution in neighboring elements. How exactly
this reconstruction is done depends on the specific limiters, but
all limiters involve two general steps:
\begin{enumerate}
\item detecting whether or not the solution in the element is ``bad''
  (troubled-cell indicators),
\item correcting the degrees of freedom/solution in the element.
\end{enumerate}
A good troubled-cell indicator (TCI) avoids triggering the limiter where the
solution is smooth while still preventing spurious unphysical
oscillations. Unfortunately, making this statement mathematically rigorous is
challenging and the last word is yet to be written on which TCIs are the
best. Since the TCI may trigger in smooth regions, ideally the limiting
procedure does not flatten local extrema when applied in such regions.  In a
companion paper~\cite{Deppe:2021DgLimiterComparison} we have experimented with
the (admittedly quite dated but very robust) minmod family of
limiters~\cite{COCKBURN198990,1990MaCom..54..545C,1998JCoPh.141..199C}, the
hierarchical limiter of Krivodonova~\cite{Krivodonova2004,Krivodonova:2007}, the
simple WENO limiter~\cite{2013JCoPh.232..397Z}, and the Hermite WENO (HWENO)
limiter~\cite{2016CCoPh..19..944Z}. While this does not include every limiter
applicable to structured meshes, it covers the common ones. We will discuss each
limiter in turn, reporting what we have found to be good and bad.

The minmod family of
limiters~\cite{COCKBURN198990,1990MaCom..54..545C,1998JCoPh.141..199C} linearize
the solution and decrease the slope if the slope is deemed to be too large. This
means that the minmod limiters quickly flatten local extrema in smooth regions,
do not provide sub-element resolution, and are not higher-than-fourth
order. While they are extremely robust and tend to do a good job of maintaining
physical realizability of the solution despite not guaranteeing it, the minmod
limiters are simply too aggressive and low-order to make DG an attractive
replacement for shock-capturing FD methods. Furthermore, generalizing the minmod
limiters to curved elements in the na\"ive manner makes them very quickly
destroy any symmetries of the domain decomposition and solution. Overall, we
find that the minmod limiters satisfy only Requirements~\ref{req:dgscl remove
  oscillations},~\ref{req:dgscl slow shocks}, and~\ref{req:dgscl moving mesh}.

The hierarchical limiter of Krivodonova~\cite{Krivodonova2004,Krivodonova:2007}
works by limiting the coefficients of the solution's modal representation,
starting with the highest coefficient then decreasing in order until no more
limiting is necessary. We find that in 1d the Krivodonova limiter
works quite well, even using fourth-order elements. However, in 2d and 3d and
for increasingly complex physical systems, the limiter fails. Furthermore,
it is nontrivial to extend to curved elements since comparing modal coefficients
assumes the Jacobian matrix of the map $x^i(\xi^{\hat{\imath}})$ is spatially
uniform. The Krivodonova limiter satisfies Requirements~\ref{req:dgscl
  smooth},~\ref{req:dgscl slow shocks}, and~\ref{req:dgscl moving mesh}. We find
that how well the Krivodonova limiter works at removing unphysical oscillations
depends on the physical system being studied.

The simple WENO~\cite{2013JCoPh.232..397Z} and the
HWENO~\cite{2016CCoPh..19..944Z} limiters are quite similar to each other. When
limiting is needed, these limiters combine the element's solution with a set of
solution estimates obtained from the neighboring elements' solutions. An
oscillation indicator is applied on each solution estimate to determine the
convex nonlinear weights for the reconstruction. Overall, the WENO limiters are,
by design, very similar to WENO reconstruction used in FV and FD methods. We
have found that the WENO limiters are generally robust for second- and
third-order DG, but start producing unphysical solutions at higher orders. The
WENO limiters satisfy our Requirements~\ref{req:dgscl smooth},
~\ref{req:dgscl remove oscillations},~\ref{req:dgscl slow shocks},
and~\ref{req:dgscl moving mesh}. When supplemented with a positivity-preserving
limiter~\cite{WANG2012653}, the WENO schemes are also able to satisfy
Requirement~\ref{req:dgscl physical}.

In short, none of the above limiters satisfy even half of our
Requirements~\ref{req:dgscl all}. Furthermore, they all have parameters that
need to be tuned for them to work well on different problems. This is
unacceptable in realistic astrophysics simulations, where a large variety of
complex fluid interactions are occurring simultaneously in different parts of
the computational domain, and it is impossible to tune parameters such that all
fluid interactions are resolved.

The subcell limiters~\cite{doi:10.1002/fld.2654, 10.1007/978-3-319-05591-6_96,
  Dumbser2014a} are much more promising and we will extend them to meet
\textit{all} the Requirements~\ref{req:dgscl all}. We will focus on the scheme
proposed in~\cite{Dumbser2014a} since it satisfies most of
Requirements~\ref{req:dgscl all}. The basic idea behind the DG-subcell scheme is
to switch to FV or, as proposed here, FD if the high-order DG solution is
inadmissible, either because of excessive oscillations or violation of physical
requirements on the solution. This idea was first presented
in~\cite{COSTA2007970}, where a spectral scheme was hybridized with a WENO
scheme. In~\cite{doi:10.1002/fld.2654, 10.1007/978-3-319-05591-6_96} the
decision whether to switch to a FV scheme is made before a time step is
taken. In contrast, the scheme presented in~\cite{Dumbser2014a} undoes the time
step \blue{(or substep if using a Runge-Kutta substep method)} and switches to
a FV scheme. The advantage of undoing the time (sub) step is that
physical realizability of the solution can be guaranteed as long as the FV or FD
scheme guarantees physical realizability. The scheme of~\cite{Dumbser2014a} is
often referred to as an \textit{a posteriori} limiting approach, where the time
step is redone using the more robust method. Given a TCI that does not allow
unphysical oscillations and a high-order positivity-preserving FV/FD method, the
subcell limiters as presented in the literature meet all Requirements
except~\ref{req:dgscl curved elements} (curved hexahedral
elements),~\ref{req:dgscl slow shocks} (slow-moving shocks), and~\ref{req:dgscl
  moving mesh} (moving mesh), limitations that we will address below. The key
feature that makes the DG-subcell scheme a very promising candidate for a
generic, robust, and high-order method is that the limiting is not based on
polynomial behavior alone but considers the \textit{physics of the problem}. By
switching to a low-order method to guarantee physical realizability, the
DG-subcell scheme guarantees that the resulting numerical solution satisfies the
governing equations, even if only at a low order locally in space and time.
Moreover, the DG-subcell scheme can guarantee that unphysical solutions such as
negative densities never appear.

\section{Discontinuous Galerkin-finite difference hybrid
  method\label{sec:dgscl DG-FD hybrid}}
In this section we present our DG-FD hybrid scheme. The method is designed
specifically to address \textit{all} Requirements~\ref{req:dgscl all}, and means
in particular that the method is a robust high-order shock-capturing method.  We
first discuss how to switch between the DG and FD grids. Then we explain how
neighboring elements communicate flux information if one element is using DG
while the other is using FD. Next we review the \textit{a posteriori} idea and
discuss the TCIs we use, when we apply them, and how we handle communication
between elements. Finally, we discuss the number of subcells to use and provide
a new perspective on the DG-FD hybrid scheme that makes the attractiveness of
such a scheme clear. In~\ref{sec:dgscl curved moving mesh} we provide an example
of how curved hexahedral elements can be handled.

\subsection{Projection and reconstruction between DG and FD
  grids\label{sec:dgscl dg fd projection}}
We will denote the solution on the DG grid by $u_{\breve{\imath}}$ and the
solution on the FD grid by $u_{\underline{i}}$. We need to determine how
to project the solution from the DG grid to the FD grid and how to reconstruct
the DG solution from the FD solution. For simplicity, we assume an isotropic
number of DG collocation points $(N+1)^d$ and FD cells $(N_s)^d$. Since FD
schemes evolve the solution value at the cell-center, one method of projecting
the DG solution to the FD grid is to use interpolation. However, interpolation
is not conservative and so we opt for an $L_2$ projection\blue{, which is conservative
if projecting to a grid with equal or more degrees of freedom. That is, we
assume that $N_s \ge N+1$.} The $L_2$ projection minimizes the integral
\begin{eqnarray}
  \label{eq:dgscl fd subcell projection minimization}
  \int_{-1}^{1}\left(u-\underline{u}\right)^2\,dx =
  \int_{-1}^{1}\left(u-\underline{u}\right)^2J\,d\xi
\end{eqnarray}
with respect to $\underline{u}$, where $\underline{u}$ is the solution on the FD
subcells.  While we derive the projection matrix in 1d, generalizing to 2d and
3d is straightforward for our tensor product basis. Substituting the nodal basis
expansion into~\eref{eq:dgscl fd subcell projection minimization} we obtain
\begin{eqnarray}
  \label{eq:dgscl fd subcell projection minimization basis}
  \int_{-1}^{1}\left[u_{\breve{\imath}}\ell_{\breve{\imath}}(\xi)
    u_{\breve{\jmath}}\ell_{\breve{\jmath}}(\xi)
    +u_{\underline{i}}\ell_{\underline{i}}(\xi)
    u_{\underline{j}}\ell_{\underline{j}}(\xi)
    -2u_{\underline{i}}\ell_{\underline{i}}(\xi)
    u_{\breve{\imath}}\ell_{\breve{\imath}}(\xi)
    \right]J\,d\xi,
\end{eqnarray}
where $\ell_{\underline{j}}(\xi)$ are the Lagrange interpolating polynomials on
the subcells
(i.e.~$\ell_{\underline{j}}(\xi_{\underline{i}})=\delta_{\underline{j}\underline{i}}$).
Varying~\eref{eq:dgscl fd subcell projection minimization basis}
with respect to the coefficients $u_{\underline{i}}$ and setting the result
equal to zero we get
\begin{eqnarray}
  \label{eq:dgscl fd subcell projection minimization basis with variation}
  \int_{-1}^{1}\left[
  u_{\underline{j}}\ell_{\underline{i}}(\xi)\ell_{\underline{j}}(\xi)
  -u_{\breve{\imath}}\ell_{\underline{i}}(\xi)\ell_{\breve{\imath}}(\xi)
  \right]\delta u_{\underline{i}}J\,d\xi=0.
\end{eqnarray}
Since~\eref{eq:dgscl fd subcell projection minimization basis with variation}
must be true for all variations $\delta u_{\underline{i}}$ we see that
\begin{eqnarray}
  \label{eq:dgscl fd subcell projection minimization basis without variation}
  \int_{-1}^{1}\left[
  u_{\underline{j}}\ell_{\underline{i}}(\xi)\ell_{\underline{j}}(\xi)
  -u_{\breve{\imath}}\ell_{\underline{i}}(\xi)\ell_{\breve{\imath}}(\xi)
  \right]J\,d\xi=0.
\end{eqnarray}
By expanding the determinant of the Jacobian on the basis we can
simplify~\eref{eq:dgscl fd subcell projection minimization basis without
  variation} to get
\begin{eqnarray}
  \label{eq:dgscl fd subcell projection integral form}
  u_{\underline{i}} J_{\underline{i}}
  \int_{-1}^{1}\ell_{\underline{i}}(\xi)\ell_{\underline{j}}(\xi)\,d\xi
  =u_{ \breve{\imath}}J_{\breve{\imath}}
  \int_{-1}^{1}\ell_{\breve{\imath}}(\xi)\ell_{\underline{j}}(\xi)\,d\xi.
\end{eqnarray}
Note that expanding $u J$ on the basis instead of $u$ creates some
decrease in accuracy and can cause aliasing if $u J$ is not fully
resolved by the basis functions. However, this procedure
allows us to cache the projection
matrices to make the method more efficient. Furthermore, expanding the Jacobian
on the basis means interpolation and projection are equal when $N_s\ge N+1$. We
solve for $u_{\underline{i}} J_{\underline{i}}$ in~\eref{eq:dgscl fd
  subcell projection integral form} by inverting the matrix
$\int_{-1}^{1}\ell_{\underline{i}}(\xi)\ell_{\underline{j}}(\xi)\,d\xi$ and find
that
\begin{eqnarray}
  \label{eq:dgscl fd subcell projection}
  u_{ \underline{i}} J_{\underline{i}}
  &= \left(\int_{-1}^{1}\ell_{\underline{i}}(\xi)\ell_{\underline{j}}(\xi)
    \,d\xi\right)^{-1}
    \int_{-1}^{1}\ell_{\breve{l}}(\xi)\ell_{\underline{j}}(\xi)\,d\xi
    u_{ \breve{l}} J_{\breve{l}} \nonumber \\
  &= \ell_{\breve{l}}(\xi_{\underline{i}})
    u_{ \breve{l}} J_{\breve{l}} =
    \mathcal{P}_{\underline{i}\breve{l}} u_{\breve{l}} J_{\breve{l}},
\end{eqnarray}
where $\mathcal{P}_{\underline{i}\breve{l}}$ is the $L_2$ projection matrix.

Reconstructing the DG solution from the FD solution is a bit more
involved. Denoting the projection operator by $\mathcal{P}$ and the
reconstruction operator by $\mathcal{R}$, we desire the property
\begin{eqnarray}
  \label{eq:dgscl reconstruction is pseudo-inverse}
  \mathcal{R}(\mathcal{P}(u_{\breve{\imath}}
J_{\breve{\imath}}))=u_{\breve{\imath}} J_{\breve{\imath}}.
\end{eqnarray}
We also
require that the integral of the conserved variables over the subcells is equal
to the integral over the DG element. That is,
\begin{eqnarray}
  \label{eq:fd subcell constraint}
  \int_{\Omega}u \,d^3x
  =\int_{\Omega} \underline{u} \,d^3x \Longrightarrow
  \int_{\Omega}u J \,d^3\xi
  =\int_{\Omega} \underline{u} J \,d^3\xi.
\end{eqnarray}
Since $N_s\ge N+1$ we need to solve a constrained linear least squares
problem.

We will denote the weights used to numerically evaluate the integral over the
subcells by $R_{\underline{i}}$ and the weights for the integral over the DG
element by $w_l$. To find the reconstruction operator we need to solve the
system
\begin{eqnarray}
  \label{eq:subcell fd unconstrained reconstruction system}
  \sum_{\breve{l}} \mathcal{P}_{\underline{i}\breve{l}} u_{  \breve{l}}
  J_{\breve{l}}=
  &u_{\underline{i}} J_{\underline{i}},
\end{eqnarray}
subject to the constraint
\begin{eqnarray}
  \label{eq:subcell constraint}
  \sum_{\breve{l}} w_{\breve{l}} u_{ \breve{l}} J_{\breve{l}} =
  &\sum_{\underline{i}} R_{\underline{i}} u_{\underline{i}}
    J_{\underline{i}}.
\end{eqnarray}
We do so by using the method of Lagrange multipliers. Denoting the Lagrange
multiplier by $\lambda$, we must minimize the functional
\begin{eqnarray}
  \label{eq:dgscl Lagrange multiplier functional}
  f=
  \left(\mathcal{P}_{\underline{i}\breve{l}} u_{ \breve{l}} J_{\breve{l}}
  - u_{\underline{i}} J_{\underline{i}}\right)
  \left(\mathcal{P}_{\underline{i}\breve{\jmath}} u_{ \breve{\jmath}}
  J_{\breve{\jmath}}
  - u_{\underline{i}} J_{\underline{i}}\right)
  - \lambda \left(w_{\breve{l}} u_{ \breve{l}} J_{\breve{l}} -
  R_{\underline{i}}u_{ \underline{i}}J_{\underline{i}}\right)
\end{eqnarray}
with respect to $u_{ \breve{l}} J_{\breve{l}}$ and $\lambda$. Doing so we
obtain the Euler-Lagrange equations
\begin{eqnarray}
  \label{eq:fd subcell constrained least squares}
  \left(\begin{array}{cc}
    2\mathcal{P}_{\underline{i} \breve{l}}\mathcal{P}_{\underline{i}
      \breve{\jmath}} & -w_{\breve{l}} \\
    w_{\breve{l}} \delta_{\breve{l}\breve{\jmath}} & 0
  \end{array}\right)\left(\begin{array}{c}
    u_{\breve{\jmath}} J_{\breve{\jmath}} \\
    \lambda
  \end{array}\right)
  =
  \left(\begin{array}{c}
    2\mathcal{P}_{\underline{i} \breve{l}} \\
    R_{\underline{i}}
  \end{array}\right)
  \left(\begin{array}{c}
    u_{\underline{i}} J_{\underline{i}}
  \end{array}\right).
\end{eqnarray}
Inverting the matrix on the left side of~\eref{eq:fd subcell constrained
  least squares}, we obtain
\begin{eqnarray}
  \label{eq:fd subcell constrained least squares inverted}
  \left(\begin{array}{c}
    u_{ \breve{\jmath}} J_{\breve{\jmath}} \\
    \lambda
  \end{array}\right)
  =
  \left(\begin{array}{cc}
    2\mathcal{P}_{\underline{i} \breve{l}}\mathcal{P}_{\underline{i}
      \breve{\jmath}} & -w_{\breve{l}} \\
    w_{\breve{l}} \delta_{\breve{l}\breve{\jmath}} & 0
  \end{array}\right)^{-1}
  \left(\begin{array}{c}
    2\mathcal{P}_{\underline{i} \breve{l}} \\
    R_{\underline{i}}
  \end{array}\right)
  \left(\begin{array}{c}
    u_{\underline{i}} J_{\underline{i}}
  \end{array}\right).
\end{eqnarray}
To make the notation less cumbersome we suppress indices by writing
$w_{\breve{l}}$ as $\vec{w}$ and $w_{\breve{l}}\delta_{\breve{l}\breve{\jmath}}$
as $\mathbf{w}$.  Treating the matrix as a partitioned matrix, we invert it to
find
\begin{eqnarray}
  \label{eq:dgscl reconstruction matrix first part inverse}
  \left(\begin{array}{cc}
    2\mathcal{P}\mathcal{P} & -\vec{w} \\
    \mathbf{w} & 0
  \end{array}\right)^{-1}
  =
  \left(\begin{array}{cc}
    \Pi - \Pi
    \vec{w}
      \mathcal{W}\mathbf{w}\Pi
     &
     \Pi\vec{w}
      \mathcal{W} \\
    -\mathcal{W}
    \mathbf{w}\Pi
    & \mathcal{W}
  \end{array}\right).
\end{eqnarray}
Here we have defined
\begin{equation}
\Pi=(2 \mathcal{P}\mathcal{P})^{-1},\qquad
\mathcal{W} = \left[\mathbf{w}(2 \mathcal{P}\mathcal{P})^{-1}\vec{w}\right]^{-1}
\end{equation}
Substituting~\eref{eq:dgscl reconstruction matrix first part inverse}
into~\eref{eq:fd subcell constrained least squares inverted} and performing the
matrix multiplication we get
\begin{eqnarray}
  \label{eq:dgscl full reconstruction equation}
  \left(\begin{array}{c}
    u_{ \breve{\jmath}} J_{\breve{\jmath}} \\
    \lambda
  \end{array}\right)
  =
  \left(\begin{array}{c}
    \Pi 2\mathcal{P} -
    \Pi \vec{w}
      \mathcal{W}\mathbf{w}\Pi
    2\mathcal{P}
    +\Pi\vec{w}
      \mathcal{W}\vec{R} \\
    -\mathcal{W}
    \mathbf{w}\Pi 2\mathcal{P}
    + \mathcal{W}\vec{R}
  \end{array}\right)_{\breve{\jmath}\underline{i}}
  u_{\underline{i}} J_{\underline{i}},
\end{eqnarray}
where $\vec{R}$ is short for $R_{\underline{i}}$. We can see that the first row
of~\eref{eq:dgscl full reconstruction equation} gives
\begin{eqnarray}
  \label{eq:dgscl reconstruction multiplied out}
  u_{ \breve{\jmath}} J_{\breve{\jmath}}
  =\left\{\Pi 2\mathcal{P} -
  \Pi\vec{w}
  \mathcal{W}\mathbf{w}
  \Pi 2\mathcal{P}
  + \Pi \vec{w}
  \mathcal{W}\vec{R}
  \right\}_{\breve{\jmath}\underline{i}}
  u_{\underline{i}} J_{\underline{i}},
\end{eqnarray}
and so the reconstruction matrix used to obtain the DG solution from the FD
solution is given by
\begin{eqnarray}
  \label{eq:dgscl reconstruction matrix}
  R_{\breve{\jmath}\underline{i}}
  =\left\{\Pi 2\mathcal{P} -
  \Pi\vec{w}
  \mathcal{W}\mathbf{w}
  \Pi 2\mathcal{P}
  + \Pi\vec{w}
  \mathcal{W}\vec{R}
  \right\}_{\breve{\jmath}\underline{i}}.
\end{eqnarray}

To show that the reconstruction matrix~\eref{eq:dgscl reconstruction matrix}
satisfies~\eref{eq:dgscl reconstruction is pseudo-inverse} we start by
substituting~\eref{eq:dgscl reconstruction matrix} into~\eref{eq:dgscl
  reconstruction is pseudo-inverse}:
\begin{eqnarray}
  \label{eq:dgscl proof reconstruction is inverse}
  &\mathcal{R}\mathcal{P}uJ \nonumber \\
  &= \left\{\Pi 2\mathcal{P} -
    \Pi\vec{w}
    \mathcal{W}\mathbf{w}
    \Pi 2\mathcal{P}
    + \Pi\vec{w}
    \mathcal{W}\vec{R}
    \right\}\mathcal{P}uJ \nonumber \\
  &= \left\{\mathbb{1} -
    \Pi\vec{w}
    \mathcal{W}\mathbf{w}
    + \Pi\vec{w}
    \mathcal{W}\vec{R}\mathcal{P}
    \right\}uJ \nonumber \\
  &= \left\{\mathbb{1} - \Pi
     \vec{w}
     \mathcal{W}\mathbf{w}
     + \Pi\vec{w}
     \mathcal{W}\mathbf{w}
     \right\}uJ \nonumber \\
  &= uJ,
\end{eqnarray}
where we used the constraint
$\mathbf{w}u J=\vec{R}\mathcal{P}u J$. Thus, the matrix given
in~\eref{eq:dgscl reconstruction matrix} is the reconstruction matrix for
obtaining the DG solution from the FD solution on the subcells and is the
pseudo-inverse of the projection matrix. Note that since the reconstruction
matrices also only depend on the reference coordinates, they can be precomputed
for all elements and cached.

We now turn to deriving the integration weights $R_{\underline{i}}$ on the
subcells. One simple option is using the \textit{extended midpoint rule}:
\begin{eqnarray}
  \label{eq:extended midpoint rule}
  \int_{\Omega}\underline{u}\,d^3x\approx \Delta \xi\Delta \eta\Delta
  \zeta \sum_{\underline{i}}\underline{u}_{\underline{i}}
  J_{\underline{i}},
\end{eqnarray}
which means $R_{\underline{i}}=\Delta \xi\Delta \eta\Delta \zeta$.  However,
this formula is only second-order accurate. To obtain a higher-order
approximation, we need to find weights $R_{\underline{i}}$ that approximate the
integral
\begin{equation*}
  \int_a^b f(x)\,dx\approx\sum_{\underline{i}=0}^n R_{\underline{i}}
  f(x_{\underline{i}}).
\end{equation*}
We provide the weights $R_{\underline{i}}$ in \ref{sec:fd integration weights}.

\subsection{Intercell fluxes\label{sec:dgscl intercell fluxes}}
One approach to dealing with the intercell fluxes is to use the mortar
method~\cite{1989ddcm.proc..392M,KOPRIVA1996475,Kopriva2002,Thanh2012}. In the
mortar method, the boundary correction terms and numerical fluxes are computed
on a new mesh whose resolution is the greater of the two elements sharing the
boundary. In practice, we have found this not to be necessary to achieve a
stable scheme. This can be understood by noting that from a shock capturing
perspective, violating conservation is only an issue at
discontinuities. Wherever the solution is smooth, conservation violations
converge away. Since the hybrid scheme switches from DG to FD \textit{before} a
shock enters an element by retaking the time (sub) step, and since
discontinuities are inevitably always somewhat smeared in any shock capturing
scheme, we have found that exact conservation is not required between a DG and
FD grid. \blue{The lack of conservation arises from reconstructing the FD
  variables to the DG element's interface before computing $G$, rather than
  computing $G$ on the FD cell faces and then reconstructing $G$. Note that not
  enforcing exact conservation at boundaries is merely an implementation
  convenience.}

First, let us describe the element using FD. In this case, the neighbor input
data to the boundary correction from the DG grid is projected onto the FD grid
on the interface. Then the Riemann solver computes the boundary correction $G$,
which is then used in the FD scheme. On the DG grid the FD scheme is used to
reconstruct the neighboring data on the common interface from the subcell
data. The reconstructed FD data is then reconstructed to the DG grid, that is,
it is transferred from the FD to the DG grid on the interface. Finally, the
boundary correction is computed on the DG grid. It is the reordering of the
reconstruction and projection with the Riemann solver that violates conservation
at the truncation error level. Note that the DG and FD solvers must use the same
Riemann solver.

\subsection{The a posteriori idea\label{sec:dgscl a posteriori}}
In this section we will discuss how the \textit{a posteriori} idea is
implemented. For now, we will not concern ourselves with which TCI is used, just
that one is used to detect troubled cells. \blue{We have several criteria that
  drive the design decision. Specifically,
\begin{itemize}
\item only one communication between nearest neighbors is necessary per time
  (sub) step;
\item switching between DG and FD does not require additional communication and
  neighbor information;
\item exact conservation between neighboring elements can be enforced;
\item both substep (Runge-Kutta) and multi-step (Adams-Bashforth) time
  integrators are supported;
\item physical realizability of the solution can be guaranteed.
\end{itemize}
We present a schematic of our DG-FD hybrid scheme in figure~\ref{fig:dgscl
    schematic}. The schematic has the unlimited DG loop on the left and the
  positivity-preserving FD loop on the right. Between them are the projection
  and reconstruction operations that allow the two schemes to work together and
  communicate data back and forth. The scheme starts in the ``Unlimited DG
  Loop'' in the top left with a computation of the volume candidate. If the TCI
  finds the solution admissible the ``Passed'' branch is taken, otherwise the
  ``Failed'' branch is taken.}

\begin{figure}
  \begin{center}
    \begin{tikzpicture}[scale=0.85, every node/.style={transform shape}]
      \tikzstyle{AlgorithmStep}=[rectangle, draw=black, rounded corners,
      fill=white, drop shadow, text centered, anchor=center, text=black, text
      width=105pt] \tikzstyle{PassFail}=[rectangle, text centered,
      anchor=center, text=black]

      \draw[thick,-{Latex[length=3mm]}] (-3, -7.125) -- (-3, -8.35);
      \node[AlgorithmStep, dashed, fill=lightgray] at (-3, -7.125)
      {Exchange ghost cells and fluxes};

      \draw[thick,-{Latex[length=3mm]}] (-3, -8.75) -- (-3, -9.6);
      \node[AlgorithmStep] at (-3, -8.75) {Compute
        $u^{\star,n+1}_{\breve{\imath}}$};

      \draw[thick,-{Latex[length=3mm]}] (-2., -10.375) --
      (-2, -11.2) -- (0.25, -11.2);
      \draw[thick,-{Latex[length=3mm]}] (-3, -10.375) -- (-3, -13.45);

      \node[PassFail, color=Maroon] at (-1.3, -11.0) {Failed};
      \node[PassFail, color=OliveGreen] at (-3.75, -11.0) {Passed};

      \node[AlgorithmStep] at (-3, -10.0) {
        $\mathrm{TCI}\left(u^{\star,n+1}_{\breve{\imath}}\right)$};

      \draw[thick,-{Latex[length=3mm]}] (-3, -13.8) -- (-3, -15.5)
      -- (-5.25, -15.5) -- (-5.25, -6.1) -- (-3, -6.1) -- (-3, -6.6);

      \node[AlgorithmStep] at (-3, -13.8) {
        $u^{n+1}_{\breve{\imath}}
        =u^{\star,n+1}_{\breve{\imath}}$};


      \draw[thick,-{Latex[length=3mm]}] (2.25, -11.2) --
      (7.0, -11.2) -- (7.0, -12.2);

      \node[AlgorithmStep] at (2.25, -11.3)
      {$\mathcal{P}\left(u^{n}_{\breve{\imath}}\right)$,
        $\mathcal{P}\left(F_{\breve{\imath}}^{i,n}\right)$,
        $\mathcal{P}\left(S_{\breve{\imath}}^{n}\right)$};

      \draw[thick,-] (2.25, -16.875) -- (-5.25, -16.875) -- (-5.25, -15.5);

      \node[AlgorithmStep] at (2.25, -16.875)
      {$\mathcal{R}\left(u^{n+1}_{\underline{i}}\right)$};


      \draw[thick,-{Latex[length=3mm]}] (7.5, -7.125) -- (7.5, -12.2);
      \node[AlgorithmStep, dashed, fill=lightgray] at (7.5, -7.125) {Exchange
        ghost cells};

      \draw[thick,-{Latex[length=3mm]}] (7.5, -12.5) -- (7.5, -13.4);
      \node[AlgorithmStep] at (7.5, -12.5) {FD reconstruction};

      \draw[thick,-{Latex[length=3mm]}] (7.5, -13.8) -- (7.5, -14.85);
      \node[AlgorithmStep] at (7.5, -13.8) {Compute
        $u^{n+1}_{\underline{i}}$};

      \draw[thick,-{Latex[length=3mm]}] (8.5, -15.25) -- (8.5, -16.875) --
      (9.8, -16.875) -- (9.8, -6.1) -- (7.5, -6.1) -- (7.5, -6.6);

      \draw[thick,-{Latex[length=3mm]}] (7.5, -15.25) -- (7.5, -16.875) --
      (4.25, -16.875);

      \node[PassFail, color=Maroon] at (9.2, -16.65) {Failed}; \node[PassFail,
      color=OliveGreen] at (6.75, -16.65) {Passed};

      \node[AlgorithmStep] at (7.5, -15.25) {
        $\mathrm{TCI}\left(u^{n+1}_{\underline{i}}\right)$};

      \draw[blue, very thick, dotted] (5.075, -5.0) rectangle (10.1, -17.5);
      \draw[blue, very thick, dashed] (-5.45, -5.0) rectangle (-0.575, -17.5);
      \draw[blue, very thick, dash dot] (-0.175,-5.0) rectangle (4.675,-17.5);

      \node[anchor=north,align=center] at (-3, -5.)  {\textbf{Unlimited DG
          Loop}}; \node[anchor=north,text width=110pt,align=center] at (2.25,
      -5.)  {\textbf{Projection and Reconstruction}};
      \node[anchor=north,align=center] at (7.5, -5.)  {\textbf{FD Loop}};
    \end{tikzpicture}
  \end{center}
  \caption{A schematic description of the proposed DG-FD hybrid method. We use
    superscripts $n$ and $n+1$ to denote variables at time $t^{n}$ and
    $t^{n+1}$. The unlimited DG loop, projection to and reconstructions from the
    FD subcells, and the FD loop are boxed to highlight how the hybrid scheme
    can be split into the unlimited DG and FD schemes with \blue{transformations
      (projection and reconstruction) that allow switching between the two
      methods. Steps that exchange data with neighboring elements are
      highlighted in light gray and have a dashed border. Specifically, these
      are ``Exchange ghost cells and fluxes'' and ``Exchange ghost
      cells''.}\label{fig:dgscl schematic}}
\end{figure}

The algorithm proceeds as follows. We first compute a candidate solution
$u^{\star}(t^{n+1})$ at time $t^{n+1}$ using an unlimited DG scheme. The TCI is
then used to check whether or not the candidate solution $u^{\star}(t^{n+1})$ is
admissible. The TCI may depend on the candidate solution, the solution at the
current time $u(t^{n})$ within the element, and the solution in neighboring
elements at time $t^n$. In order to minimize communication between elements, the
TCI may not depend on the candidate solution in neighboring elements. If the
candidate solution is found to be admissible by the TCI, we use it as the
solution at $t^{n+1}$. That is, $u(t^{n+1})=u^{\star}(t^{n+1})$. If the
candidate solution is inadmissible, then we redo the time step using the FD
subcells. In this case, the solution at $t^n$ \blue{and the time stepper history (the
time derivatives $\partial_t u(t^{n-1})$, etc.) are} projected onto the subcells, FD
reconstruction is performed, data for the boundary correction/Riemann solver at
the element boundaries is overwritten by projecting the DG solution to the FD
grid on the element boundaries, and the FD scheme takes the time
step. Overwriting the FD reconstructed data
$u_{\mathrm{FD}}^{\mathrm{interface}}$ with the projected DG solution
$\mathcal{P}(u_{\mathrm{DG}}^{\mathrm{interface}})$ on the interfaces makes the
scheme conservative when retaking the time step.  Since the scheme is switching
from DG to FD, it is likely a discontinuity is present and conservation is
important. \blue{This ultimately means that neighboring elements are not aware
  that the element switched from DG to FD between times $t^n$ and $t^{n+1}$
  until boundary data for time $t^{n+1}$ is exchanged. Since the DG solution at
  time $t^n$ is admissible, projecting it to the FD interface grid will be
  acceptable in nearly all cases. In cases where this projection leads to an
  unphysical solution, all elements sharing the interface can detect this and
  switch to FD; however, we have not yet implemented this.} We now describe in
detail how the algorithm is implemented in terms of communication patterns and
parallelization.

First consider an element using DG. We start by computing the local
contributions to the time derivative, the fluxes, source terms, non-conservative
terms, and flux divergence. We store $\partial_t u$, compute local contributions
to the boundary correction $G$, and then send our contributions to the boundary
correction as well as the ghost cells of the primitive variables used for FD
reconstruction to neighboring elements, \blue{as well as the interface mesh used
  to inform the neighbor that we are using DG}. By sending both the inputs to
the boundary correction and the data for FD reconstruction, we reduce the number
of times communication is necessary. This is important since generally it is the
number of times data is communicated not the amount of data communicated that
causes a bottleneck. Once all contributions to the Riemann problem are received
from neighboring elements, we compute the boundary correction and compute the
candidate solution $u^{\star}(t^{n+1})$. We then apply the troubled-cell
indicator described in $\S$\ref{sec:dgscl TCI} below. If the cell is marked as
troubled we undo the last timestep/substep and retake the timestep/substep using the
FD method\footnote{\blue{Note that only the most recent substep is retaken if a
  substep time integrator is being used.}}. FD reconstruction is performed, but
the projected boundary correction from the DG solve is used to ensure
conservation between neighboring elements using FD. If the cell was not marked
as troubled, we accept the candidate solution as being valid and take the next
timestep/substep.

The FD solver starts by sending the data necessary for FD reconstruction to
neighboring elements\blue{, including the interface mesh used to inform the
  neighbor that FD is being used}. This means any neighboring elements doing DG
need to reconstruct the inputs into the boundary correction using FD
reconstruction. However, this allows us to maintain a single communication per
time step, unlike traditional limiting strategies which inherently need two
communications per time step. Once all FD reconstruction and boundary correction
data has been received from neighboring elements, a FD time step is taken. Any
DG boundary correction data is projected to the FD grid in order to reduce
conservation violations at element boundaries. With the FD time step complete,
we apply a troubled-cell indicator to see if the DG solution would be
admissible. In both Runge-Kutta and multi-step methods, care is taken so as to
not introduce discontinuities into the solution because they were
present in past
timesteps or substeps. In the case of Runge-Kutta time stepping we only switch back
to DG at the end of a complete time step in order to avoid reconstructing
discontinuities in the time stepper history to the DG grid. When multi-step
methods are used, we wait until the TCI has marked enough time steps as being
representable on the DG grid so that any discontinuities have cleared the time
stepper history. For example, when using a third-order multi-step method the TCI
needs to deem three time steps as representable on the DG grid before we switch
to DG. \blue{For the multi-step method we apply the reconstruction operator
  $\mathcal{R}$ to the time stepper history ($\partial_t u(t^{n-1})$ etc.).}

\subsection{Troubled-cell indicators\label{sec:dgscl TCI}}
One of the most important parts of the DG-FD hybrid method is the TCI that
determines when to switch from DG to FD. In~\cite{Dumbser2014a} a numerical
indicator based on the behavior of the polynomials representing the solution was
used as well as physical indicators such as the density or pressure becoming
negative. We believe that the combination of numerical and physical indicators
is crucial, since it enables the development of non-oscillatory methods that
also guarantee physical realizability of the solution. We will first outline
the numerical indicator in this section. Then we will give a detailed
description of the TCIs we use with the GRMHD system for the initial data,
determining when to switch from DG to FD, and when to switch from FD back to DG.

The numerical indicator used in~\cite{Dumbser2014a} is a relaxed discrete
maximum principle (RDMP). The RDMP is a two-time-level indicator in the sense
that it compares the candidate at $t^{n+1}$ to the solution at time $t^n$. The
RDMP requires that
\begin{eqnarray}
  \label{eq:dgscl rdmp}
  \min_{\mathcal{N}}\left[u(t^n)\right]
  - \delta
  \le
  u^\star(t^{n+1})
  \le
  \max_{\mathcal{N}}\left[u(t^n)\right] + \delta,
\end{eqnarray}
where $\mathcal{N}$ are either the Neumann or Voronoi neighbors plus the element
itself, $\delta$ is a parameter defined below that relaxes the discrete maximum
principle\blue{, and $u$ are the conserved variables}\footnote{\blue{Any choice of quantities
  can be monitored.}}. When computing $\max(u)$ and $\min(u)$ over an
element using DG, we first project the DG solution to the subcells and then
compute the maximum and minimum over \textit{both} the DG solution and the
projected subcell solution. However, when an element is using FD we compute the
maximum and minimum over the subcells only. Note that the maximum and minimum
values of $u^\star$ are computed in the same manner as those of $u$. The
parameter $\delta$ used to relax the discrete maximum principle is given by:
\begin{eqnarray}
  \label{eq:dgscl rdmp delta}
  \delta =
  \max\left(\delta_{0},\epsilon
  \left\{\max_{\mathcal{N}}\left[u(t^n)\right]
  - \min_{\mathcal{N}}\left[u(t^n)\right]\right\}\right),
\end{eqnarray}
where, as in~\cite{Dumbser2014a}, we take $\delta_{0}=10^{-7}$ and
$\epsilon=10^{-3}$.

We have found that the RDMP TCI is not able to handle slow-moving shocks. This
is precisely because it is a two-time-level TCI and measures the change in the
solution from one time step to the next. Since discontinuities are inevitably
still somewhat smeared with a FD scheme, a discontinuity moving slowly enough
gradually generates large oscillations inside the element it is entering. The
RDMP, measuring relative changes, does not react quickly enough or at all, and
so the DG method ends up being used in elements with discontinuities. We
demonstrate this below in the simple context of a 1d Burgers step solution with
the mesh moving at nearly the speed of the discontinuity.

Since using the RDMP means we are unable to satisfy Requirements~\ref{req:dgscl
  slow shocks} and~\ref{req:dgscl moving mesh}, we seek a supplementary TCI to
deal with these cases.  We use the TCI proposed in~\cite{PerssonTci}, which we
will refer to as the Persson TCI. This TCI looks at the falloff of the spectral
coefficients of the solution, effectively comparing the power in the highest
mode to the total power of the solution. Consider a discontinuity sensing
quantity $U$, which is typically a scalar but could be a tensor of any rank. Let
$U$ have the 1d spectral decomposition:
\begin{eqnarray}
  \label{eq:dgscl Persson U expansion}
  U(x)=\sum_{i=0}^{N}c_i P_i(x),
\end{eqnarray}
where in our case $P_i(x)$ are Legendre polynomials, and $c_i$ are the spectral
coefficients.\footnote{When a filter is being used to prevent aliasing-driven
  instabilities, lower modes need to be included in $\hat{U}$. $\hat{U}$ should
  generally be the highest unfiltered mode.} We then define a filtered solution
$\hat{U}$ as
\begin{eqnarray}
  \label{eq:dgscl Persson U hat}
  \hat{U}(x)=c_N P_N(x).
\end{eqnarray}
The main goal of $\hat{U}$ is to measure how much power is in the highest mode,
which is the mode most responsible for Gibbs phenomenon. In 2d and 3d we
consider $\hat{U}$ on a dimension-by-dimension basis, taking the $L_2$ norm over
the extra dimensions, reducing the discontinuity sensing problem to always being
1d.  We define the discontinuity indicator $s^\Omega$ as
\begin{eqnarray}
  \label{eq:dgscl Persson indicator}
  s^\Omega=\log_{10}\left(\frac{(\hat{U}, \hat{U})}{(U, U)}\right),
\end{eqnarray}
where $(\cdot,\cdot)$ is an inner product, which we take to be the Euclidean
$L_2$ norm (i.e.~we do not divide by the number of grid points since that
cancels out anyway).

We must now decide what values of $s^\Omega$ are large and therefore mean the DG
solution is inadmissible. For a spectral expansion, we would like the solution
to be at least continuous and so the spectral coefficients should decay at least
as $1/N^2$~\cite{Gottlieb1977}. Since our sensor depends on the square of the
coefficients, we expect at least $1/N^4$ decay for smooth solutions. With this
in mind, we have found that requiring
\begin{eqnarray}
  \label{eq:dgscl persson tci}
  s^\Omega<s^e=-\alpha_N\log_{10}(N+1),
\end{eqnarray}
with $\alpha_N=4$ works well for detecting oscillations and switching to the FD
scheme. In order to prevent rapid switching between the DG and FD schemes, we
use $\alpha_N+1$ for the TCI when deciding whether to switch back to
DG.

\subsubsection{Initial data TCI for GRMHD\label{sec:dgscl ID TCI}}
We set the initial data on the DG grid, and then check a series of conditions to
see if the initial data is representable on the DG grid. We require:
\begin{enumerate}
\item that $\min(\tilde{D})$ over both the DG grid and the subcells is above
  a user-specified threshold. This is essentially a positivity check on
  $\tilde{D}$.
\item that $\min(\tilde{\tau})$ over both the DG grid and the subcells is above
  a user-specified threshold. This is essentially a positivity check on
  $\tilde{\tau}$.
\item that for all conserved variables their max and min on the subcells
  satisfies an RDMP compared to the max and min on the DG grid. The tolerances
  chosen are typically the same as those used for the two-level RDMP during the
  evolution.
\item that $\tilde{D}$ and $\tilde{\tau}$ pass the Persson TCI.
\item that if $\max\left(\sqrt{\tilde{B}^i\delta_{ij}\tilde{B}^j}\right)$ is
  above a user-specified threshold, $\sqrt{\tilde{B}^i\delta_{ij}\tilde{B}^j}$
  satisfies the Persson TCI.
\end{enumerate}
If all requirements are met, then the DG solution is admissible.

\subsubsection{TCI on DG grid for GRMHD\label{sec:dgscl DG TCI}}
On the DG grid we require:
\begin{enumerate}
\item that the RDMP TCI passes.
\item that $\min(\tilde{D})$ is above a user-specified threshold. This is
  essentially a positivity check. This is done over both the DG and projected
  subcell solution.
\item that $\min(\tilde{\tau})$ is above a user-specified threshold. This is
  essentially a positivity check. This is done over both the DG and projected
  subcell solution.
\item that $\tilde{B}^2\le1.0 - \epsilon_B 2 \tilde{\tau}\sqrt{\gamma}$ at all
  grid points in the DG element.
\item that primitive recovery is successful.
\item that if we are in the atmosphere, we stay on DG. Since we have now
  recovered the primitive variables, we are able to say with certainty whether
  or not we are in atmosphere.
\item that $\tilde{D}$ and $\tilde{\tau}$ pass the Persson TCI.
\item that if $\max\left(\sqrt{\tilde{B}^i\delta_{ij}\tilde{B}^j}\right)$ is
  above a user-specified threshold, $\sqrt{\tilde{B}^i\delta_{ij}\tilde{B}^j}$
  satisfies the Persson TCI.
\end{enumerate}
If all requirements are met, then the DG solution is admissible.

\subsubsection{TCI on FD grid for GRMHD\label{sec:dgscl FD TCI}}
In order to switch to DG from FD, we require:
\begin{enumerate}
\item that the RDMP TCI passes.
\item that no conserved variable fixing was necessary. If the conserved
  variables needed to be adjusted in order to recover the primitive variables,
  then even the FD solution is inaccurate.
\item that $\min(\tilde{D})$ is above a user-specified threshold. This is
  essentially a positivity check.
\item that $\min(\tilde{\tau})$ is above a user-specified threshold. This is
  essentially a positivity check.
\item that $\tilde{D}$ and $\tilde{\tau}$ pass the Persson TCI.
\item that if $\max\left(\sqrt{\tilde{B}^i\delta_{ij}\tilde{B}^j}\right)$ is
  above a user-specified threshold, $\sqrt{\tilde{B}^i\delta_{ij}\tilde{B}^j}$
  satisfies the Persson TCI.
\end{enumerate}
If all the above checks are satisfied, then the numerical solution is
representable on the DG grid.

\subsection{On the number of subcells to use\label{sec:dgscl number of
    subcells}}
The only hard requirement on the number of subcells used in 1d is $N_s\ge N+1$
so that there are at least as many degrees of freedom to represent the solution
on the subcells as there are in the DG scheme. However, the more optimal choice,
as is argued in~\cite{Dumbser2014a}, is $N_s=2N+1$. This arises from comparing
the time step size allowed when using a DG method,~\eref{eq:dgscl time step
  size}, to the time step size allowed when using a FV or FD
method,~\eref{eq:dgscl time step size FD}. Choosing $N_s>2N+1$ is not desirable
since that would result in having to take smaller time steps when switching from
DG to FD. We refer the reader to \S4.5 of~\cite{Dumbser2014a} for a more
detailed discussion of the optimal number of subcells to use.

\subsection{Perspective on DG-FD hybrid method\label{sec:dgscl perspective}}
Given the complexity of the DG-FD hybrid scheme and the relative expense of FD
schemes compared to the DG scheme, the DG-FD hybrid scheme might seem like a
poor choice. We argue that this is not the case and that the hybrid scheme is
actually a good choice. Consider needing a resolution of $130^d$ (very modest)
to solve a problem using a FD scheme to a desired accuracy. The equivalent DG-FD
hybrid scheme would use ten seventh-order elements so that in the worst case,
where there are large discontinuities everywhere in the domain, the scheme is as
accurate as the FD scheme. However, wherever the solution is smooth enough to be
representable using DG, roughly $2^d$ fewer grid points are necessary. In 3d
this makes a significant difference, especially if the numerical solution is
representable using DG in much of the computational domain. For example,
consider the case where half the elements are using FD. In this case the DG-FD
hybrid scheme uses ${}\sim0.58$ times as many grid points as the equivalent FD
scheme. Furthermore, the DG scheme only needs to solve the Riemann problem on
element boundaries, and does not need to perform the expensive reconstruction
step necessary in FD and FV schemes. Thus, the decrease in the number of grid
points is a lower bound on the performance improvement the DG-FD hybrid scheme
has to offer. Ultimately, we believe that the more useful view of the DG-FD
hybrid scheme is that it is a FD scheme that uses DG as a way to compress the
representation of the solution in smooth regions in order to increase
efficiency.

\section{Numerical results\label{sec:dgscl numerical results}}

\subsection{Burgers equation: a slowly moving discontinuity\label{sec:dgscl
    burgers results}}
While extremely simple, Burgers equation allows us to easily test how well the
RDMP and Persson TCI are able to handle slowly-moving
discontinuities. \blue{Burgers'} equation is given by
\begin{eqnarray}
  \label{eq:dgscl Burgers}
  \partial_t U+\partial_x\left(\frac{U^2}{2}\right)=0.
\end{eqnarray}
Whenever we use the Persson TCI we use the evolved variable $U$ as the
discontinuity sensing quantity.

We evolve the solution
\begin{eqnarray}
  \label{eq:dgscl Burgers step solution}
  U(x,t)=\left\{
    \begin{array}{ll}
      2 & \mathrm{if} \; x\le0.25+1.5t \\
      1 & \mathrm{otherwise}
    \end{array}\right.
\end{eqnarray}
on a moving mesh. The mesh has a velocity $v_g^x=1.4$, while the discontinuity
moves at speed $1.5$. Thus, the discontinuity moves relatively slowly across the
grid, allowing us to test how well each TCI handles such discontinuities. We
integrate~\eref{eq:dgscl Burgers} using a third-order Adams-Bashforth time
stepper, on an initial domain $x\in[-1,1]$ with eight P$_5$ elements.
We compare the RDMP TCI and the Persson
TCI in figure~\ref{fig:dgscl burgers step moving mesh} at a final time of
$t_f=1.5$. The top row uses a time step of $\Delta t=2.5\times10^{-3}$ and the
bottom row uses $\Delta t=5\times10^{-4}$. In all cases a third-order weighted
compact nonlinear scheme is used for FD reconstruction. We use a Rusanov or
local Lax-Friedrichs numerical flux/boundary correction.

The leftmost plot in the top row of figure~\ref{fig:dgscl burgers step moving
  mesh} uses the Persson TCI with $\alpha_N=3$, the center plot in the top row
uses the Persson TCI with $\alpha_N=4$, and the rightmost plot in the top row
uses the RDMP TCI. We see that, in agreement with what is expected from a
convergence analysis of Legendre polynomials~\cite{Gottlieb1977}, using
$\alpha_N=4$ to switch to the FD scheme is most robust as an indicator. We see
that both the Persson TCI with $\alpha_N=3$ and the RDMP TCI struggle to switch
to the FD scheme quickly enough to prevent unphysical oscillations from entering
the solution. In the bottom row of figure~\ref{fig:dgscl burgers step moving
  mesh} we use a smaller time step size, $\Delta t=5\times10^{-4}$, to make the
relative change in $U$ from one time step to the next smaller. From left to
right we show results using the Persson TCI with $\alpha_N=4$, the RDMP TCI, and
the Persson TCI with $\alpha_N=3$ alongside the RDMP TCI. In general, the RDMP
is much better at preventing oscillations from appearing on the left of the
discontinuity, while the Persson TCI does a better job on the right of the
discontinuity. While interesting, it is unclear how this translates to more
complex systems and flows. Although we cannot completely discount the RDMP, the
Persson indicator does have an advantage in all cases, but using both TCIs
together gives the best results. We ran the Persson TCI with $\alpha_N=4$
alongside the RDMP TCI for the smaller time step case and found that no
unphysical oscillations are visible, just as in the top middle plot of
figure~\ref{fig:dgscl burgers step moving mesh}. We have verified that our
results are the same whether using the SSP RK3 time stepper or the
Adams-Bashforth time stepper.

\begin{figure}[ht]
  \begin{center}
    \begin{tabular}{ccc}
      \includegraphics[width=0.3\textwidth]{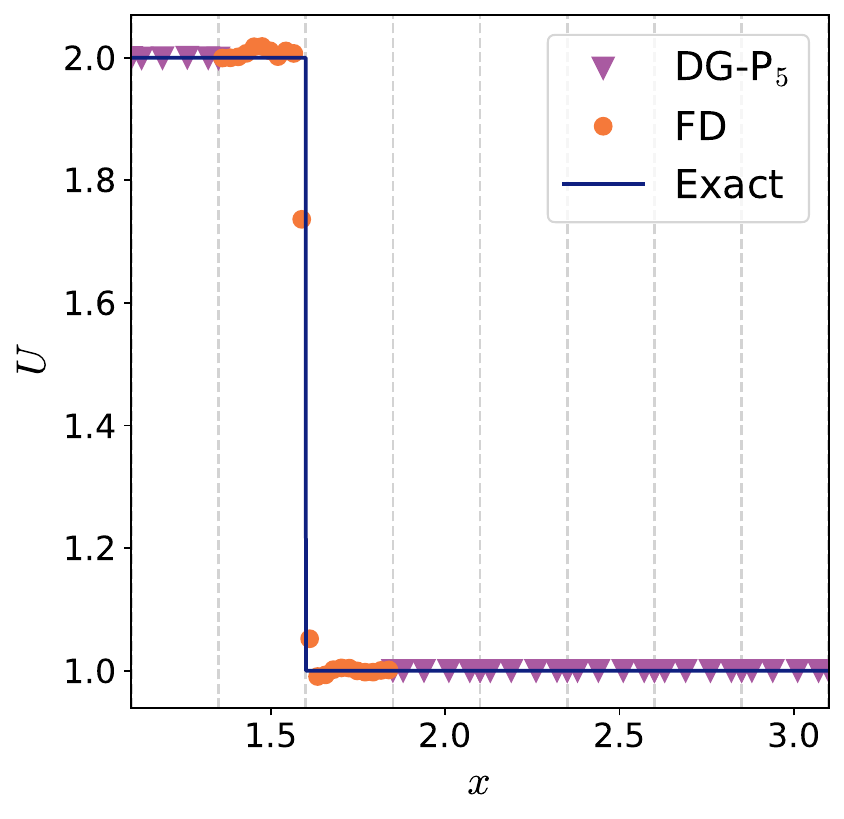}
      & \includegraphics[width=0.3\textwidth]{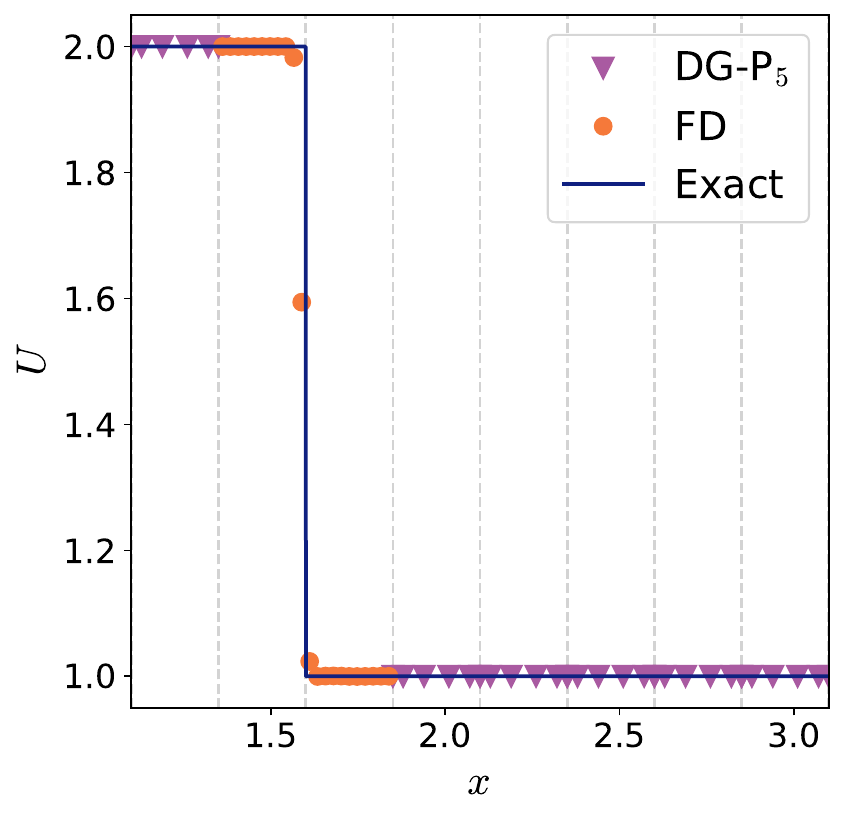}
      & \includegraphics[width=0.3\textwidth]{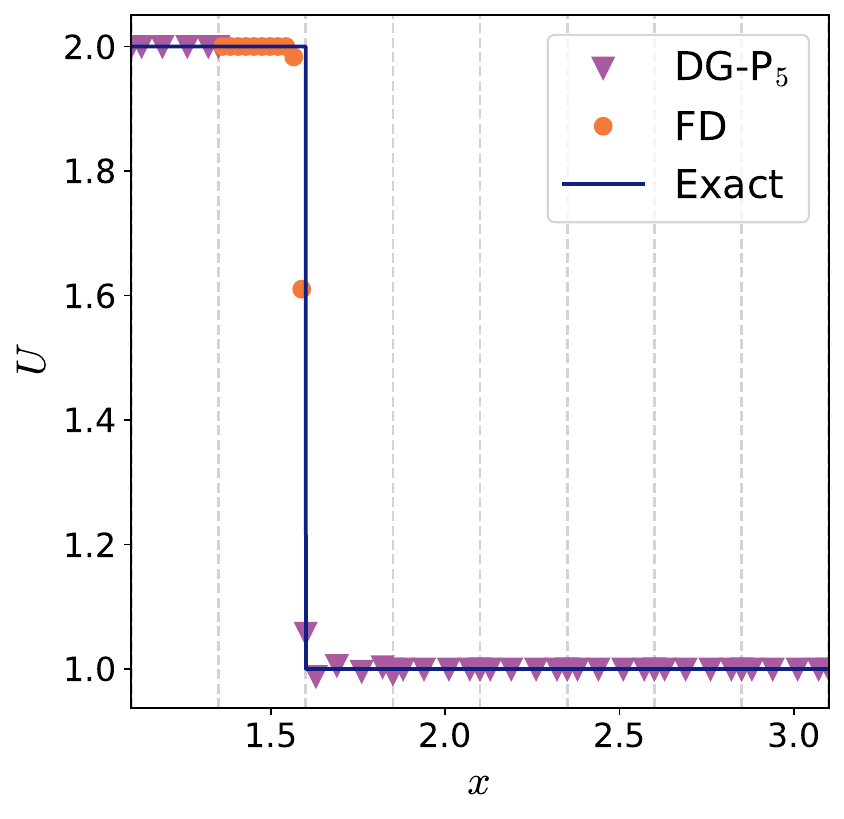}
      \\
      \includegraphics[width=0.3\textwidth]{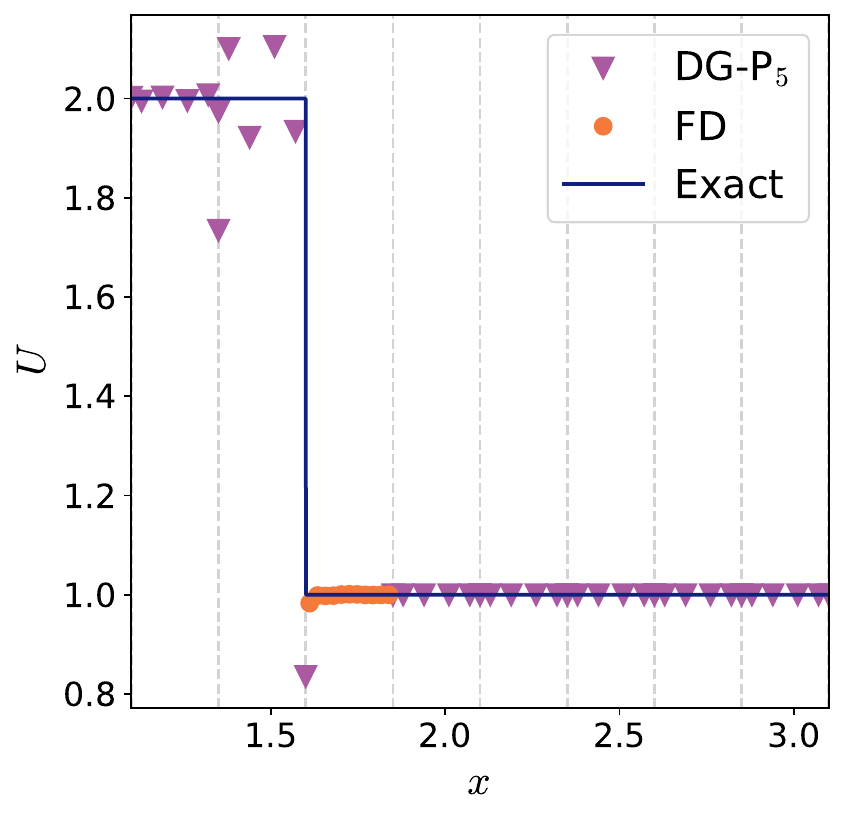}
      & \includegraphics[width=0.3\textwidth]{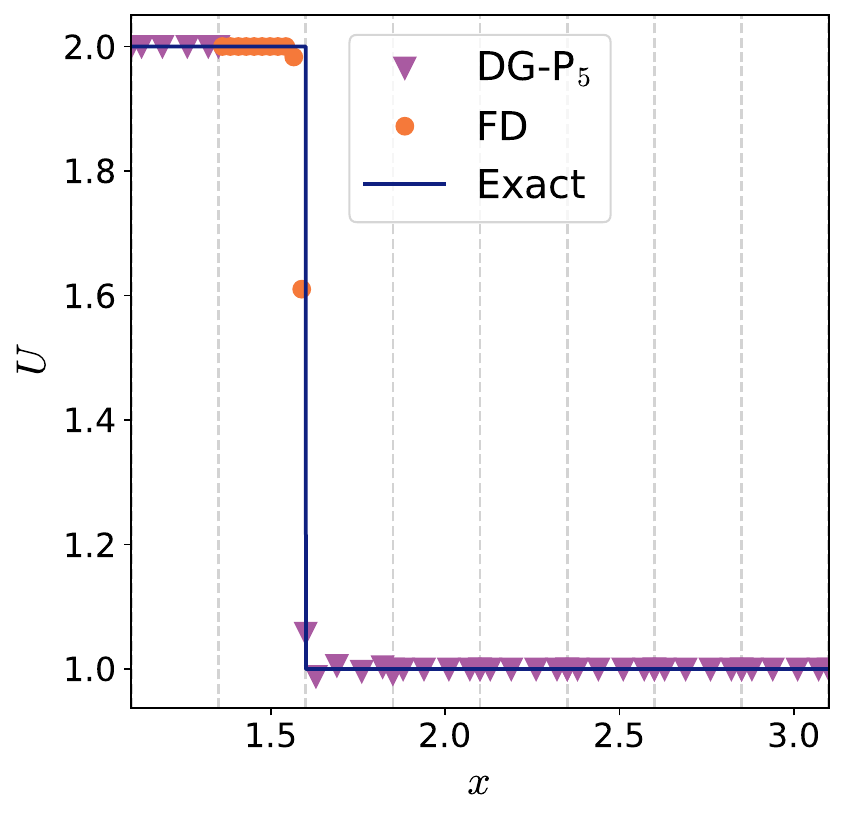}
      & \includegraphics[width=0.3\textwidth]{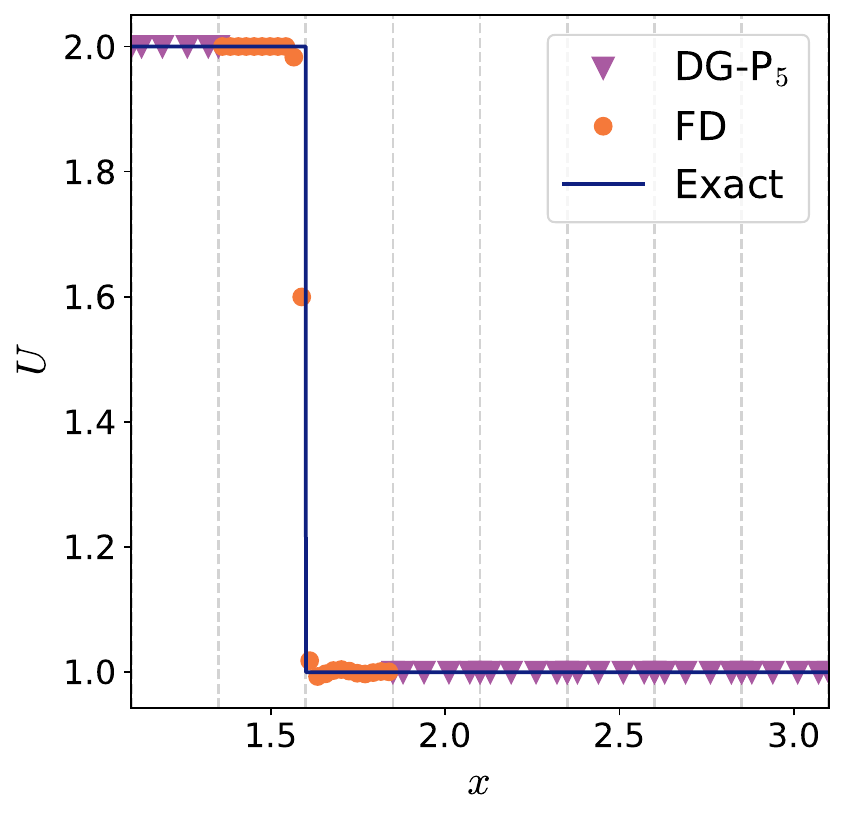}
    \end{tabular}
    \caption{The step Burgers problem at $t_f=1.5$ using a DG-P$_5$ scheme
      hybridized with a WCNS3 FD scheme. A third-order Adams-Bashforth time
      stepper is used and the mesh is moving at velocity $v_g^x=1.4$. Results in
      the top row are obtained using a time step size of
      $\Delta t=2.5\times10^{-3}$ and in the bottom row using a time step size
      of $\Delta t=5\times10^{-4}$. Going from left to right in the top row, the
      TCI used is the Persson TCI with $\alpha_N=3$, the Persson TCI with
      $\alpha_N=4$, and the RDMP TCI. Going from left to right in the bottom
      row, the TCI used is the Persson TCI with $\alpha_N=3$, the RDMP TCI, and
      Persson TCI with $\alpha_N=4$ along with the RDMP TCI.\label{fig:dgscl
        burgers step moving mesh}}
  \end{center}
\end{figure}

\subsection{General relativistic magnetohydrodynamics\label{sec:dgscl grmhd
    results}}
In this section we present results of our DG-FD hybrid scheme when applied to
various GRMHD test problems. The final test problem in this section is that of a
single magnetized neutron star, demonstrating that our hybrid scheme is capable
of simulating interesting relativistic astrophysics scenarios. We always use an
HLL Riemann solver and typically the third-order strong-stability preserving
Runge-Kutta (SSP RK3) time stepper~\cite{Hesthaven2008}. However, we also
compare a fifth-order Dormand-Prince method~\cite{DORMAND198019} to the RK3
method for some test problems. We mainly use the SSP RK3 stepper since this is a
commonly used method when comparing shock capturing schemes. We also reconstruct
the variables $\{\rho, p, Wv^i, B^i, \Phi\}$ using a monotonised central
reconstruction scheme. We choose the resolution for the different problems by
having the number of FD grid points be approximately equal to the number of grid
points used by current production FD codes. Unless stated otherwise, we do not
monitor $\tilde{B}^i$ with the Persson indicator since in most of the test cases
we look at the magnetic field has discontinuities at or near the same place the
fluid variables have discontinuities. All simulations use \texttt{SpECTRE}
v2022.04.04~\cite{deppe_nils_2021_5501002} and the input files are available as
part of the arXiv version of this paper.

\subsubsection{1d Smooth Flow\label{sec:Smooth Flow}}

We consider a simple 1d smooth flow problem to test which of the limiters and
troubled-cell indicators are able to solve a smooth problem without degrading
the order of accuracy. A smooth density
perturbation is advected across the domain with a velocity $v^i$. The analytic
solution is given by
\begin{eqnarray}
  \rho&=1 + 0.7 \sin[k^i (x^i-v^it)], \\
  v^i&=(0.8,0,0),\\
  k^i&=(1,0,0),\\
  p&=1,\\
  B^i&=(0,0,0),
\end{eqnarray}
and we close the system with an adiabatic equation of state,
\begin{eqnarray}
  \label{eq:ideal fluid eos}
  p = \rho\epsilon\left(\Gamma-1\right),
\end{eqnarray}
where $\Gamma$ is the adiabatic index, which we set to 1.4. We use a domain
given by $[0,2\pi]^3$, and apply periodic boundary conditions in all directions.
The time step size is $\Delta t = 2\pi/ 5120$ so that the spatial discretization
error is larger than the time stepping error for all resolutions that we use.

\begin{table}
  \caption{\label{tab:Smooth flow errors} The errors and local convergence order
    for the smooth flow problem using different limiting strategies. Note that
    the limiter is not applied if the troubled-cell indicator determines the DG
    solution to be valid. We observe the expected convergence rate except when
    the solution is underresolved because too few elements are used or when the
    error is no longer dominated by the truncation error of the DG scheme.}
  \begin{indented}
    \lineup
  \item[]\begin{tabular}{@{}cccc}
           \br
           Method & $N_x$ & $L_2(\mathcal{E}(\rho))$ & $L_2$ Order\\ \mr
           DG-FD P$_3$ & \02 & 3.50983e-1 & \\
                  & \04 & 1.22554e-1 & \01.52\\
                  & \08 & 3.72266e-4 & \08.36\\
                  & 16 & 1.61635e-5 & \04.53\\
                  & 32 & 9.76927e-7 & \04.05\\
           \hline
           DG-FD P$_4$ & \02 & 3.62426e-1 & \\
                  & \04 & 3.79759e-4 & \09.90\\
                  & \08 & 1.15193e-5 & \05.04\\
                  & 16 & 3.73055e-7 & \04.95\\
           \hline
           DG-FD P$_5$ & 2 & 3.45679e-01 & \\
                  & 4 & 2.23822e-05 & 13.91\\
                  & 8 & 3.18504e-07 & \06.13\\
                  & 16 & 5.08821e-09 & \05.97\\
           \br
         \end{tabular}
  \end{indented}
\end{table}

We perform convergence tests at different DG orders and present the results in
table~\ref{tab:Smooth flow errors}. We show both the $L_2$ norm of the error and
the convergence rate. The $L_2$ norm is defined as
\begin{eqnarray}
  \label{eq:L2 norm}
  L_2(u)=\sqrt{\frac{1}{M}\sum_{i=0}^{M-1}u_i^2},
\end{eqnarray}
where $M$ is the total number of grid points and $u_i$ is the value of $u$ at
grid point $i$ and the convergence order is given by
\begin{equation}
  \label{eq:convergence order}
  L_2\;\mathrm{Order} =
  \log_2\left[\frac{L_2(\mathcal{E}_{N_x/2})}{L_2(\mathcal{E}_{N_x})}\right].
\end{equation}
We find that when very few elements are used, the TCI decides the solution is
not well represented on the DG grid. Although if we disable the FD scheme
completely, we find the DG method is stable, we find it acceptable that the TCI
switches to FD in order to ensure robustness. Ultimately we observe the
expected rate of convergence for smooth problems.

\subsubsection{1d Riemann Problems}

\begin{figure}
  \centering
  \begin{minipage}{0.425\columnwidth}
    \centering
    \includegraphics[width=1\textwidth]{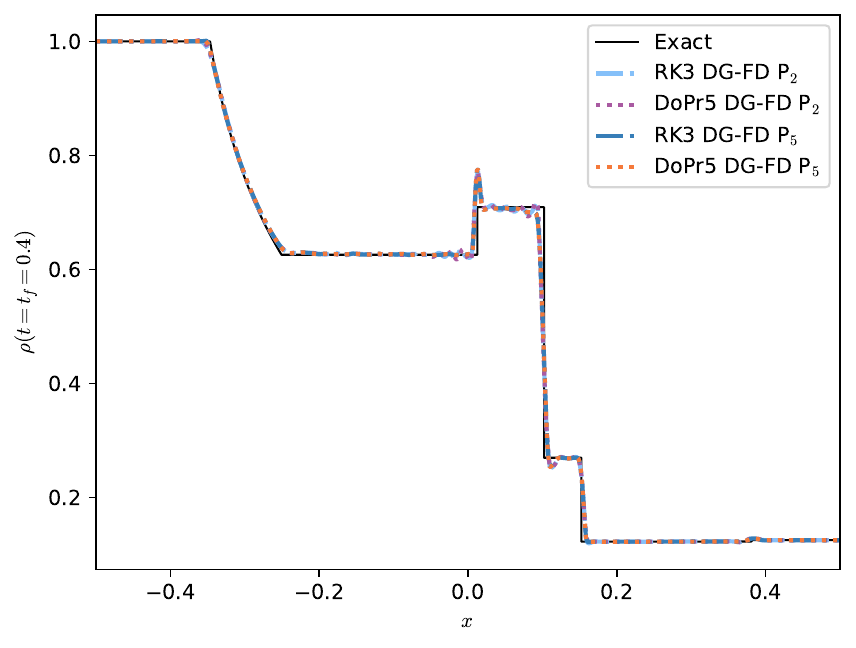} \\
    \includegraphics[width=1\textwidth]{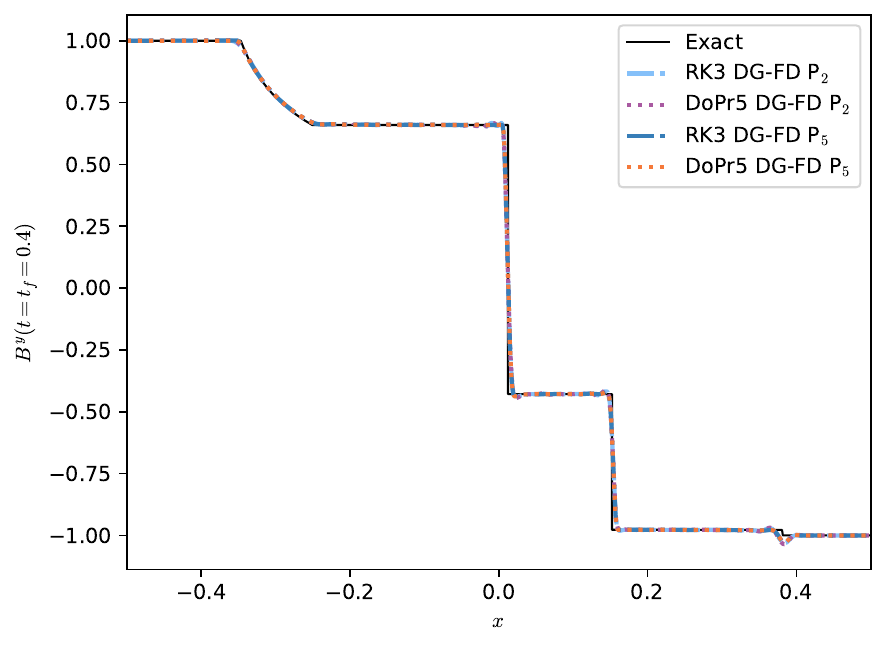}
  \end{minipage}
  \begin{minipage}{0.425\columnwidth}
    \centering
    \includegraphics[width=1\textwidth]{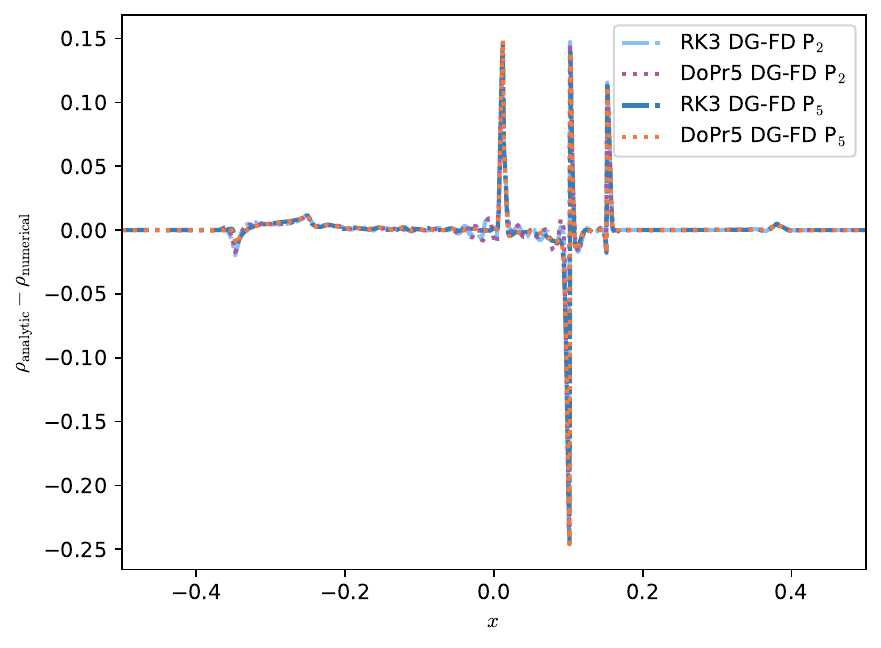} \\
    \includegraphics[width=1\textwidth]{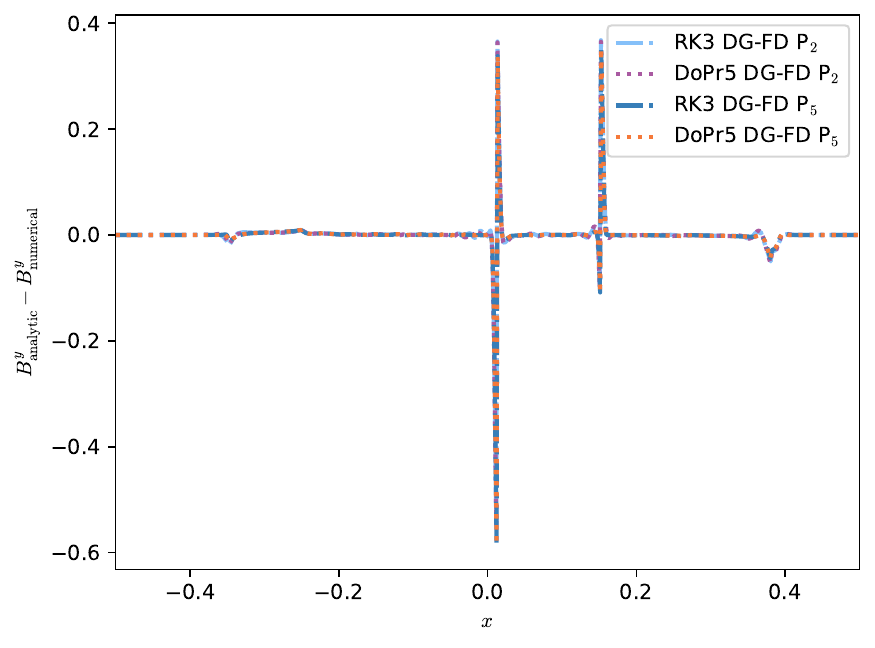}
  \end{minipage}
  \caption{\label{fig:RiemannProblem1}Results from Riemann Problem 1
    of~\cite{Balsara2001} using a P$_5$ (64 elements) and P$_2$ (128 elements)
    DG-FD hybrid scheme with an SSP RK3 and Dormand-Prince 5 time stepper.  All
    results are at the final time $t=0.4$. The top left panel shows the rest
    mass density $\rho$, the bottom left the magnetic field $B^y$, and the right
    panels show the difference between the analytic and numerical solution. The
    P$_5$ scheme is able to resolve the discontinuities just as well as the
    P$_2$ scheme, while also admitting fewer unphysical oscillations away from
    the discontinuities.}
\end{figure}

\begin{figure}
  \centering
  \begin{minipage}{0.425\columnwidth}
    \centering
    \includegraphics[width=1\textwidth]{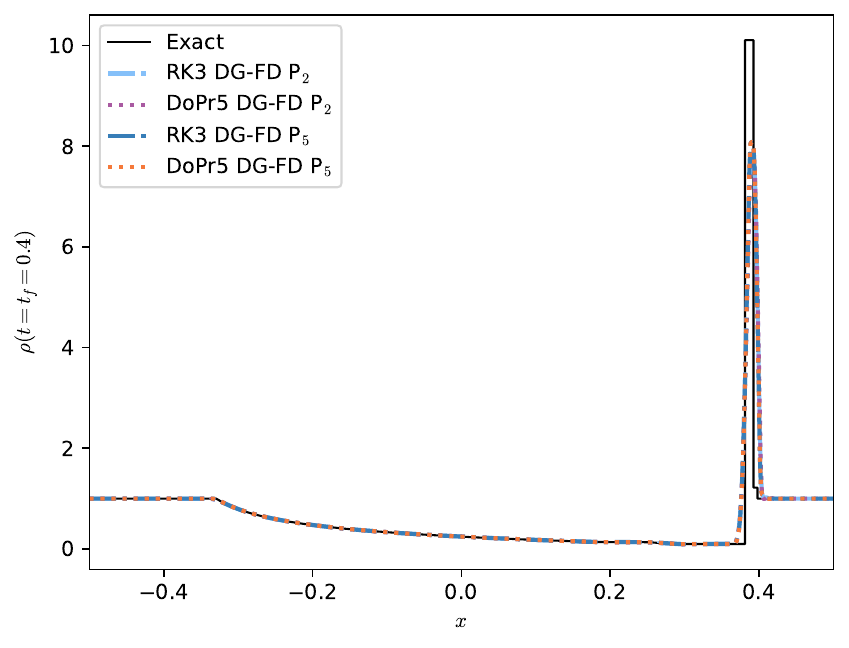} \\
    \includegraphics[width=1\textwidth]{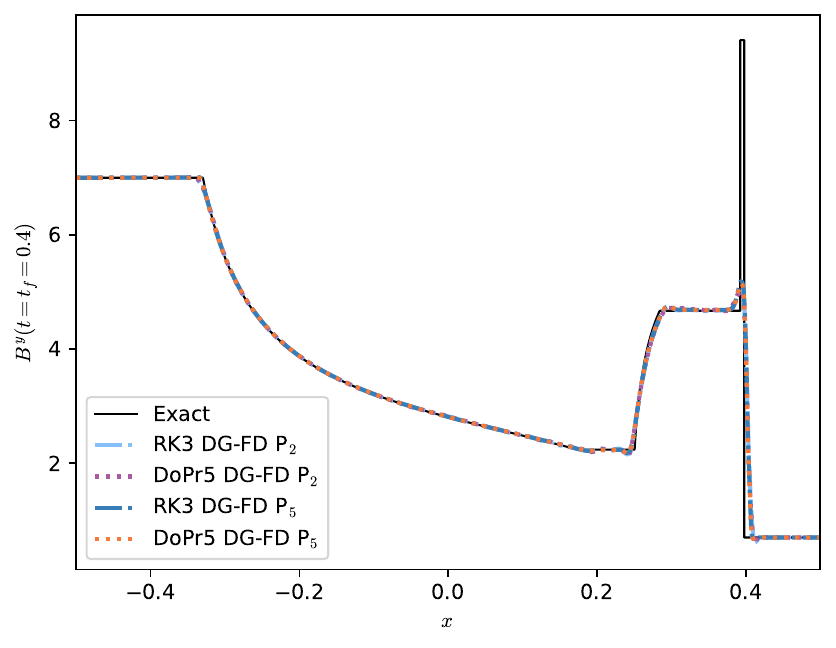}
  \end{minipage}
  \begin{minipage}{0.425\columnwidth}
    \centering
    \includegraphics[width=1\textwidth]{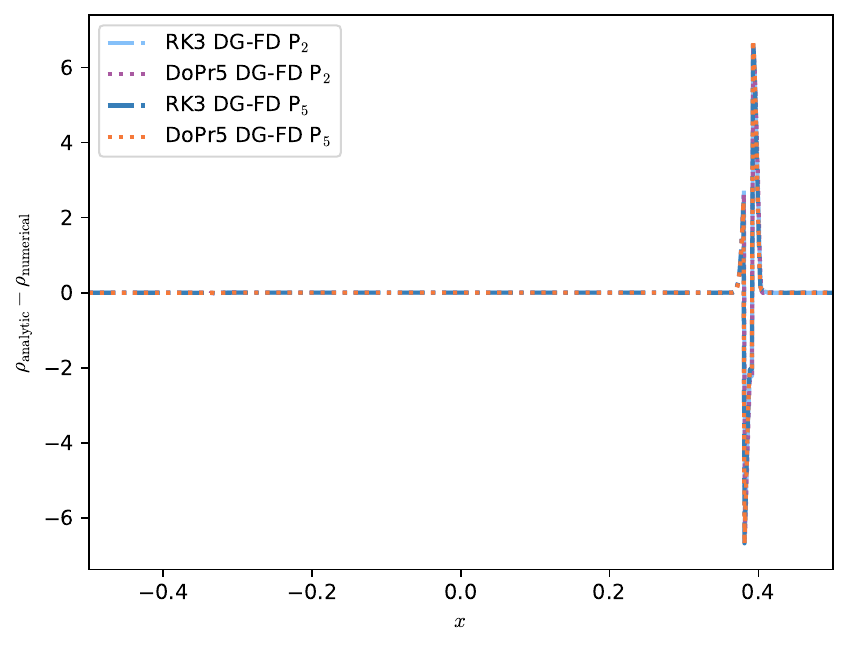} \\
    \includegraphics[width=1\textwidth]{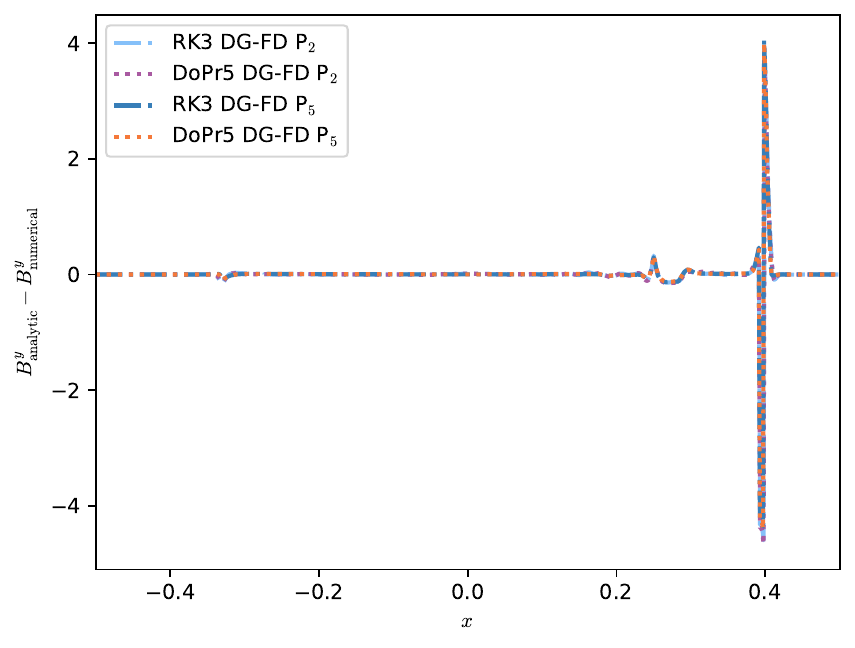}
  \end{minipage}
  \caption{\label{fig:RiemannProblem3}Results from Riemann Problem 3
    of~\cite{Balsara2001} using a P$_5$ (64 elements) and P$_2$ (128 elements)
    DG-FD hybrid scheme with an SSP RK3 and Dormand-Prince 5 time stepper.  All
    results are at the final time $t=0.4$. The top left panel shows the rest
    mass density $\rho$, the bottom left the magnetic field $B^y$, and the right
    panels show the difference between the analytic and numerical solution. The
    P$_5$ scheme is able to resolve the discontinuities just as well as the
    P$_2$ scheme, while also admitting fewer unphysical oscillations away from
    the discontinuities.}
\end{figure}

\begin{figure}
  \centering
  \begin{minipage}{0.425\columnwidth}
    \centering
    \includegraphics[width=1\textwidth]{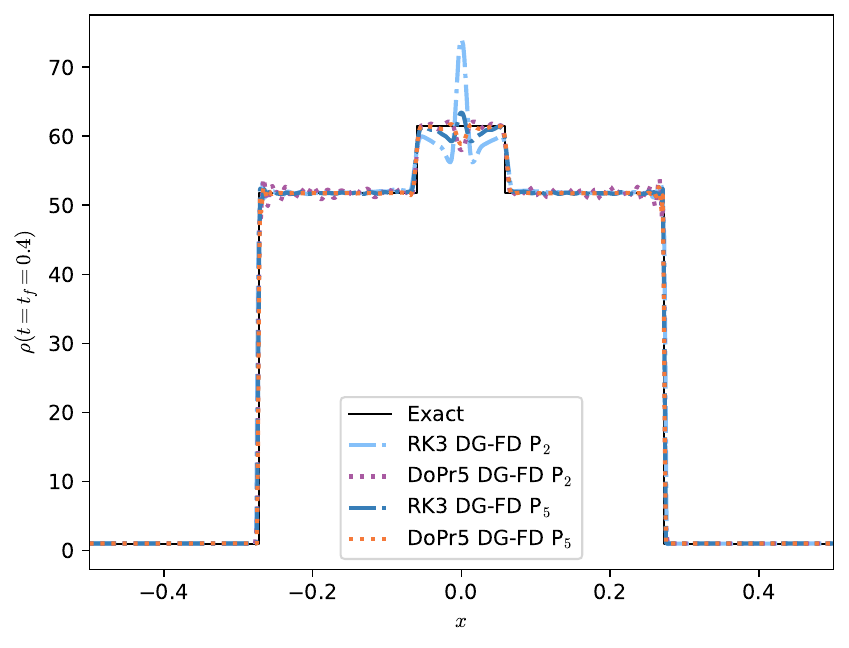} \\
    \includegraphics[width=1\textwidth]{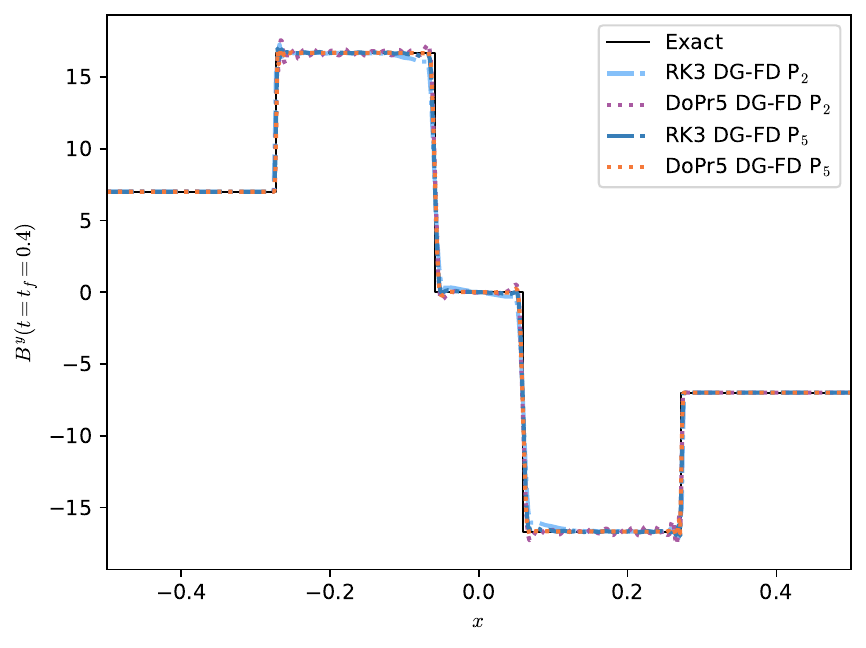}
  \end{minipage}
  \begin{minipage}{0.425\columnwidth}
    \centering
    \includegraphics[width=1\textwidth]{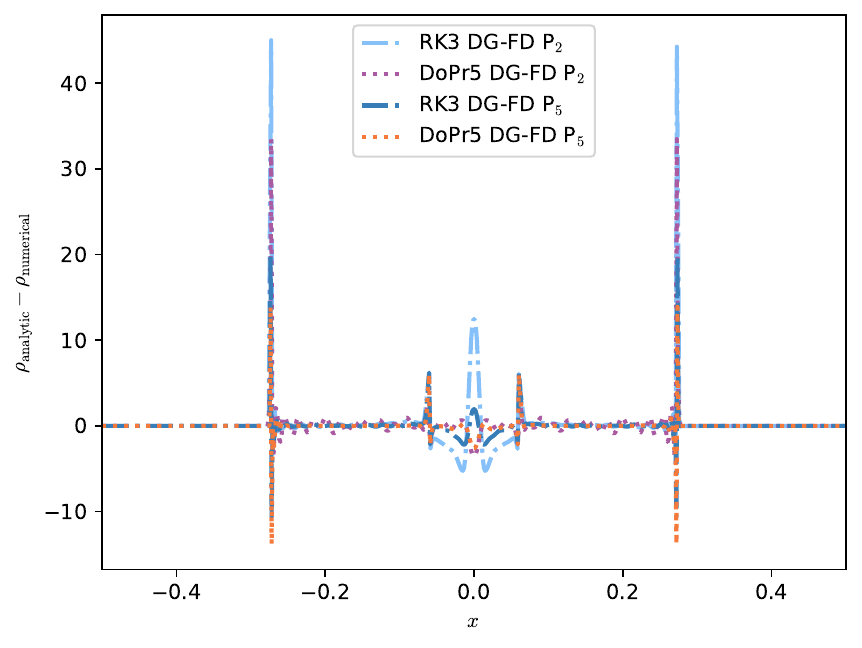} \\
    \includegraphics[width=1\textwidth]{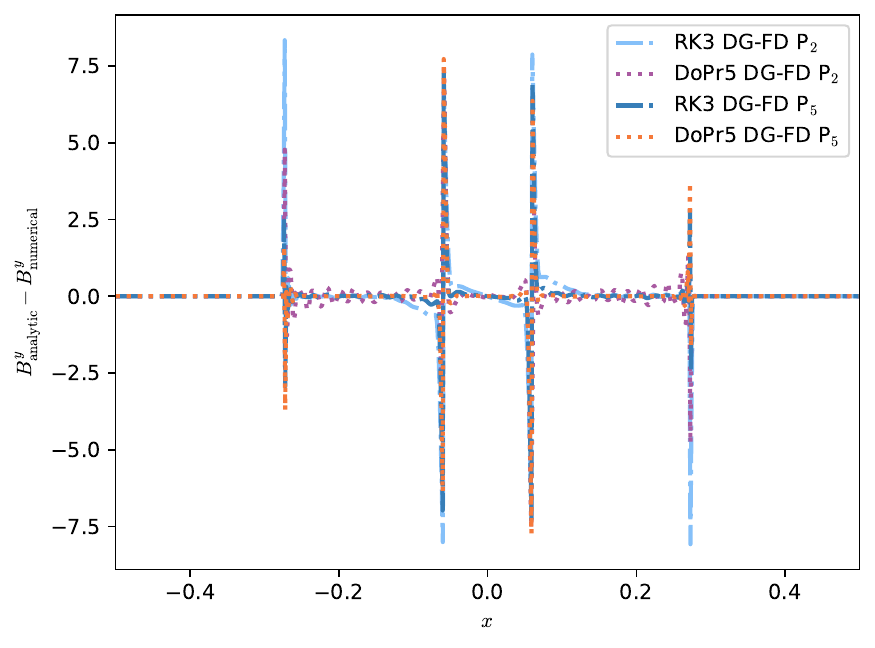}
  \end{minipage}
  \caption{\label{fig:RiemannProblem4}Results from Riemann Problem 4
    of~\cite{Balsara2001} using a P$_5$ (64 elements) and P$_2$ (128 elements)
    DG-FD hybrid scheme with an SSP RK3 and Dormand-Prince 5 time stepper.  All
    results are at the final time $t=0.4$. The top left panel shows the rest
    mass density $\rho$, the bottom left the magnetic field $B^y$, and the right
    panels show the difference between the analytic and numerical solution. The
    P$_5$ scheme is able to resolve the discontinuities just as well as the
    P$_2$ scheme, while also admitting fewer unphysical oscillations away from
    the discontinuities.}
\end{figure}

One-dimensional Riemann problems are a standard test for any scheme that must be
able to handle shocks. We will focus on the first, third, and fourth Riemann
problems (RP1, RP3, RP4) of~\cite{Balsara2001}. The setup is given in
table~\ref{tab:Rp1 conditions}. We perform simulations using an SSP RK3 and a
Dormand-Prince 5 method with $\Delta t=5\times10^{-4}$. In the top left panels
of figures~\ref{fig:RiemannProblem1}-\ref{fig:RiemannProblem4} we show the rest
mass density $\rho$ at $t_f=0.4$, the bottom left panels show $B^y$, while the
right panels show the difference between the analytic and numerical solution.
The thin black curve is the analytic solution obtained using the Riemann solver
of~\cite{Giacomazzo2006}. An ideal fluid equation of state~\eref{eq:ideal fluid
  eos} is used.

\begin{table}
  \caption{\label{tab:Rp1 conditions}The initial conditions for Riemann Problems
    1, 3, and 4 of~\cite{Balsara2001}. The domain is $x\in[-0.5,0.5]$, the final
    time is $t_f=0.4$, and an ideal fluid equation of state is used with an
    adiabatic index of 2 for Riemann Problem 1 and $5/3$ for Riemann Problems 3
    and 4.}
  \begin{indented}
    \lineup
  \item[]
    \begin{tabular}{@{}lccccc} \br Problem &  & $\rho$ & $p$ & $v^i$ & $B^i$ \\
      \mr
      RP 1 & $x < 0$ & 1.0\0\0 & \0\0\01.0 & $(\phantom{-}0.0\0\0,0,0)$ & $(\00.5,\phantom{-}1.0,\phantom{-}0.0)$ \\
       & $x \ge 0$ & 0.125 & \0\0\00.1 & $(\phantom{-}0.0\0\0,0,0)$ & $(\00.5,-1.0,\phantom{-}0.0)$ \\ \mr
      RP 3 & $x < 0$ & 1.0\0\0 & 1000.0 & $(\phantom{-}0.0\0\0,0,0)$ & $(10.0,\phantom{-}7.0,\phantom{-}7.0)$ \\
       & $x \ge 0$ & 1.0\0\0 & \0\0\00.1 & $(\phantom{-}0.0\0\0,0,0)$ & $(10.0,\phantom{-}0.7,\phantom{-}0.7)$ \\ \mr
      RP 4 & $x < 0$ & 1.0\0\0 & \0\0\00.1 & $(\phantom{-}0.999,0,0)$ & $(10.0,\phantom{-}7.0,\phantom{-}7.0)$ \\
       & $x \ge 0$ & 1.0\0\0 & \0\0\00.1 & $(-0.999,0,0)$ & $(10.0,-7.0,-7.0)$ \\
      \br
    \end{tabular}
  \end{indented}
\end{table}

Impressively, the DG-FD hybrid scheme actually has fewer oscillations when going
to higher order. In the right panels of
figures~\ref{fig:RiemannProblem1}-\ref{fig:RiemannProblem4} we plot the error of
the numerical solution using a P$_2$ DG-FD scheme with 128 elements and a P$_5$
DG-FD scheme with 64 elements. We see that the P$_5$ hybrid scheme actually has
fewer oscillations than the P$_2$ scheme, while resolving the discontinuities
equally well. We attribute this to the troubled-cell indicators triggering
earlier when a higher polynomial degree is used since discontinuities entering
an element rapidly dump energy into the high modes. While the optimal order is
almost certainly problem-dependent, given that current numerical relativity
codes are mostly second order, achieving sixth order in the smooth regions is
promising. The SSP RK3 time stepper seems to generally result in fewer
oscillations than the Dormand-Prince 5 time stepper. This could stem from the
Dormand-Prince stepper not being strong stability preserving. We leave a
detailed comparison of different time integration schemes to future work.

\subsubsection{2d Cylindrical Blast Wave}

A standard test problem for GRMHD codes is the cylindrical blast
wave~\cite{2005A&A...436..503L,DelZanna:2007pk} where a magnetized fluid
initially at rest in a constant magnetic field along the $x$-axis is
evolved. The fluid obeys the ideal fluid equation of state with $\Gamma=4/3$.
The fluid begins in a cylindrically symmetric configuration, with hot, dense
fluid in the region with cylindrical radius $r < 0.8$ surrounded by a cooler,
less dense fluid in the region $r > 1$. The initial density $\rho$ and pressure
$p$ of the fluid are
\begin{eqnarray}
  \rho(r < 0.8) & = 10^{-2}, \\
  \rho(r > 1.0) & = 10^{-4}, \\
  p(r < 0.8) & = 1, \\
  p(r > 1.0) & = 5 \times 10^{-4}.
\end{eqnarray}
In the region $0.8 \leq r \leq 1$, the solution transitions continuously and
exponentially (i.e., transitions such that the logarithms of the pressure and
density are linear functions of $r$).  The fluid begins threaded with a uniform
magnetic field with Cartesian components
\begin{equation}
  (B^x, B^y, B^z) = (0.1, 0, 0).
\end{equation}
The magnetic field causes the blast wave to expand non-axisymmetrically.  For
all simulations we use a time step size $\Delta t=10^{-2}$ and an SSP RK3 time
integrator.

\begin{figure}
  \centering
  \begin{minipage}{0.4\columnwidth}
    \centering
    \includegraphics[width=1.0\textwidth]{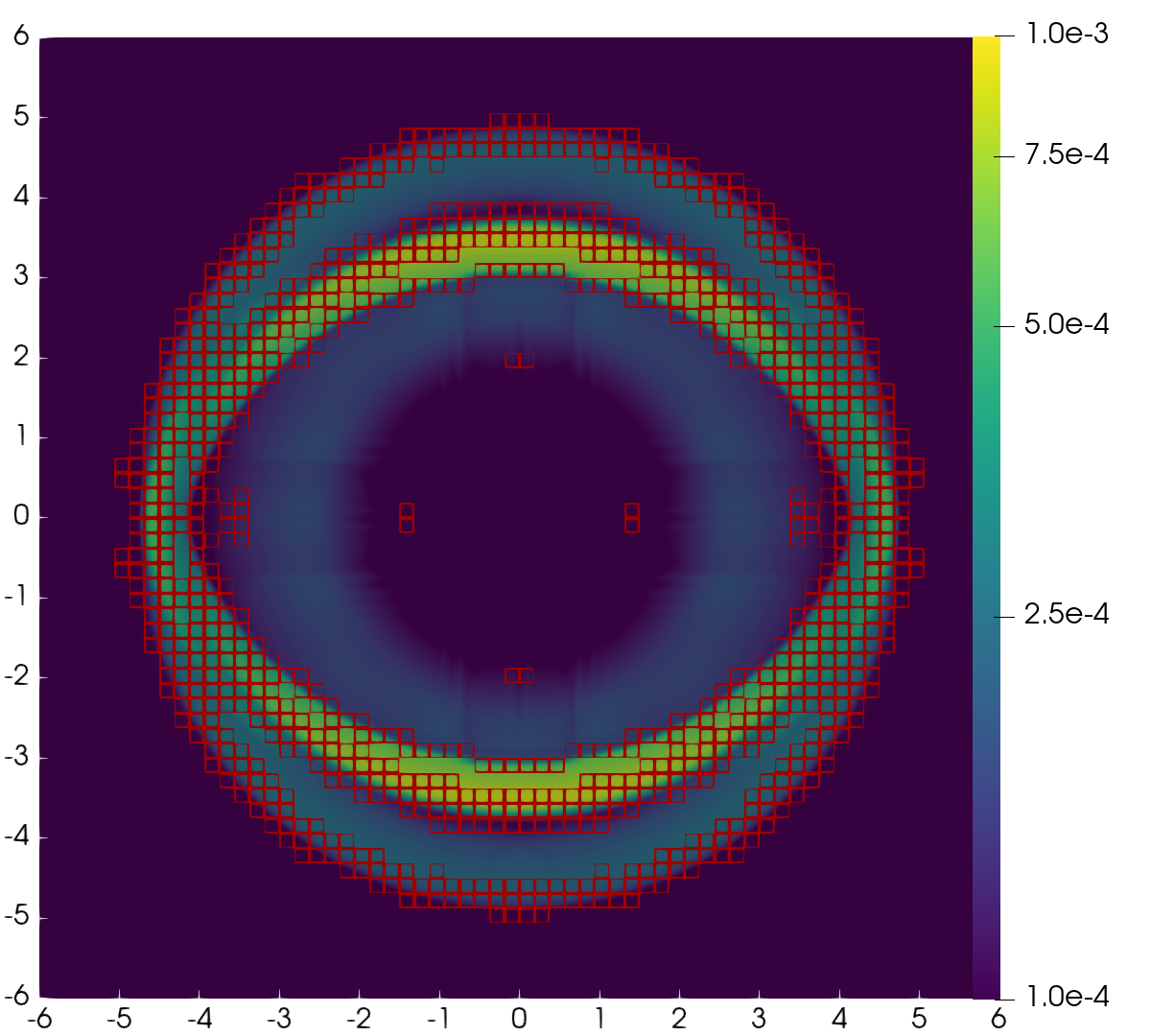}
    \\
    DG-FD, P$_2$, $64^2$ elements
  \end{minipage}
  \begin{minipage}{0.4\columnwidth}
    \centering
    \includegraphics[width=1.0\textwidth]{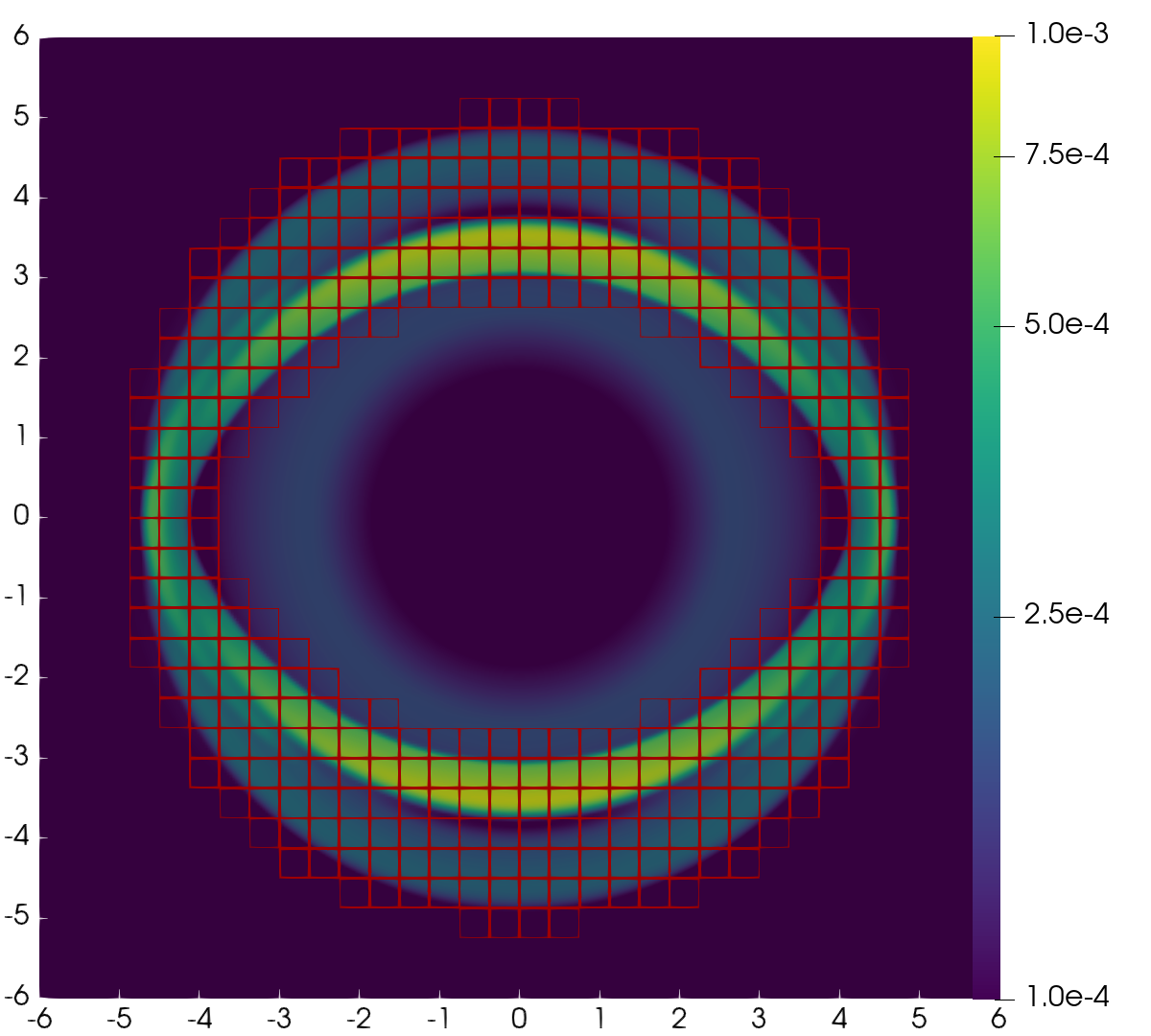}
    \\
    DG-FD, P$_5$, $32^2$ elements
  \end{minipage}
  \caption{Cylindrical blast wave $\rho$ at $t=4$ showing the results of the
    using the DG-FD hybrid scheme with $64\times64$ P$_2$ elements (left) and
    $32\times32$ P$_5$ elements (right). The regions surrounded by maroon
    squares have switched from DG to FD.}
  \label{fig:BlastWave}
\end{figure}

We evolve the blast wave to time $t=4.0$ on a grid of $64\times 64 \times 1$
elements covering a cube of extent $[-6, 6]^3$ using a P$_2$ DG-FD scheme and on
a grid of $32\times32\times1$ using a P$_5$ DG-FD scheme. With these choices the
resolution when using FD everywhere is comparable to what FD codes use for this
test.  We apply periodic boundary conditions in all directions, since the
explosion does not reach the outer boundary by
$t=4.0$. Figure~\ref{fig:BlastWave} shows the logarithm of the rest-mass density
at time $t=4.0$, at the end of evolutions using the P$_2$ (left) and P$_5$
(right) DG-FD schemes. The increased resolution of a high-order scheme is clear
when comparing the P$_2$ and P$_5$ solutions in the interior region of the blast
wave. It is not clear that going to even higher order would be useful in this
problem since to maintain the same time step size we would need to decrease the
number of elements. Furthermore, as we can already see by comparing the P$_2$
and P$_5$ schemes, a greater area of the P$_5$ solution is using FD, though it
is difficult to determine what overall effect this has, especially since
high-order FD schemes could be used. \blue{We show the percentage of elements using FD
instead of DG at the final time in table~\ref{tab:blast wave subcell
  percentage}. As expected, the percentage of elements using FD decreases as the
number of elements is increased.}

\begin{table}
  \caption{The percentage of elements using FD at the final time of the
    cylindrical blast wave simulation\label{tab:blast wave subcell percentage}}
  \begin{indented}
    \lineup
  \item[]\begin{tabular}{@{}ccc}
    \br
    Method & Domain & Percent Using FD \\ \mr
DG-FD P$_2$ & \032$^2$ & \036.7\%\\
 & \064$^2$ & \022.9\%\\
 & \0128$^2$  & \07.9\%\\
\hline
DG-FD P$_5$ & \016$^2$ & \057.8\%\\
 & \032$^2$ & \035.9\%\\
 & \064$^2$ & \021.8\%\\
           \hline
    \br
  \end{tabular}
  \end{indented}
\end{table}

\subsubsection{2d Magnetic Rotor}

The second 2-dimensional test problem we study is the magnetic rotor problem
originally proposed for non-relativistic MHD~\cite{1999JCoPh.149..270B,
  2000JCoPh.161..605T} and later generalized to the relativistic
case~\cite{2010PhRvD..82h4031E, 2003A&A...400..397D}. A rapidly rotating dense
fluid cylinder is inside a lower density fluid, with a uniform pressure and
magnetic field everywhere. The magnetic braking will slow down the rotor over
time, with an approximately 90 degree rotation by the final time $t=0.4$. We use
a domain of $[-0.5,0.5]^3$ and a time step size $\Delta t=10^{-3}$ and an SSP
RK3 time integrator. An ideal fluid equation of state with $\Gamma=5/3$ is used,
and the following initial conditions are imposed:
\begin{eqnarray}
  p&=1 \\
  B^i&=(1,0,0) \\
  v^i&=\left\{
       \begin{array}{ll}
         (-y\Omega, x\Omega, 0),
         & \mathrm{if} \; r \le R_{\mathrm{rotor}}=0.1 \\
         (0,0,0), & \mathrm{otherwise},
       \end{array}\right. \\
  \rho&=\left\{
        \begin{array}{ll}
          10,
          & \mathrm{if} \; r \le R_{\mathrm{rotor}}=0.1 \\
          1, & \mathrm{otherwise},
        \end{array}\right.
\end{eqnarray}
with angular velocity $\Omega = 9.95$. The choice of $\Omega$ and
$R_{\mathrm{rotor}}=0.1$ guarantees that the maximum velocity of the fluid
(0.995) is less than the speed of light. We impose periodic boundary conditions.

\begin{figure}
  \centering
  \begin{minipage}{0.4\columnwidth}
    \centering
    \includegraphics[width=1\textwidth]{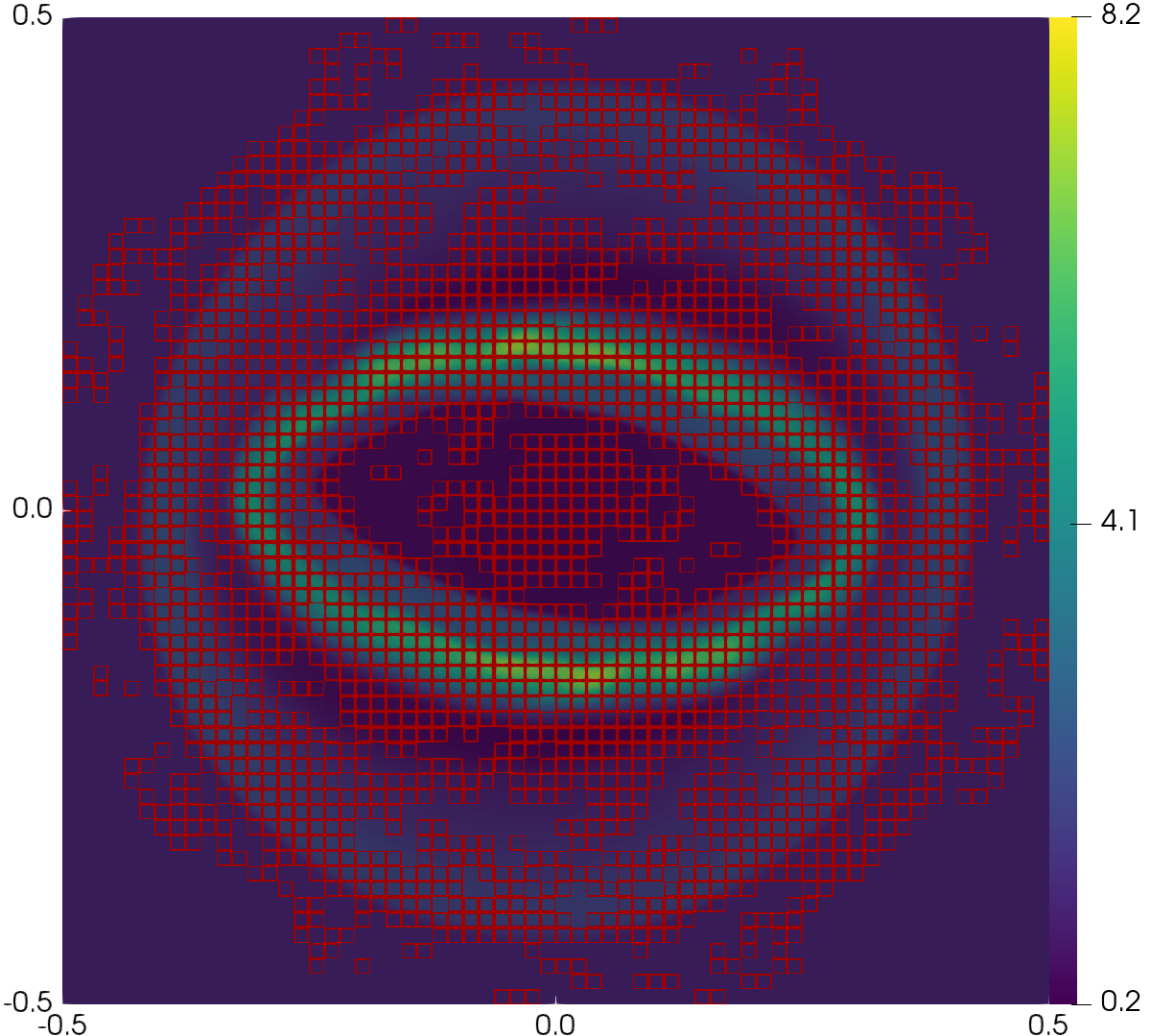}
    \\
    DG-FD, P$_2$, $64^2$ elements
  \end{minipage}
  \begin{minipage}{0.4\columnwidth}
    \centering
    \includegraphics[width=1\textwidth]{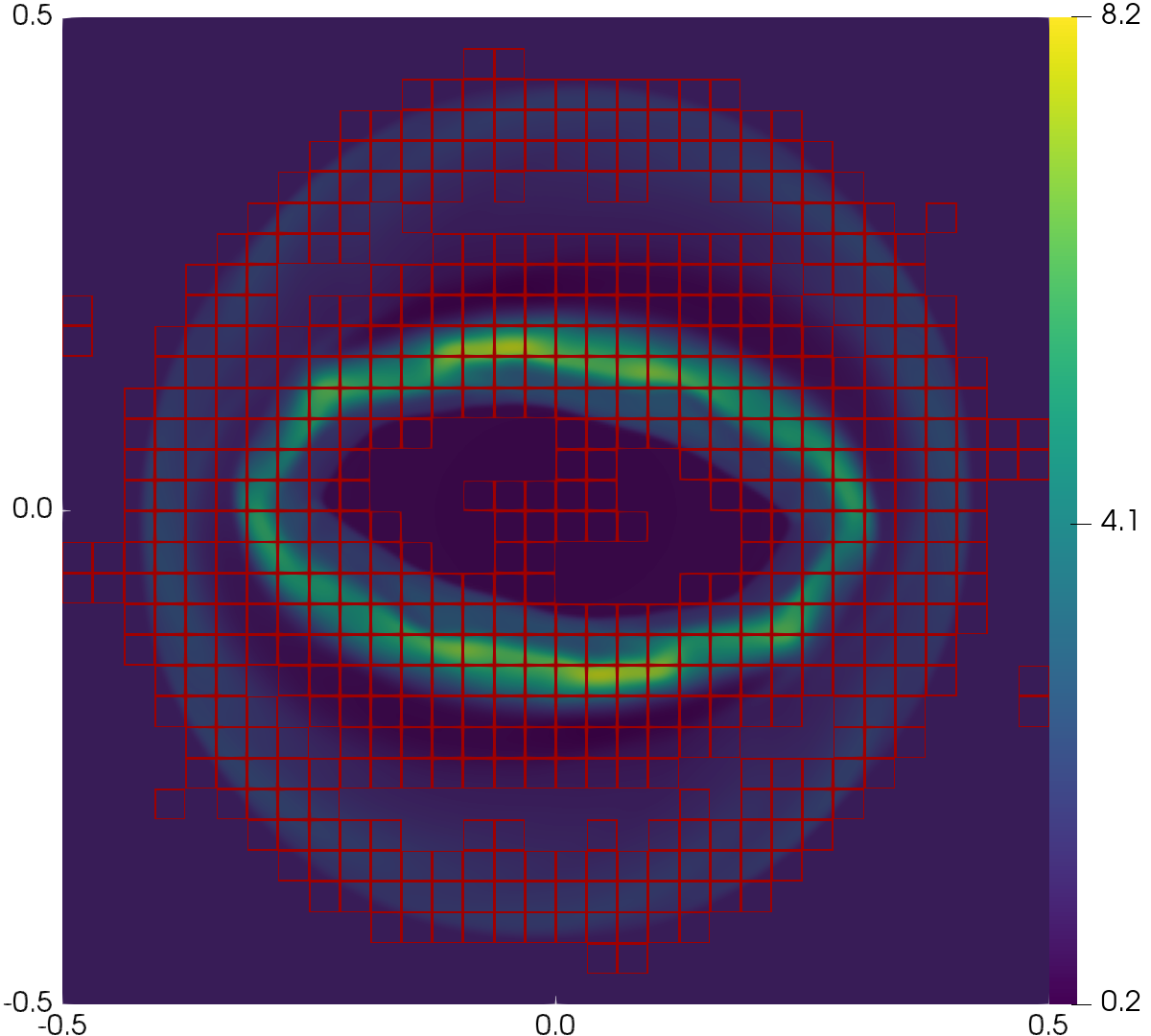}
    \\
    DG-FD, P$_5$, $32^2$ elements
  \end{minipage}
  \caption{Magnetic rotor $\rho$ at $t=0.4$ showing the results of the using the
    DG-FD hybrid scheme with $64\times64$ P$_2$ elements (left) and $32\times32$
    P$_5$ elements (right). The regions surrounded by maroon squares have
    switched from DG to FD.}
  \label{fig:MagneticRotor}
\end{figure}

We show the results of our evolutions using $64\times64$ P$_2$ elements (left)
and $32\times32$ P$_5$ elements (right) in
figure~\ref{fig:MagneticRotor}. Again, the DG-FD hybrid scheme is robust and
accurate, though a fairly large number of cells end up being marked as troubled
in this problem. However, using FD in more elements is not something we view as
inherently bad, since we favor robustness in realistic simulations. The process
of tweaking parameters and restarting simulations is both time consuming and
frustrating, and so giving up some efficiency for robustness \blue{is preferable
  in general}.

\subsubsection{2d Magnetic Loop Advection}

The last 2-dimensional test problem we study is magnetic loop advection problem
\cite{1991JCoPh..92..142D}. A magnetic loop is advected through the domain until
it returns to its starting position. We use an initial configuration very
similar to~\cite{Mosta:2013gwu, 2011ApJS..193....6B, 2005JCoPh.205..509G,
  2008ApJS..178..137S}, where
\begin{eqnarray}
  \rho&=1 \\
  p&=3 \\
  v^i &= (1/1.2, 1/2.4, 0) \\
  B^x &= \left\{
        \begin{array}{ll}
          -A_{\mathrm{loop}}y/R_{\mathrm{in}},
          & \mathrm{if} \; r \le R_{\mathrm{in}} \\
          -A_{\mathrm{loop}}y/r,
          & \mathrm{if} \; R_{\mathrm{in}}<r<R_{\mathrm{loop}} \\
          0, & \mathrm{otherwise},
        \end{array}\right. \\
  B^y &= \left\{
        \begin{array}{ll}
          A_{\mathrm{loop}}x / R_{\mathrm{in}},
          & \mathrm{if} \; r \le R_{\mathrm{in}} \\
          A_{\mathrm{loop}}x/r,
          & \mathrm{if} \; R_{\mathrm{in}}<r<R_{\mathrm{loop}} \\
          0, & \mathrm{otherwise},
        \end{array}\right.
\end{eqnarray}
with $R_{\mathrm{loop}}=0.3$, $R_{\mathrm{in}}=0.001$, and an ideal gas equation
of state with $\Gamma=5/3$. The computational domain is $[-0.5,0.5]^3$
with $64\times64\times1$ elements and periodic boundary conditions being applied
everywhere. The final time for one period is $t=2.4$. For all simulations we use
a time step size $\Delta t=10^{-3}$ and an SSP RK3 time integrator. Since the
fluid variables are smooth in this problem, we apply the Persson TCI to the
Euclidean magnitude of $\tilde{B}^i$ in elements where the maximum value of the
magnitude is above $10^{-5}$.

\begin{figure}
  \centering
  \begin{minipage}{0.4\columnwidth}
    \centering
    \includegraphics[width=1\textwidth]{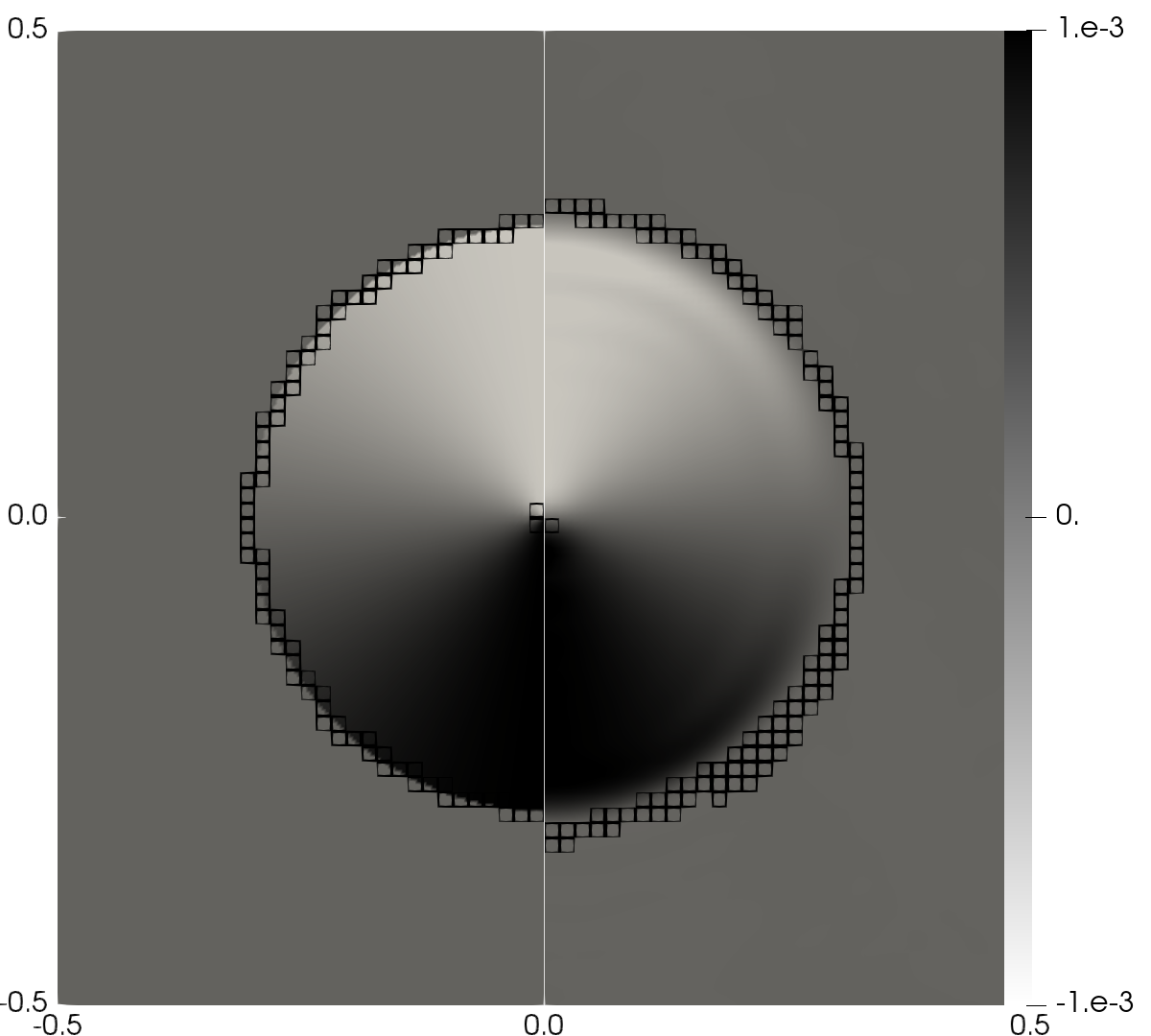}
    \\
    DG-FD, P$_2$, $64^2$ elements
  \end{minipage}
  \begin{minipage}{0.4\columnwidth}
    \centering
    \includegraphics[width=1\textwidth]{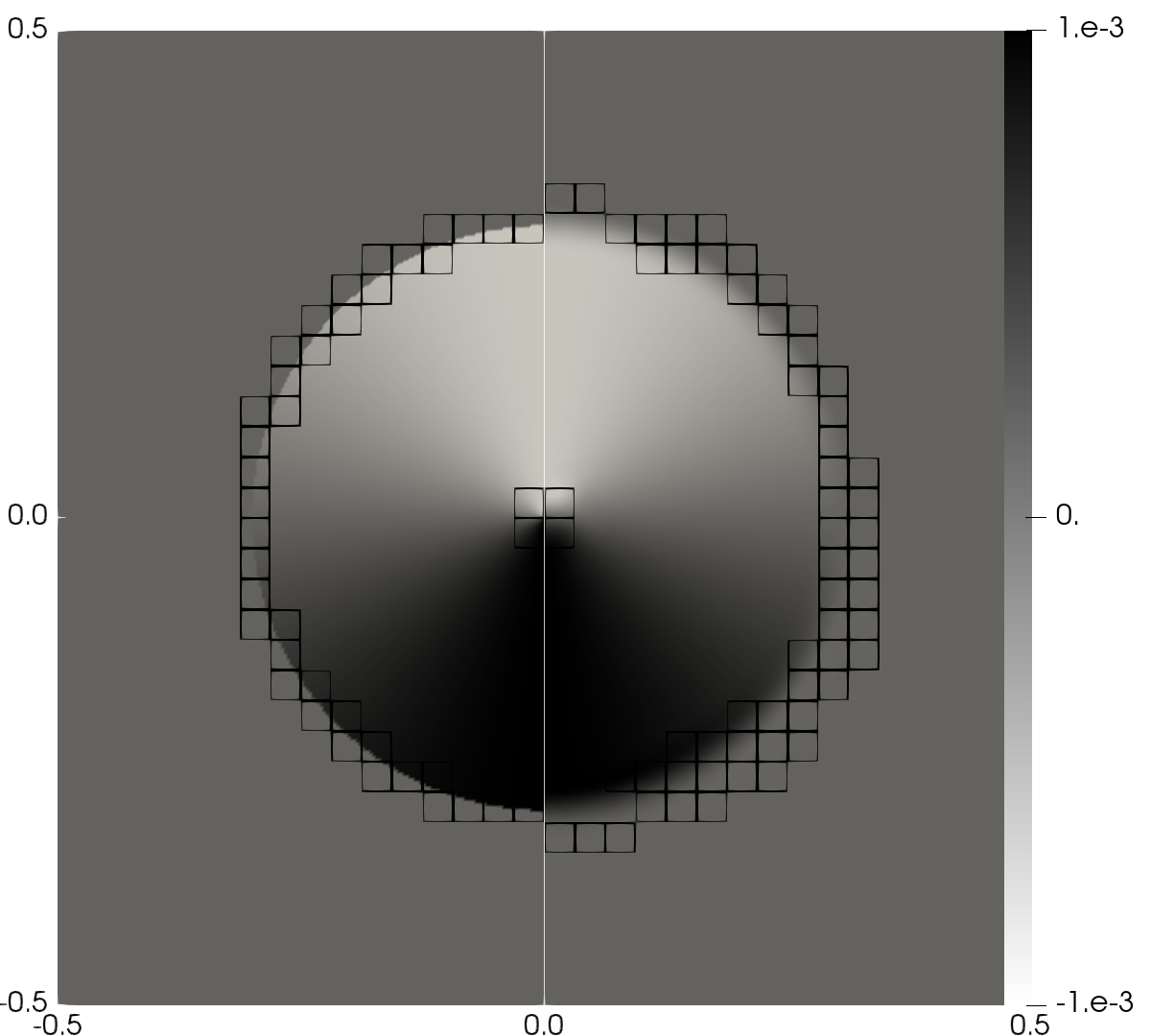}
    DG-FD, P$_5$, $32^2$ elements
  \end{minipage}
  \caption{$B^x$ for the magnetic loop advection problem. The left half of each
    plot is at the initial time, while the right half is after one period
    ($t=2.4$). We show the results of the using the DG-FD hybrid scheme with
    $64\times64$ P$_2$ elements (left) and $32\times32$ P$_5$ elements (right).
    The regions surrounded by black squares have switched from DG to FD.}
  \label{fig:MagneticLoopBx}
\end{figure}

\begin{figure}
  \centering
  \begin{minipage}{0.4\columnwidth}
    \centering
    \includegraphics[width=1\textwidth]{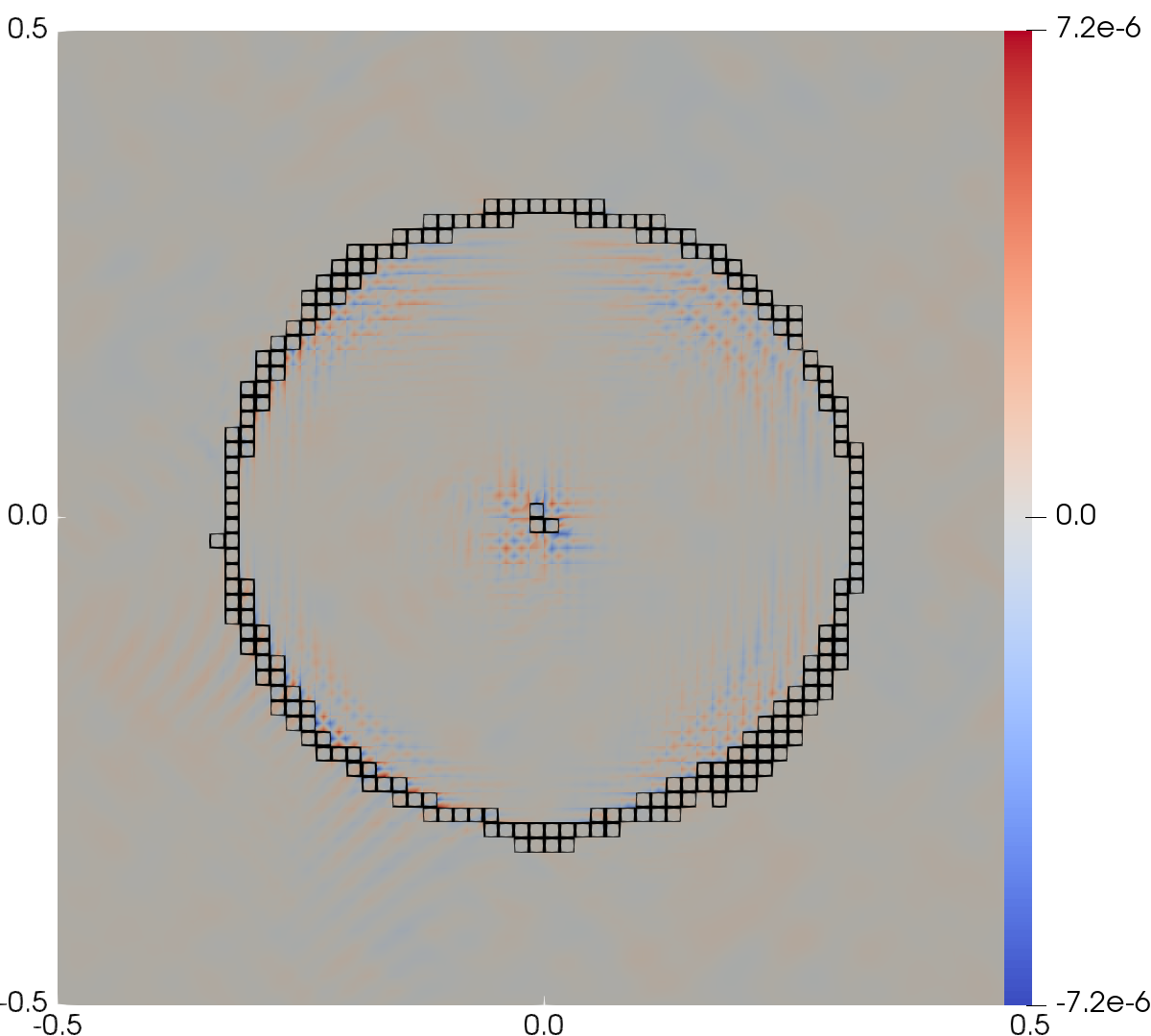}
    \\
    DG-FD, P$_2$, $64^2$ elements
  \end{minipage}
  \begin{minipage}{0.4\columnwidth}
    \centering
    \includegraphics[width=1\textwidth]{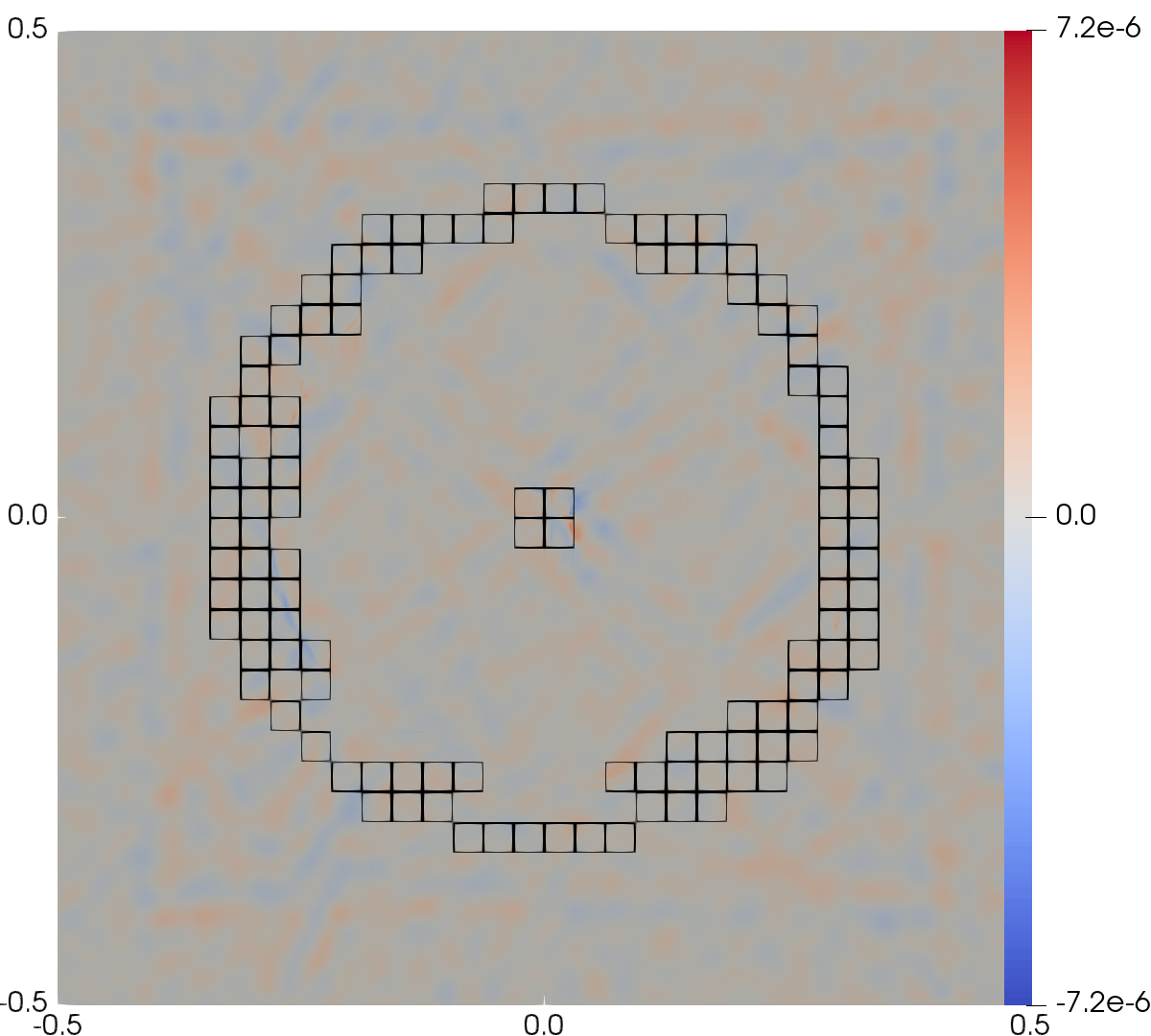}
    DG-FD, P$_5$, $32^2$ elements
  \end{minipage}
  \caption{The divergence cleaning field $\Phi$ for the magnetic loop advection
    problem after one period ($t=2.4$). We show the results of the using the
    DG-FD hybrid scheme with $64\times64$ P$_2$ elements (left) and $32\times32$
    P$_5$ elements (right). The regions surrounded by black squares have
    switched from DG to FD.}
  \label{fig:MagneticLoopPhi}
\end{figure}

In figure~\ref{fig:MagneticLoopBx} we plot the magnetic field component $B^x$ at
$t=0$ on the left half of each plot and after one period $t=2.4$ on the right
half of each plot. In the left panel of figure~\ref{fig:MagneticLoopBx} we show
the result using a P$_2$ DG-FD scheme and in the right panel of
figure~\ref{fig:MagneticLoopBx} using a P$_5$ DG-FD scheme. The P$_5$ scheme
resolves the smooth parts of the solution more accurately than the P$_2$ scheme,
as is to be expected. Finally, in figure~\ref{fig:MagneticLoopPhi} we plot the
divergence cleaning field $\Phi$ at the final time $t=2.4$. We do not observe
any artifacts appearing in the divergence cleaning field at the interfaces
between the DG and FD solvers, demonstrating that the divergence cleaning
properties of the system are not adversely affected by using two different
numerical methods.

\subsubsection{TOV star\label{sec:TOV star}}

A rigorous 3d test case in general relativity is the evolution of a
Tolman-Oppenheimer-Volkoff (TOV) star~\cite{Tolman:1939jz,
  Oppenheimer:1939ne}. In this section we study evolutions of both
non-magnetized and magnetized TOV stars. We adopt the same configuration as
in~\cite{Cipolletta:2019geh}. Specifically, we use a polytropic equation of
state,
\begin{eqnarray}
  \label{eq:polytropic EOS}
  p(\rho)=K \rho^\Gamma
\end{eqnarray}
with the polytropic exponent $\Gamma=2$, polytropic constant $K=100$, and a
central density $\rho_c=1.28\times10^{-3}$. For the magnetized case, we choose a
magnetic field given by a vector potential
\begin{eqnarray}
  \label{eq:TOV vector potential}
  A_\phi=A_b\left(x^2+y^2\right)\max\left(p-p_{\mathrm{cut}},0\right)^{n_s},
\end{eqnarray}
with $A_b=2500$, $p_{\mathrm{cut}}=0.04p_{\max}$, and $n_s=2$. This
configuration yields a magnetic field strength in CGS units
\begin{eqnarray}
  \label{eq:2}
  |B_{\mathrm{CGS}}|=\sqrt{b^2}\times8.352\times10^{19}\,\mathrm{G},
\end{eqnarray}
of $|B_{\mathrm{CGS}}|=1.03\times10^{16}\,\mathrm{G}$. The magnetic field is
only a perturbation to the dynamics of the star, since
$(p_{\mathrm{mag}}/p)(r=0)\sim5\times10^{-5}$. However, evolving the field stably
and accurately can be challenging.  The magnetic field corresponding to the
vector potential in~\eref{eq:TOV vector potential} in the magnetized region is
given by
\begin{eqnarray}
  \label{eq:TOV magnetic field}
  B^x&=\frac{1}{\sqrt{\gamma}}\frac{xz}{r}
       A_bn_s(p-p_{\mathrm{cut}})^{n_s-1}\partial_rp, \\
  B^y&=\frac{1}{\sqrt{\gamma}}\frac{yz}{r}
       A_bn_s(p-p_{\mathrm{cut}})^{n_s-1}\partial_rp, \\
  B^z&=-\frac{A_b}{\sqrt{\gamma}}\left[
       2(p-p_{\mathrm{cut}})^{n_s}
       +\frac{x^2+y^2}{r}
       n_s(p-p_{\mathrm{cut}})^{n_s-1}\partial_r p
       \right],
\end{eqnarray}
and at $r=0$ is
\begin{eqnarray}
  \label{eq:TOV magnetic field origin}
  B^x&=0, \\
  B^y&=0, \\
  B^z&=-\frac{A_b}{\sqrt{\gamma}}
       2(p-p_{\mathrm{cut}})^{n_s}.
\end{eqnarray}

We perform all evolutions in full 3d with no symmetry assumptions and in the
Cowling approximation, i.e.,~we do not evolve the spacetime. To match the
resolution usually used in FD/FV numerical relativity codes, we use a domain
$[-20,20]^3$ with a base resolution of six P$_5$ DG elements. This choice means
we have approximately 32 FD grid points covering the star's diameter at the
lowest resolution, 64 when using twelve P$_5$ elements, and 128 grid points when
using 24 P$_5$ elements. In all cases we set $\rho_{\mathrm{atm}}=10^{-15}$ and
$\rho_{\mathrm{cutoff}}=1.01\times10^{-15}$. We do not run any simulations using
a P$_2$ DG-FD hybrid scheme since the P$_5$ scheme has proven to be more
accurate and robust in all test cases so far.

\begin{figure}
  \centering
  \begin{minipage}{0.45\columnwidth}
    \centering
    \includegraphics[width=1\textwidth]{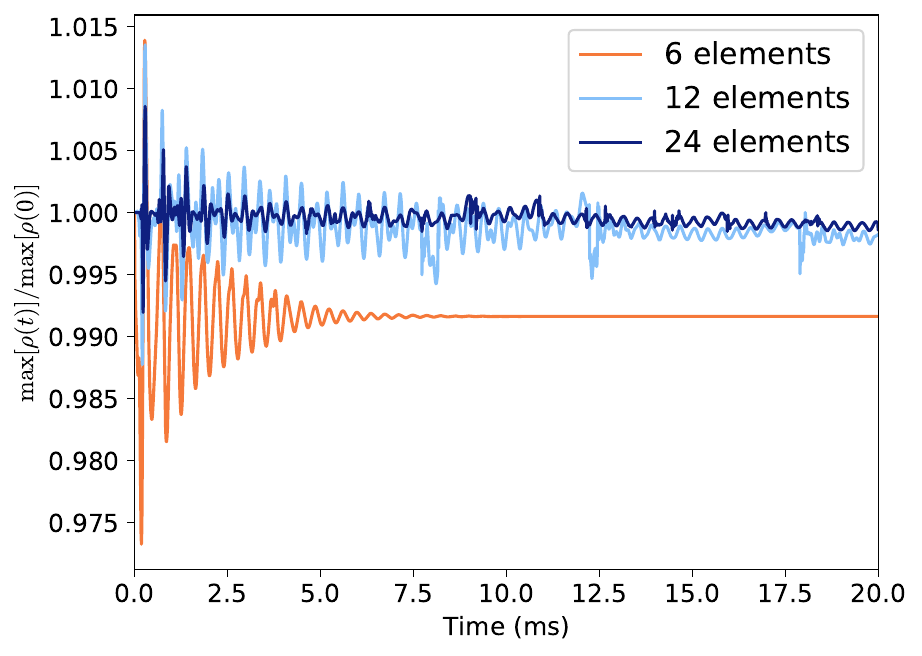}
  \end{minipage}
  \begin{minipage}{0.41\columnwidth}
    \centering
    \includegraphics[width=1\textwidth]{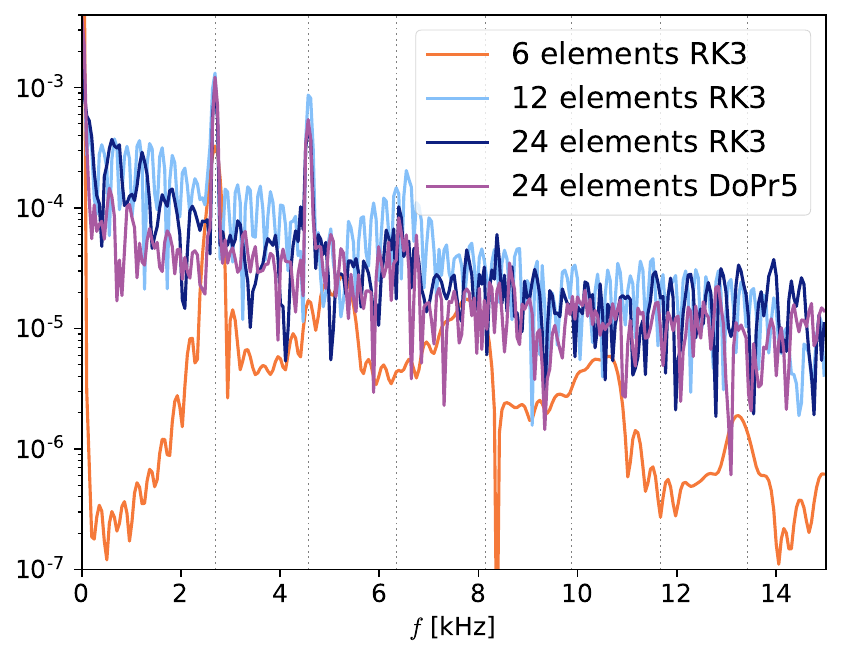}
  \end{minipage}
  \caption{\label{fig:GrmhdTovStar}A plot of $\max[\rho(t)]/\max[\rho(0)]$ at
    three different resolution (left panel) for the non-magnetized TOV star.
    The 6-element simulation uses FD throughout the interior of the star, while
    12- and 24-element simulations use DG. The maximum density in the 6-element
    case drifts down at early times because of the low resolution and the
    relatively low accuracy of using FD at the center. The power spectrum of the
    maximum density for the three different resolution is plotted in the right
    panel. The vertical dashed lines correspond to the known frequencies in the
    Cowling approximation~\cite{2002PhRvD..65h4024F}. When the high-order DG
    scheme is used, more oscillation frequencies are resolved. We also show the
    spectrum from a simulation using a fifth-order Dormand-Prince time stepper.}
\end{figure}

In the left panel of figure~\ref{fig:GrmhdTovStar} we show the maximum rest mass
density over the grid divided by the maximum density at $t=0$ for the
non-magnetized TOV star. The 6-element simulation uses FD throughout the
interior of the star because the corners of the inner elements are in vacuum. In
comparison, the 12- and 24-element simulations use the unlimited P$_5$ DG solver
throughout the star interior. The increased ``noise'' in the 12- and 24-element
data actually stems from the higher oscillation modes in the
star~\cite{2002PhRvD..65h4024F} that are induced by numerical error. In the
right panel of figure~\ref{fig:GrmhdTovStar} we plot the power spectrum using
data at the three different resolutions. The 6-element simulation only has one
mode resolved, while 12 elements resolve two modes well, and the 24-element
simulation resolves three modes well. Additionally, we plot the power spectrum
from a 24-element simulation using a fifth-order Dormand-Prince time stepper
instead of the strong stability preserving third-order Runge-Kutta
method. Increasing the time stepper order does not increase the number of radial
modes resolved, demonstrating that it is the spatial resolution that is the
limiting factor.

\begin{figure}
  \centering
  \begin{minipage}{0.45\columnwidth}
    \centering
    \includegraphics[width=1\textwidth]{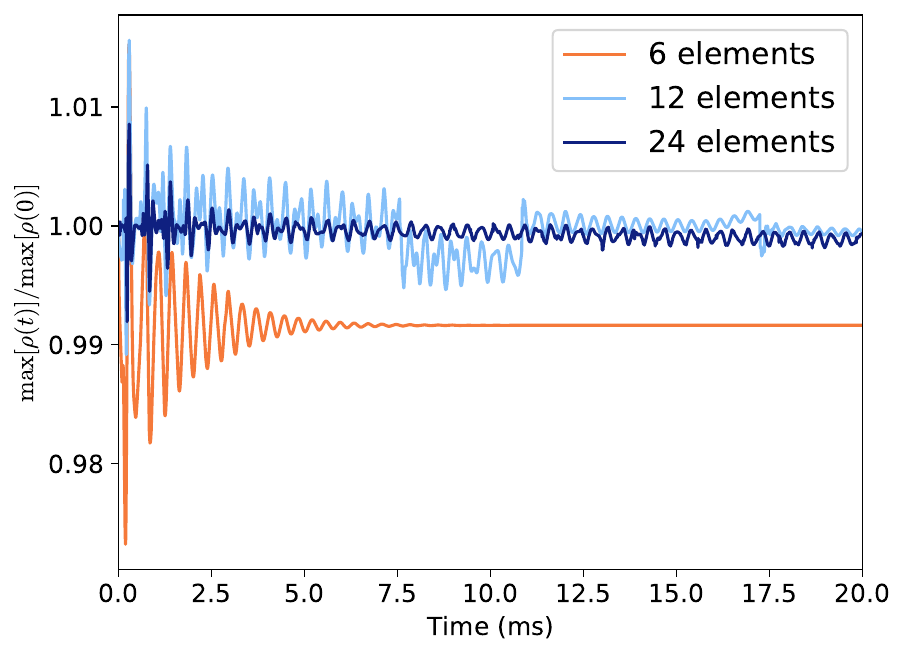}
  \end{minipage}
  \begin{minipage}{0.41\columnwidth}
    \centering
    \includegraphics[width=1\textwidth]{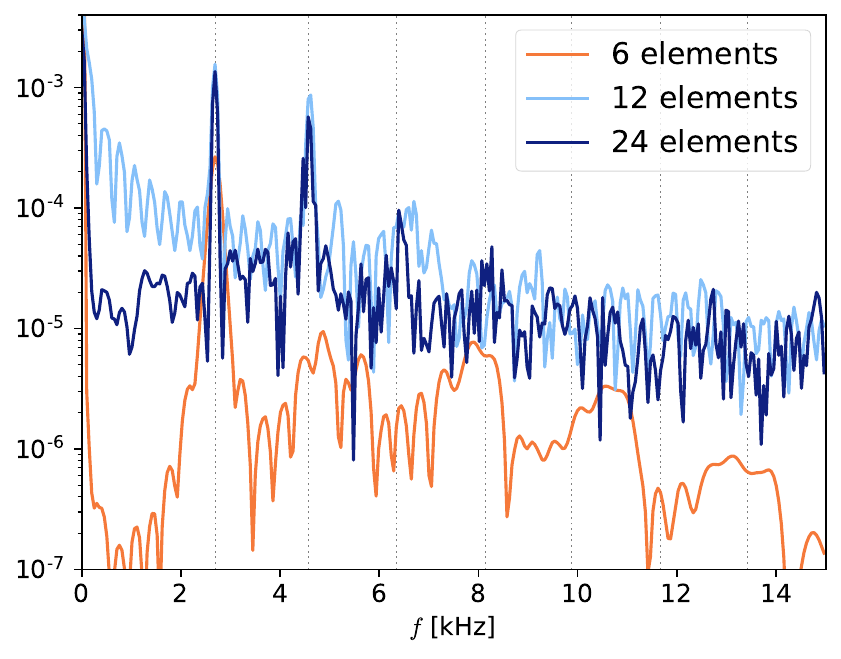}
  \end{minipage}
  \caption{\label{fig:GrmhdMagnetizedTovStar}A plot of
    $\max[\rho(t)]/\max[\rho(0)]$ at three different resolution (left panel) for
    the magnetized TOV star.  The 6-element simulation uses FD throughout the
    interior of the star, while 12- and 24-element simulations use DG.  The
    maximum density in the 6-element case drifts down at early times because of
    the low resolution and the relatively low accuracy of using FD at the
    center. The power spectrum of the maximum density for the three different
    resolution is plotted in the right panel. The vertical dashed lines
    correspond to the known frequencies in the Cowling approximation (which are
    the same as the non-magnetized case as the magnetic field is a small
    perturbation on the dynamics). When the high-order DG scheme is used, more
    oscillation frequencies are resolved.}
\end{figure}

We show the normalized maximum rest mass density over the grid for the
magnetized TOV star in the left panel of
figure~\ref{fig:GrmhdMagnetizedTovStar}. Overall the results are nearly
identical to the non-magnetized case. One notable difference is the decrease in
the 12-element simulation between 7.5ms and 11ms, which occurs because the code
switches from DG to FD at the center of the star at 7.5ms and back to DG at
11ms. Nevertheless, the frequencies are resolved just as well for the magnetized
star as for the non-magnetized case, as can be seen in the right panel of
figure~\ref{fig:GrmhdMagnetizedTovStar} where we plot the power
spectrum. Specifically, we are able to resolve the three largest modes with our
P$_5$ DG-FD hybrid scheme. To the best of our knowledge, these are the first
simulations of a magnetized neutron star using high-order DG methods.

\subsubsection{Rotating neutron star\label{sec:rotating neutron star}}
\begin{figure}
  \centering
  \begin{minipage}{0.45\columnwidth}
    \centering
    \includegraphics[width=1\textwidth]{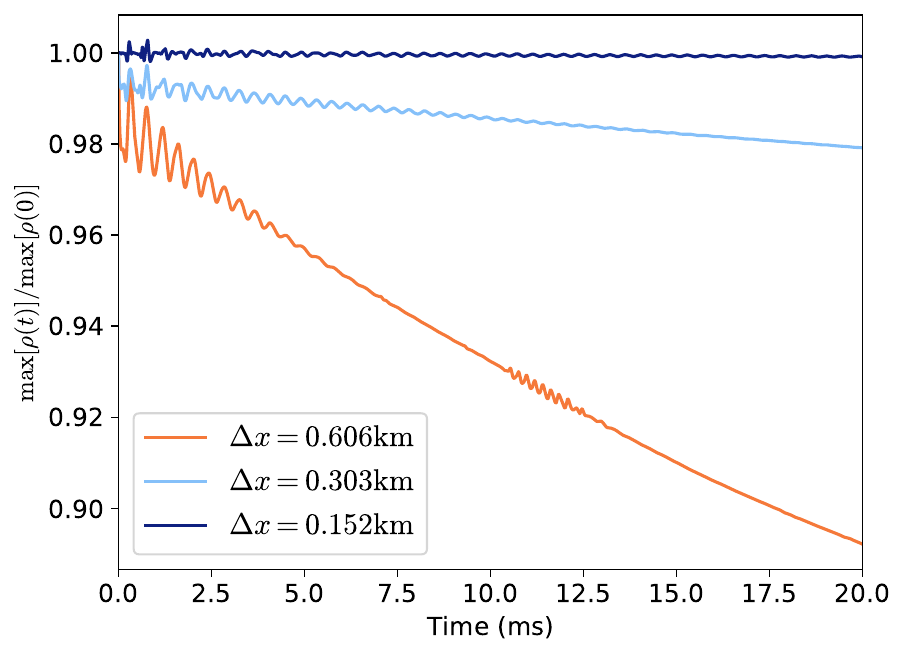}
  \end{minipage}
  \begin{minipage}{0.41\columnwidth}
    \centering
    \includegraphics[width=1\textwidth]{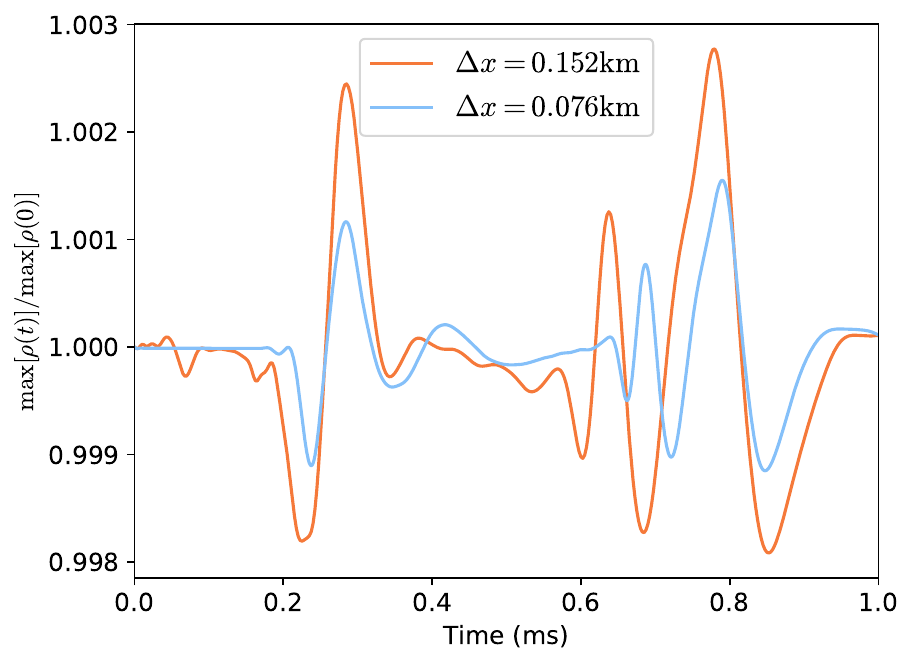}
  \end{minipage}
  \caption{\label{fig:GrmhdRotatingStar}The left panel shows
    $\max[\rho(t)]/\max[\rho(0)]$ at three different resolution for the
    uniformly rotating neutron star.  The 6-element simulation uses FD
    throughout the interior of the star, the 12-element simulation uses mostly
    DG in the stellar interior, and the 24-element simulations use DG everywhere
    except the stellar surface. \blue{The right panel shows
    $\max[\rho(t)]/\max[\rho(0)]$ for a shorter period of time but at very high
    resolutions, 24 elements and 48 elements. Increasing from 24 to 48 elements
    further reduces oscillations in the maximum density.}}
\end{figure}

As a final test case we simulate a uniformly rotating neutron star with ratio of
polar to equatorial radius of 0.7, similar to that
of~\cite{2002PhRvD..65h4024F}. The initial data is constructed using the method
described in~\cite{1992ApJ...398..203C, 1994ApJ...424..823C}. We use the same
polytropic equation of state as for the TOV star evolution. In \blue{the left
  panel of}
figure~\ref{fig:GrmhdRotatingStar} we show the maximum of the rest mass density
for the same three resolutions used for the TOV star simulations. The lowest
resolution uses FD throughout the star and sees a rapid decay in the
density. The 12-element simulation uses DG throughout most of the interior of
the star and so sees significant less decay in the density. The 24-element
simulation uses DG everywhere in the interior of the star and sees less than a
0.1\% decay in the density over 20ms. \blue{To further test convergence we plot
  the maximum density of a 1ms long 24-element and 48-element simulation in the
  right panel of figure~\ref{fig:GrmhdRotatingStar}. The oscillations continue
  to decrease with increasing resolution. We attribute the decay in the maximum
  density at lower resolutions to the dissipative nature of FD schemes,
  consistent with the rapid reduction in decay of the maximum density when
  switching to DG and increasing the resolution.} These are also the first
simulations of a rotating neutron star using high-order DG methods.

\section{Conclusions\label{sec:dgscl conclusions}}
In this paper we gave a detailed description of our DG-FD hybrid method that can
successfully solve challenging relativistic astrophysics test problems like the
simulation of a magnetized or rotating neutron star. Our method combines
an unlimited DG solver with a conservative FD solver. Alternatively, this can be
thought of as taking a standard FD code in numerical relativity and compressing
the data to a DG grid wherever the solution is smooth. The DG solver is more
efficient than the FD solver since no reconstruction is necessary and fewer
Riemann problems need to be solved. In theory a speedup of about eight is
achievable, though we have not optimized our code
\texttt{SpECTRE}~\cite{deppe_nils_2021_5501002} enough and so we find in
practice a speedup of about two to three when comparing the hybrid method to
using FD everywhere. The basic idea of the hybrid scheme is similar
to~\cite{doi:10.1002/fld.2654, 10.1007/978-3-319-05591-6_96, Dumbser2014a,
  BOSCHERI2017449}. An unlimited DG solver is used wherever a troubled-cell
indicator deems the DG solution admissible, while a FD solver is used
elsewhere. Unlike classical limiting strategies like WENO which attempt to
filter out unphysical oscillations, the hybrid scheme prevents spurious
oscillations from entering the solution. This is achieved by retaking any time
step using a robust high-resolution shock-capturing conservative FD where the DG
solution was inadmissible, either because the DG scheme produced unphysical
results like negative densities or because a numerical criterion like the
percentage of power in the highest modes deemed the DG solution bad. Our DG-FD
hybrid scheme was used to perform what is to the best of our knowledge the first
ever simulations of a magnetized TOV star and rotating neutron star using
DG methods. In the future we plan to extend the hybrid scheme to curved meshes,
simulations in full general relativity where the metric is evolved, and to use
positivity-preserving adaptive-order FD methods in order to maintain the highest
order possible even when using FD instead of DG.

\ack

Charm++/Converse~\cite{laxmikant_kale_2020_3972617} was developed by the Parallel
Programming Laboratory in the Department of Computer Science at the University
of Illinois at Urbana-Champaign. The figures in this article were produced with
\texttt{matplotlib}~\cite{Hunter:2007, thomas_a_caswell_2020_3948793},
\texttt{TikZ}~\cite{tikz} and \texttt{ParaView}~\cite{paraview,
  paraview2}. Computations were performed with the Wheeler cluster at
Caltech. This work was supported in part by the Sherman Fairchild Foundation and
by NSF Grants No.~PHY-2011961, No.~PHY-2011968, and No.~OAC-1931266 at Caltech,
and NSF Grants No.~PHY- 1912081 and No.~OAC-1931280 at Cornell.

\appendix
\section{Curved hexahedral elements and moving meshes\label{sec:dgscl curved
    moving mesh}}
We have not yet implemented support for curved hexahedral meshes into
SpECTRE. However, we have given careful consideration on how they could be
implemented. In this appendix we discuss two possible implementations, one that
requires many additional ghost cells with dimension-by-dimension reconstruction,
and one that requires multidimensional reconstruction but no additional ghost
cells.

Support for curved hexahedral or rectangular meshes can be achieved by combining
the DG scheme with a multipatch or multidomain FD scheme. We will discuss only
the 2d case, since the 3d case has more tedious bookkeeping, but otherwise is a
straightforward extension. As a concrete example, we consider a 2d disk made out
of a square surrounded by four wedges as shown in figure~\ref{fig:dgscl
  multipatch wedge}. We focus on an element at the top right corner of the
central square and its neighbors, highlighted by the dashed squared in
figure~\ref{fig:dgscl multipatch wedge}. We will first discuss how to handle the
boundaries when a pair of neighboring elements are using the FD scheme, and then
consider the case when one element is using DG and the other FD.

\begin{figure}[h]
  \centering
  \begin{tikzpicture}[ hatch distance/.store in=\hatchdistance, hatch
    distance=30pt, scale=1.0]
    \draw[-,very thick] (0,0) circle (2.5);
    \draw[-,very thick] (-1,-1) rectangle (1,1);
    \draw[-,very thick] (1,1) -- (1.7677669529663687, 1.7677669529663687);
    \draw[-,very thick] (1,-1) -- (1.7677669529663687, -1.7677669529663687);
    \draw[-,very thick] (-1,1) -- (-1.7677669529663687, 1.7677669529663687);
    \draw[-,very thick] (-1,-1) -- (-1.7677669529663687, -1.7677669529663687);

    \draw[-,very thick,draw=red,dashed] (0.7,0.7) rectangle (1.3,1.3);
  \end{tikzpicture}
  \caption{A 2d disk made out of a central square surrounded by four wedges. In
    the text we describe the method of handling intercell fluxes for the
    elements inside the dashed square.\label{fig:dgscl multipatch wedge}}
\end{figure}

In figure~\ref{fig:dgscl multipatch} we illustrate the domain setup, showing the
subcell center points as circles in the two elements of interest. The diamonds
in left panel of figure~\ref{fig:dgscl multipatch} represent the ghost cells needed for
reconstruction to the element boundary in the element on the right. We use
diagonal dotted lines to trace out lines of constant reference coordinates in
the element on the right and dashed lines in the element on the left. Notice
that the dashed and dotted lines intersect on the element boundary. This is
because the mapping from the reference frame is continuous across element
boundaries and allows us to have a conservative scheme using centered stencils
even in the multipatch case.

\begin{figure}[h]
  \centering
  \begin{minipage}{0.3\textwidth}
    \begin{center}
    \begin{tikzpicture}[ hatch distance/.store in=\hatchdistance, hatch
      distance=30pt, scale=1.9]

      \draw[-, very thick] (1,-1) -- (1,1); \draw[-, very thick] (-0.6,1) --
      (1,1); \draw[-, very thick] (-0.6,-1) -- (1,-1);

      \node[circle,draw=black, fill=blue, inner sep=0pt,minimum size=5pt] (b) at
      (-0.4, -0.8) {}; \node[circle,draw=black, fill=blue, inner sep=0pt,minimum
      size=5pt] (b) at (-0.4, -0.4) {}; \node[circle,draw=black, fill=blue,
      inner sep=0pt,minimum size=5pt] (b) at (-0.4, 0.0) {};
      \node[circle,draw=black, fill=blue, inner sep=0pt,minimum size=5pt] (b) at
      (-0.4, 0.4) {}; \node[circle,draw=black, fill=blue, inner sep=0pt,minimum
      size=5pt] (b) at (-0.4, 0.8) {};

      \node[circle,draw=black, fill=blue, inner sep=0pt,minimum size=5pt] (b) at
      (0.0, -0.8) {}; \node[circle,draw=black, fill=blue, inner sep=0pt,minimum
      size=5pt] (b) at (0.0, -0.4) {}; \node[circle,draw=black, fill=blue, inner
      sep=0pt,minimum size=5pt] (b) at (0.0, 0.0) {}; \node[circle,draw=black,
      fill=blue, inner sep=0pt,minimum size=5pt] (b) at (0.0, 0.4) {};
      \node[circle,draw=black, fill=blue, inner sep=0pt,minimum size=5pt] (b) at
      (0.0, 0.8) {};

      \node[circle,draw=black, fill=blue, inner sep=0pt,minimum size=5pt] (b) at
      (0.4, -0.8) {}; \node[circle,draw=black, fill=blue, inner sep=0pt,minimum
      size=5pt] (b) at (0.4, -0.4) {}; \node[circle,draw=black, fill=blue, inner
      sep=0pt,minimum size=5pt] (b) at (0.4, 0.0) {}; \node[circle,draw=black,
      fill=blue, inner sep=0pt,minimum size=5pt] (b) at (0.4, 0.4) {};
      \node[circle,draw=black, fill=blue, inner sep=0pt,minimum size=5pt] (b) at
      (0.4, 0.8) {};

      \node[circle,draw=black, fill=blue, inner sep=0pt,minimum size=5pt] (b) at
      (0.8, -0.8) {}; \node[circle,draw=black, fill=blue, inner sep=0pt,minimum
      size=5pt] (b) at (0.8, -0.4) {}; \node[circle,draw=black, fill=blue, inner
      sep=0pt,minimum size=5pt] (b) at (0.8, 0.0) {}; \node[circle,draw=black,
      fill=blue, inner sep=0pt,minimum size=5pt] (b) at (0.8, 0.4) {};
      \node[circle,draw=black, fill=blue, inner sep=0pt,minimum size=5pt] (b) at
      (0.8, 0.8) {};

      \draw[-, very thick] (1,1) -- (2.1,2.1); \draw[-, very thick] (1,-1) --
      (2.1,-0.2);

      \draw[very thick, dotted, draw=red] (2.100, 1.790) -- (0.530, 0.377);
      \draw[very thick, dotted, draw=red] (2.100, 1.379) -- (0.530, -0.054);
      \draw[very thick, dotted, draw=red] (2.100, 0.968) -- (0.530, -0.414);
      \draw[very thick, dotted, draw=red] (2.100, 0.557) -- (0.530, -0.809);
      \draw[very thick, dotted, draw=red] (2.100, 0.124) -- (0.530, -1.195);

      \draw[very thick, dashed, draw=blue] (0.0, 0.8) -- (1.4, 0.8); \draw[very
      thick, dashed, draw=blue] (0.0, 0.4) -- (1.4, 0.4); \draw[very thick,
      dashed, draw=blue] (0.0, 0.0) -- (1.4, 0.0); \draw[very thick, dashed,
      draw=blue] (0.0, -0.4) -- (1.4, -0.4); \draw[very thick, dashed,
      draw=blue] (0.0, -0.8) -- (1.4, -0.8);
      \node[circle,draw=black, fill=blue, inner sep=0pt,minimum size=5pt] (b) at
      (1.180, 0.962) {}; \node[circle,draw=black, fill=blue, inner
      sep=0pt,minimum size=5pt] (b) at (1.620, 1.358) {};
      \node[circle,draw=black, fill=blue, inner sep=0pt,minimum size=5pt] (b) at
      (2.100, 1.790) {}; \node[circle,draw=black, fill=blue, inner
      sep=0pt,minimum size=5pt] (b) at (1.180, 0.560) {};
      \node[circle,draw=black, fill=blue, inner sep=0pt,minimum size=5pt] (b) at
      (1.620, 0.952) {}; \node[circle,draw=black, fill=blue, inner
      sep=0pt,minimum size=5pt] (b) at (2.100, 1.379) {};
      \node[circle,draw=black, fill=blue, inner sep=0pt,minimum size=5pt] (b) at
      (1.180, 0.158) {}; \node[circle,draw=black, fill=blue, inner
      sep=0pt,minimum size=5pt] (b) at (1.620, 0.546) {};
      \node[circle,draw=black, fill=blue, inner sep=0pt,minimum size=5pt] (b) at
      (2.100, 0.968) {}; \node[circle,draw=black, fill=blue, inner
      sep=0pt,minimum size=5pt] (b) at (1.180, -0.243) {};
      \node[circle,draw=black, fill=blue, inner sep=0pt,minimum size=5pt] (b) at
      (1.620, 0.139) {}; \node[circle,draw=black, fill=blue, inner
      sep=0pt,minimum size=5pt] (b) at (2.100, 0.557) {};
      \node[circle,draw=black, fill=blue, inner sep=0pt,minimum size=5pt] (b) at
      (1.180, -0.649) {}; \node[circle,draw=black, fill=blue, inner
      sep=0pt,minimum size=5pt] (b) at (1.620, -0.279) {};
      \node[circle,draw=black, fill=blue, inner sep=0pt,minimum size=5pt] (b) at
      (2.100, 0.124) {};

      \node[diamond,draw=black, fill=red, inner sep=0pt,minimum size=7pt] (b) at
      (0.530, 0.377) {}; \node[diamond,draw=black, fill=red, inner
      sep=0pt,minimum size=7pt] (b) at (0.840, 0.656) {};
      \node[diamond,draw=black, fill=red, inner sep=0pt,minimum size=7pt] (b) at
      (0.530, -0.018) {}; \node[diamond,draw=black, fill=red, inner
      sep=0pt,minimum size=7pt] (b) at (0.840, 0.258) {};
      \node[diamond,draw=black, fill=red, inner sep=0pt,minimum size=7pt] (b) at
      (0.530, -0.414) {}; \node[diamond,draw=black, fill=red, inner
      sep=0pt,minimum size=7pt] (b) at (0.840, -0.141) {};
      \node[diamond,draw=black, fill=red, inner sep=0pt,minimum size=7pt] (b) at
      (0.530, -0.809) {}; \node[diamond,draw=black, fill=red, inner
      sep=0pt,minimum size=7pt] (b) at (0.840, -0.539) {};
      \node[diamond,draw=black, fill=red, inner sep=0pt,minimum size=7pt] (b) at
      (0.530, -1.195) {}; \node[diamond,draw=black, fill=red, inner
      sep=0pt,minimum size=7pt] (b) at (0.840, -0.934) {};
    \end{tikzpicture}
  \end{center}
  {\footnotesize
    An illustration of the ghost points needed for the FD scheme where
    neighboring elements do not have aligned coordinate axes in their reference
    frames. Circles denote the cell-center FD points in the elements, and
    diamonds denote the ghost cells needed for reconstruction in the element on
    the right. The diagonal dotted lines trace out lines of constant reference
    coordinates in the element on the right, and dashed lines in the element on
    the left. Notice that the dashed and dotted lines intersect on the element
    boundary.}
  \end{minipage}\hfill
  \begin{minipage}{0.3\textwidth}
    \begin{center}
    \begin{tikzpicture}[ hatch distance/.store in=\hatchdistance, hatch
      distance=30pt, scale=1.9]

      \draw[-, very thick] (1,-1) -- (1,1); \draw[-, very thick] (-0.6,1) --
      (1,1); \draw[-, very thick] (-0.6,-1) -- (1,-1);

      \node[circle,draw=black, fill=blue, inner sep=0pt,minimum size=5pt] (b) at
      (-0.4, -0.8) {}; \node[circle,draw=black, fill=blue, inner sep=0pt,minimum
      size=5pt] (b) at (-0.4, -0.4) {}; \node[circle,draw=black, fill=blue,
      inner sep=0pt,minimum size=5pt] (b) at (-0.4, 0.0) {};
      \node[circle,draw=black, fill=blue, inner sep=0pt,minimum size=5pt] (b) at
      (-0.4, 0.4) {}; \node[circle,draw=black, fill=blue, inner sep=0pt,minimum
      size=5pt] (b) at (-0.4, 0.8) {};

      \node[circle,draw=black, fill=blue, inner sep=0pt,minimum size=5pt] (b) at
      (0.0, -0.8) {}; \node[circle,draw=black, fill=blue, inner sep=0pt,minimum
      size=5pt] (b) at (0.0, -0.4) {}; \node[circle,draw=black, fill=blue, inner
      sep=0pt,minimum size=5pt] (b) at (0.0, 0.0) {}; \node[circle,draw=black,
      fill=blue, inner sep=0pt,minimum size=5pt] (b) at (0.0, 0.4) {};
      \node[circle,draw=black, fill=blue, inner sep=0pt,minimum size=5pt] (b) at
      (0.0, 0.8) {};

      \node[circle,draw=black, fill=blue, inner sep=0pt,minimum size=5pt] (b) at
      (0.4, -0.8) {}; \node[circle,draw=black, fill=blue, inner sep=0pt,minimum
      size=5pt] (b) at (0.4, -0.4) {}; \node[circle,draw=black, fill=blue, inner
      sep=0pt,minimum size=5pt] (b) at (0.4, 0.0) {}; \node[circle,draw=black,
      fill=blue, inner sep=0pt,minimum size=5pt] (b) at (0.4, 0.4) {};
      \node[circle,draw=black, fill=blue, inner sep=0pt,minimum size=5pt] (b) at
      (0.4, 0.8) {};

      \node[circle,draw=black, fill=blue, inner sep=0pt,minimum size=5pt] (b) at
      (0.8, -0.8) {}; \node[circle,draw=black, fill=blue, inner sep=0pt,minimum
      size=5pt] (b) at (0.8, -0.4) {}; \node[circle,draw=black, fill=blue, inner
      sep=0pt,minimum size=5pt] (b) at (0.8, 0.0) {}; \node[circle,draw=black,
      fill=blue, inner sep=0pt,minimum size=5pt] (b) at (0.8, 0.4) {};
      \node[circle,draw=black, fill=blue, inner sep=0pt,minimum size=5pt] (b) at
      (0.8, 0.8) {};


      \node[regular polygon,regular polygon sides=3,draw=black, fill=violet,
      inner sep=0pt,minimum size=8pt] (b) at (1.2, -0.8) {}; \node[regular
      polygon,regular polygon sides=3,draw=black, fill=violet, inner
      sep=0pt,minimum size=8pt] (b) at (1.2, -0.4) {}; \node[regular
      polygon,regular polygon sides=3,draw=black, fill=violet, inner
      sep=0pt,minimum size=8pt] (b) at (1.2, 0.0) {}; \node[regular
      polygon,regular polygon sides=3,draw=black, fill=violet, inner
      sep=0pt,minimum size=8pt] (b) at (1.2, 0.4) {}; \node[regular
      polygon,regular polygon sides=3,draw=black, fill=violet, inner
      sep=0pt,minimum size=8pt] (b) at (1.2, 0.8) {};

      \node[regular polygon,regular polygon sides=3,draw=black, fill=violet,
      inner sep=0pt,minimum size=8pt] (b) at (1.6, -0.8) {}; \node[regular
      polygon,regular polygon sides=3,draw=black, fill=violet, inner
      sep=0pt,minimum size=8pt] (b) at (1.6, -0.4) {}; \node[regular
      polygon,regular polygon sides=3,draw=black, fill=violet, inner
      sep=0pt,minimum size=8pt] (b) at (1.6, 0.0) {}; \node[regular
      polygon,regular polygon sides=3,draw=black, fill=violet, inner
      sep=0pt,minimum size=8pt] (b) at (1.6, 0.4) {}; \node[regular
      polygon,regular polygon sides=3,draw=black, fill=violet, inner
      sep=0pt,minimum size=8pt] (b) at (1.6, 0.8) {};

      \node[regular polygon,regular polygon sides=3,draw=black, fill=violet,
      inner sep=0pt,minimum size=8pt] (b) at (-0.4, 1.2) {}; \node[regular
      polygon,regular polygon sides=3,draw=black, fill=violet, inner
      sep=0pt,minimum size=8pt] (b) at (0.0, 1.2) {}; \node[regular
      polygon,regular polygon sides=3,draw=black, fill=violet, inner
      sep=0pt,minimum size=8pt] (b) at (0.4, 1.2) {}; \node[regular
      polygon,regular polygon sides=3,draw=black, fill=violet, inner
      sep=0pt,minimum size=8pt] (b) at (0.8, 1.2) {};

      \node[regular polygon,regular polygon sides=3,draw=black, fill=violet,
      inner sep=0pt,minimum size=8pt] (b) at (-0.4, 1.6) {}; \node[regular
      polygon,regular polygon sides=3,draw=black, fill=violet, inner
      sep=0pt,minimum size=8pt] (b) at (0.0, 1.6) {}; \node[regular
      polygon,regular polygon sides=3,draw=black, fill=violet, inner
      sep=0pt,minimum size=8pt] (b) at (0.4, 1.6) {}; \node[regular
      polygon,regular polygon sides=3,draw=black, fill=violet, inner
      sep=0pt,minimum size=8pt] (b) at (0.8, 1.6) {};

      \node[regular polygon,regular polygon sides=3,draw=black, fill=violet,
      inner sep=0pt,minimum size=8pt] (b) at (1.2, 1.2) {}; \node[regular
      polygon,regular polygon sides=3,draw=black, fill=violet, inner
      sep=0pt,minimum size=8pt] (b) at (1.6, 1.2) {}; \node[regular
      polygon,regular polygon sides=3,draw=black, fill=violet, inner
      sep=0pt,minimum size=8pt] (b) at (1.2, 1.6) {}; \node[regular
      polygon,regular polygon sides=3,draw=black, fill=violet, inner
      sep=0pt,minimum size=8pt] (b) at (1.6, 1.6) {};

      \draw[-, very thick] (1,1) -- (2.1,2.1); \draw[-, very thick] (1,-1) --
      (2.1,-0.2);

      \node[circle,draw=black, fill=blue, inner sep=0pt,minimum size=5pt] (b) at
      (1.180, 0.962) {}; \node[circle,draw=black, fill=blue, inner
      sep=0pt,minimum size=5pt] (b) at (1.620, 1.358) {};
      \node[circle,draw=black, fill=blue, inner sep=0pt,minimum size=5pt] (b) at
      (2.100, 1.790) {}; \node[circle,draw=black, fill=blue, inner
      sep=0pt,minimum size=5pt] (b) at (1.180, 0.560) {};
      \node[circle,draw=black, fill=blue, inner sep=0pt,minimum size=5pt] (b) at
      (1.620, 0.952) {}; \node[circle,draw=black, fill=blue, inner
      sep=0pt,minimum size=5pt] (b) at (2.100, 1.379) {};
      \node[circle,draw=black, fill=blue, inner sep=0pt,minimum size=5pt] (b) at
      (1.180, 0.158) {}; \node[circle,draw=black, fill=blue, inner
      sep=0pt,minimum size=5pt] (b) at (1.620, 0.546) {};
      \node[circle,draw=black, fill=blue, inner sep=0pt,minimum size=5pt] (b) at
      (2.100, 0.968) {}; \node[circle,draw=black, fill=blue, inner
      sep=0pt,minimum size=5pt] (b) at (1.180, -0.243) {};
      \node[circle,draw=black, fill=blue, inner sep=0pt,minimum size=5pt] (b) at
      (1.620, 0.139) {}; \node[circle,draw=black, fill=blue, inner
      sep=0pt,minimum size=5pt] (b) at (2.100, 0.557) {};
      \node[circle,draw=black, fill=blue, inner sep=0pt,minimum size=5pt] (b) at
      (1.180, -0.649) {}; \node[circle,draw=black, fill=blue, inner
      sep=0pt,minimum size=5pt] (b) at (1.620, -0.279) {};
      \node[circle,draw=black, fill=blue, inner sep=0pt,minimum size=5pt] (b) at
      (2.100, 0.124) {};
    \end{tikzpicture}
  \end{center}
  {\footnotesize
    An illustration of extending the FD element by additional cells in order to
    support high-order reconstruction to arbitrary points inside the element, as
    discussed in the text. The additional cells for the central element are
    shown as purple triangles. These additional cells are evolved alongside the
    cells inside the element.}
  \end{minipage}\hfill
  \begin{minipage}{0.3\textwidth}
    \begin{center}
    \begin{tikzpicture}[ hatch distance/.store in=\hatchdistance, hatch
      distance=30pt, scale=1.9]

      \draw[-, very thick] (1,-1) -- (1,1); \draw[-, very thick] (-0.6,1) --
      (1,1); \draw[-, very thick] (-0.6,-1) -- (1,-1);

      \node[circle,draw=black, fill=blue, inner sep=0pt,minimum size=5pt] (b) at
      (-0.4, -0.8) {}; \node[circle,draw=black, fill=blue, inner sep=0pt,minimum
      size=5pt] (b) at (-0.4, -0.4) {}; \node[circle,draw=black, fill=blue,
      inner sep=0pt,minimum size=5pt] (b) at (-0.4, 0.0) {};
      \node[circle,draw=black, fill=blue, inner sep=0pt,minimum size=5pt] (b) at
      (-0.4, 0.4) {}; \node[circle,draw=black, fill=blue, inner sep=0pt,minimum
      size=5pt] (b) at (-0.4, 0.8) {};

      \node[circle,draw=black, fill=blue, inner sep=0pt,minimum size=5pt] (b) at
      (0.0, -0.8) {}; \node[circle,draw=black, fill=blue, inner sep=0pt,minimum
      size=5pt] (b) at (0.0, -0.4) {}; \node[circle,draw=black, fill=blue, inner
      sep=0pt,minimum size=5pt] (b) at (0.0, 0.0) {}; \node[circle,draw=black,
      fill=blue, inner sep=0pt,minimum size=5pt] (b) at (0.0, 0.4) {};
      \node[circle,draw=black, fill=blue, inner sep=0pt,minimum size=5pt] (b) at
      (0.0, 0.8) {};

      \node[circle,draw=black, fill=blue, inner sep=0pt,minimum size=5pt] (b) at
      (0.4, -0.8) {}; \node[circle,draw=black, fill=blue, inner sep=0pt,minimum
      size=5pt] (b) at (0.4, -0.4) {}; \node[circle,draw=black, fill=blue, inner
      sep=0pt,minimum size=5pt] (b) at (0.4, 0.0) {}; \node[circle,draw=black,
      fill=blue, inner sep=0pt,minimum size=5pt] (b) at (0.4, 0.4) {};
      \node[circle,draw=black, fill=blue, inner sep=0pt,minimum size=5pt] (b) at
      (0.4, 0.8) {};

      \node[circle,draw=black, fill=blue, inner sep=0pt,minimum size=5pt] (b) at
      (0.8, -0.8) {}; \node[circle,draw=black, fill=blue, inner sep=0pt,minimum
      size=5pt] (b) at (0.8, -0.4) {}; \node[circle,draw=black, fill=blue, inner
      sep=0pt,minimum size=5pt] (b) at (0.8, 0.0) {}; \node[circle,draw=black,
      fill=blue, inner sep=0pt,minimum size=5pt] (b) at (0.8, 0.4) {};
      \node[circle,draw=black, fill=blue, inner sep=0pt,minimum size=5pt] (b) at
      (0.8, 0.8) {};


      \node[regular polygon,regular polygon sides=3,draw=black, fill=violet,
      inner sep=0pt,minimum size=8pt] (b) at (1.2, -0.8) {}; \node[regular
      polygon,regular polygon sides=3,draw=black, fill=violet, inner
      sep=0pt,minimum size=8pt] (b) at (1.2, -0.4) {}; \node[regular
      polygon,regular polygon sides=3,draw=black, fill=violet, inner
      sep=0pt,minimum size=8pt] (b) at (1.2, 0.0) {}; \node[regular
      polygon,regular polygon sides=3,draw=black, fill=violet, inner
      sep=0pt,minimum size=8pt] (b) at (1.2, 0.4) {}; \node[regular
      polygon,regular polygon sides=3,draw=black, fill=violet, inner
      sep=0pt,minimum size=8pt] (b) at (1.2, 0.8) {};

      \node[regular polygon,regular polygon sides=3,draw=black, fill=violet,
      inner sep=0pt,minimum size=8pt] (b) at (1.6, -0.8) {}; \node[regular
      polygon,regular polygon sides=3,draw=black, fill=violet, inner
      sep=0pt,minimum size=8pt] (b) at (1.6, -0.4) {}; \node[regular
      polygon,regular polygon sides=3,draw=black, fill=violet, inner
      sep=0pt,minimum size=8pt] (b) at (1.6, 0.0) {}; \node[regular
      polygon,regular polygon sides=3,draw=black, fill=violet, inner
      sep=0pt,minimum size=8pt] (b) at (1.6, 0.4) {}; \node[regular
      polygon,regular polygon sides=3,draw=black, fill=violet, inner
      sep=0pt,minimum size=8pt] (b) at (1.6, 0.8) {};

      \node[regular polygon,regular polygon sides=3,draw=black, fill=violet,
      inner sep=0pt,minimum size=8pt] (b) at (-0.4, 1.2) {}; \node[regular
      polygon,regular polygon sides=3,draw=black, fill=violet, inner
      sep=0pt,minimum size=8pt] (b) at (0.0, 1.2) {}; \node[regular
      polygon,regular polygon sides=3,draw=black, fill=violet, inner
      sep=0pt,minimum size=8pt] (b) at (0.4, 1.2) {}; \node[regular
      polygon,regular polygon sides=3,draw=black, fill=violet, inner
      sep=0pt,minimum size=8pt] (b) at (0.8, 1.2) {};

      \node[regular polygon,regular polygon sides=3,draw=black, fill=violet,
      inner sep=0pt,minimum size=8pt] (b) at (-0.4, 1.6) {}; \node[regular
      polygon,regular polygon sides=3,draw=black, fill=violet, inner
      sep=0pt,minimum size=8pt] (b) at (0.0, 1.6) {}; \node[regular
      polygon,regular polygon sides=3,draw=black, fill=violet, inner
      sep=0pt,minimum size=8pt] (b) at (0.4, 1.6) {}; \node[regular
      polygon,regular polygon sides=3,draw=black, fill=violet, inner
      sep=0pt,minimum size=8pt] (b) at (0.8, 1.6) {};

      \node[regular polygon,regular polygon sides=3,draw=black, fill=violet,
      inner sep=0pt,minimum size=8pt] (b) at (1.2, 1.2) {}; \node[regular
      polygon,regular polygon sides=3,draw=black, fill=violet, inner
      sep=0pt,minimum size=8pt] (b) at (1.6, 1.2) {}; \node[regular
      polygon,regular polygon sides=3,draw=black, fill=violet, inner
      sep=0pt,minimum size=8pt] (b) at (1.2, 1.6) {}; \node[regular
      polygon,regular polygon sides=3,draw=black, fill=violet, inner
      sep=0pt,minimum size=8pt] (b) at (1.6, 1.6) {};

      \draw[-, very thick] (1,1) -- (2.1,2.1); \draw[-, very thick] (1,-1) --
      (2.1,-0.2);

      \node[circle,draw=black, fill=blue, inner sep=0pt,minimum size=5pt] (b) at
      (1.180, 0.962) {}; \node[circle,draw=black, fill=blue, inner
      sep=0pt,minimum size=5pt] (b) at (1.620, 1.358) {};
      \node[circle,draw=black, fill=blue, inner sep=0pt,minimum size=5pt] (b) at
      (2.100, 1.790) {}; \node[circle,draw=black, fill=blue, inner
      sep=0pt,minimum size=5pt] (b) at (1.180, 0.560) {};
      \node[circle,draw=black, fill=blue, inner sep=0pt,minimum size=5pt] (b) at
      (1.620, 0.952) {}; \node[circle,draw=black, fill=blue, inner
      sep=0pt,minimum size=5pt] (b) at (2.100, 1.379) {};
      \node[circle,draw=black, fill=blue, inner sep=0pt,minimum size=5pt] (b) at
      (1.180, 0.158) {}; \node[circle,draw=black, fill=blue, inner
      sep=0pt,minimum size=5pt] (b) at (1.620, 0.546) {};
      \node[circle,draw=black, fill=blue, inner sep=0pt,minimum size=5pt] (b) at
      (2.100, 0.968) {}; \node[circle,draw=black, fill=blue, inner
      sep=0pt,minimum size=5pt] (b) at (1.180, -0.243) {};
      \node[circle,draw=black, fill=blue, inner sep=0pt,minimum size=5pt] (b) at
      (1.620, 0.139) {}; \node[circle,draw=black, fill=blue, inner
      sep=0pt,minimum size=5pt] (b) at (2.100, 0.557) {};
      \node[circle,draw=black, fill=blue, inner sep=0pt,minimum size=5pt] (b) at
      (1.180, -0.649) {}; \node[circle,draw=black, fill=blue, inner
      sep=0pt,minimum size=5pt] (b) at (1.620, -0.279) {};
      \node[circle,draw=black, fill=blue, inner sep=0pt,minimum size=5pt] (b) at
      (2.100, 0.124) {};

      \draw[very thick, dotted, draw=red] (0.530, 1.6) -- (0.530, -0.8);

      \node[rectangle,draw=black, fill=red, inner sep=0pt,minimum size=5pt] (b)
      at (0.530, 1.6) {}; \node[rectangle,draw=black, fill=red, inner
      sep=0pt,minimum size=5pt] (b) at (0.530, 1.2) {};
      \node[rectangle,draw=black, fill=red, inner sep=0pt,minimum size=5pt] (b)
      at (0.530, 0.8) {}; \node[rectangle,draw=black, fill=red, inner
      sep=0pt,minimum size=5pt] (b) at (0.530, 0.4) {};
      \node[rectangle,draw=black, fill=red, inner sep=0pt,minimum size=5pt] (b)
      at (0.530, 0.0) {}; \node[rectangle,draw=black, fill=red, inner
      sep=0pt,minimum size=5pt] (b) at (0.530, -0.4) {};
      \node[rectangle,draw=black, fill=red, inner sep=0pt,minimum size=5pt] (b)
      at (0.530, -0.8) {};
    \end{tikzpicture}
  \end{center}
  {\footnotesize An illustration of the first stage of the reconstruction to the
    ghost cells needed by the neighboring element on the right. The central
    element reconstructs the solution to a line in the reference coordinates,
    followed by a second reconstruction to the ghost cells that fall on the line
    (not shown for simplicity).}
\end{minipage}
\caption{An illustration of the multipatch or multidomain FD reconstruction
  needed to support curved meshes. We show a 2d example for simplicity. The 3d
  case is a tedious but otherwise straightforward
  generalization.\label{fig:dgscl multipatch}}
\end{figure}

Since we are unable to interpolate to the ghost cells shown in the left panel of
figure~\ref{fig:dgscl multipatch} with centered stencils, one option is to use
non-centered stencils. Using non-centered stencils was explored in
reference~\cite{sebastian2003multidomain}, which did not find any instabilities
from the use of such stencils in their test cases. Another option is to use
reconstruction methods for unstructured meshes (see, for example,~\cite{
  doi:10.1002/fld.1469, TSOUTSANIS2014254, Tsoutsanis2019, FARMAKIS2020112921,
  DUMBSER2007204, doi:10.2514/6.2019-2955} and references therein), though this
adds significant conceptual and technical overhead. Another option is adding
additional subcells that overlap with the neighboring elements to allow the use
of centered reconstruction schemes to interpolate to the ghost cells. These
additional subcells are shown as triangles in the middle panel of
figure~\ref{fig:dgscl multipatch}. We can now do two reconstructions to
reconstruct the ghost cells. First, we reconstruct along one reference axis of
the central element as shown by the squares in the right panel of
figure~\ref{fig:dgscl multipatch}. Next we reconstruct along the other
direction, which is illustrated by the dotted vertical line in the right panel
of figure~\ref{fig:dgscl multipatch}.

In order to maintain conservation between elements, we need to define a unique
left and right state at the boundary of the elements. A unique state can be
obtained by using the average of the reconstructed variables from the diagonal
and horizontal stencils in figure~\ref{fig:dgscl multipatch}. That is, we use the
average of the result obtained from reconstruction in each element for the right
and left states when updating any subcells that need the numerical flux on the
element boundaries. Recall that when using a second-order FD derivative the
semi-discrete evolution equations are (we only show 1d for simplicity since it
is sufficient to illustrate our point)
\begin{eqnarray}
  \label{eq:dgscl second-order FD derivative}
  \partial_t u +
  \frac{\partial\xi}{\partial
  x}\left(
  \frac{\hat{F}^x_{\underline{i}+1/2,\underline{j}} -
  \hat{F}^{x}_{\underline{i}-1/2},\underline{j}}
  {\Delta \xi}\right)=S.
\end{eqnarray}
Thus, as long as all cells that share the boundary on which the numerical fluxes
are defined use the same numerical flux, the scheme is conservative. When using
higher-order derivative approximations the fluxes away from the cell boundaries
are also needed. In the case of the element boundaries we are considering, we do
not have a unique solution in the region of overlap (e.g.~the region covered by
the purple triangles in the middle panel of figure~\ref{fig:dgscl multipatch})
where we compute the fluxes. As a result, we do not know if using high-order FD
derivatives would violate conservation at the element boundaries. However, if
the solution is smooth in this region, small violations of conservation are not
detrimental, and if a discontinuity is passing through the boundary a
second-order FD derivative should be used anyway.

Another method of doing reconstruction at locations where the coordinate axes do
not align is described in~\cite{doi:10.2514/6.2017-0845} for finite-volume
methods. This same approach should be applicable to FD methods. Whether adding
ghost zones or using unstructured mesh reconstruction is easier to implement and
more efficient is unclear and will need to be tested.

\section{Integration weights\label{sec:fd integration weights}}
The standard weights available in textbooks assume the abscissas are distributed
at the boundaries of the subcells, not the subcell centers, and so do not
apply. The weights $R_{\underline{i}}$ are given by integrals over Lagrange
polynomials:
\begin{eqnarray}
  \label{eq:integral coefficients Lagrange polynomials}
  R_{\underline{i}}=\int_{a}^b
  \prod _{\underline{j}=0\atop\underline{j}\ne \underline{i}}^n
  \frac{(x-x_{\underline{j}})}{(x_{\underline{i}}-x_{\underline{j}})}\,dx.
\end{eqnarray}
The integration coefficients are not unique since there are choices on how to
handle points near the boundaries and how to stitch the interior solution
together. Rather than using one-sided or low-order centered stencils near the
boundaries, we choose to integrate from $0$ to $3\Delta x$ for the fourth-order
stencil and from $0$ to $5\Delta x$ for the sixth-order stencils. The
fourth-order stencil at the boundary is
\begin{eqnarray}
  \label{eq:dgscl FD recons integral 3 point coeffs}
  \int_{0}^{3\Delta x}f(x)dx\approx \Delta x\left(\frac{9}{8}f_{1/2}
  +\frac{3}{4}f_{3/2}+\frac{9}{8}f_{5/2}\right),
\end{eqnarray}
and the sixth-order stencil is
\begin{eqnarray}
  \label{eq:dgscl FD five point boundary}
  \int_{0}^{5\Delta x}f(x)dx
  &\approx \Delta x\left(
    \frac{1375}{1152}f_{1/2}
    +\frac{125}{288}f_{3/2}
    +\frac{335}{192}f_{5/2} \right.\nonumber \\
  &\left.+\frac{125}{288}f_{7/2}
    +\frac{1375}{1152}f_{9/2}
    \right).
\end{eqnarray}
If we have more than three (five) points we need to stitch the formulas
together. We do this by integrating from $x_k$ to $x_{k+1}$. For the
fourth-order stencil we get
\begin{eqnarray}
  \label{eq:dgscl FD three point interior}
  \int_{x_k}^{x_{k+1}}f(x)dx\approx
  \Delta x\left(\frac{1}{24}f_{k-1/2}+\frac{11}{12}f_{k+1/2} +
  \frac{1}{24}f_{k+3/2}\right).
\end{eqnarray}
and for the sixth-order stencil we get
\begin{eqnarray}
  \label{eq:dgscl FD five point interior}
  \int_{x_k}^{x_{k+1}}f(x)dx
  &\approx\Delta x\left(
  \frac{-17}{5760}f_{k-3/2}+\frac{308}{5760}f_{k-1/2}
  +\frac{5178}{5760}f_{k+1/2}\right. \nonumber \\
  &\left.+\frac{308}{5760}f_{k+3/2}
    -\frac{17}{5760}f_{k+5/2}\right).
\end{eqnarray}
We present the weights for a fourth-order approximation to the integral in
table~\ref{tab:dgscl fourth order subcell integral weights} and for a
sixth-order approximation to the integral in table~\ref{tab:dgscl sixth order
  subcell integral weights}. The weights are obtained by using~\eref{eq:dgscl FD
  recons integral 3 point coeffs} and~\eref{eq:dgscl FD five point boundary} at
the boundaries and~\eref{eq:dgscl FD three point interior} and~\eref{eq:dgscl FD
  five point interior} on the interior. The stencils are symmetric about the
center and so only half the coefficients are shown.

\begin{table}
  \caption{Weights for a fourth-order approximation to an integral using
    stencils symmetric about the center. Only the first half of the coefficients
    are shown, the second half are such that the stencil is symmetric. The
    number of points in the stencil is shown in the first
    column.\label{tab:dgscl fourth order subcell integral weights}}
  \begin{indented}
    \lineup
  \item[] \def\arraystretch{1.5} \begin{tabular}{cccccc}
    \br
    Number of cells & $x_{1/2}$ & $x_{3/2}$ & $x_{5/2}$ & $x_{7/2}$
           & $x_{9/2}$ \\ \mr
    3 & $\frac{9}{8}$ & $\frac{3}{4}$ & --- & --- & --- \\
    4 & $\frac{13}{12}$ & $\frac{11}{12}$ & --- & --- & --- \\
    5 & $\frac{13}{12}$ & $\frac{21}{24}$ & $\frac{13}{12}$ & --- & --- \\
    6 & $\frac{9}{8}$ & $\frac{3}{4}$ & $\frac{9}{8}$ & --- & --- \\
    7 & $\frac{9}{8}$ & $\frac{3}{4}$ & $\frac{7}{6}$ & $\frac{11}{12}$ & --- \\
    8 & $\frac{9}{8}$ & $\frac{3}{4}$ & $\frac{7}{6}$ & $\frac{23}{24}$ & --- \\
    9+ & $\frac{9}{8}$ & $\frac{3}{4}$ & $\frac{7}{6}$ & $\frac{23}{24}$ & 1 \\
    \br
  \end{tabular}
  \end{indented}
\end{table}
\begin{table}
  \caption{Weights for a sixth-order approximation to an integral using
    stencils symmetric about the center. Only the first half of the coefficients
    are shown, the second half are such that the stencil is symmetric. The
    number of points in the stencil is shown in the first
    column.\label{tab:dgscl sixth order subcell integral weights}}
  \begin{indented}
    \lineup
  \item[]   \def\arraystretch{1.5} \begin{tabular}{@{}ccccccccc}
    \br
    Number of cells & $x_{1/2}$ & $x_{3/2}$ & $x_{5/2}$ & $x_{7/2}$ & $x_{9/2}$
    & $x_{11/2}$ & $x_{13/2}$ & $x_{15/2}$ \\ \mr
    5 & $\frac{1375}{1152}$ & $\frac{125}{288}$ & $\frac{335}{192}$
    & --- & --- & --- & --- & --- \\
    6 & $\frac{741}{640}$ & $\frac{417}{640}$ & $\frac{381}{320}$
    & --- & --- & --- & --- & --- \\
    7 & $\frac{741}{640}$ & $\frac{3547}{5760}$ & $\frac{8111}{5760}$
    & $\frac{611}{960}$ & --- & --- & --- & --- \\
    8 & $\frac{1663}{1440}$ & $\frac{227}{360}$ & $\frac{323}{240}$
    & $\frac{139}{160}$ & --- & --- & --- & --- \\
    9 & $\frac{1663}{1440}$ & $\frac{227}{360}$ & $\frac{1547}{1152}$
    & $\frac{245}{288}$ & $\frac{3001}{2880}$ & --- & --- & --- \\
    10 & $\frac{1375}{1152}$ & $\frac{125}{288}$ & $\frac{335}{192}$
    & $\frac{125}{288}$ & $\frac{1375}{1152}$ & --- & --- & --- \\
    11 & $\frac{1375}{1152}$ & $\frac{125}{288}$ & $\frac{335}{192}$
    & $\frac{2483}{5760}$ & $\frac{7183}{5760}$ & $\frac{863}{960}$
    & --- & --- \\
    12 & $\frac{1375}{1152}$ & $\frac{125}{288}$ & $\frac{335}{192}$
    & $\frac{2483}{5760}$ & $\frac{3583}{2880}$ & $\frac{2743}{2880}$
    & --- & --- \\
    13 & $\frac{1375}{1152}$ & $\frac{125}{288}$ & $\frac{335}{192}$
    & $\frac{2483}{5760}$ & $\frac{3583}{2880}$ & $\frac{1823}{1920}$
    & $\frac{2897}{2880}$ & --- \\
    14 & $\frac{1375}{1152}$ & $\frac{125}{288}$ & $\frac{335}{192}$
    & $\frac{2483}{5760}$ & $\frac{3583}{2880}$ & $\frac{1823}{1920}$
    & $\frac{5777}{5760}$ & --- \\
    15+ & $\frac{1375}{1152}$ & $\frac{125}{288}$ & $\frac{335}{192}$
    & $\frac{2483}{5760}$ & $\frac{3583}{2880}$ & $\frac{1823}{1920}$
    & $\frac{5777}{5760}$ & 1 \\
    \br
  \end{tabular}
  \end{indented}
\end{table}

\newcommand\aap{Astron.~Astrophys.~}
\newcommand\mnras{Mon.~Not.~R.~Astron.~Soc.~}
\newcommand\prd{Phys.~Rev.~D }

\section*{References}
\bibliographystyle{unsrt}
\bibliography{refs}

\begin{thebibliography}{10}

\bibitem{reed1973triangular}
William~H Reed and TR~Hill.
\newblock Triangular mesh methods for the neutron transport equation.
\newblock Technical report, Los Alamos Scientific Lab., N. Mex.(USA), 1973.

\bibitem{Cockburn1989tvb}
Bernardo Cockburn and Chi-Wang Shu.
\newblock {TVB Runge-Kutta} local projection discontinuous {Galerkin} finite
  element method for conservation laws. {II. General framework}.
\newblock {\em Mathematics of Computation}, 52(186):411--435, 1989.

\bibitem{COCKBURN198990}
Bernardo Cockburn, San-Yih Lin, and Chi-Wang Shu.
\newblock {TVB Runge-Kutta} local projection discontinuous {Galerkin} finite
  element method for conservation laws {III: One-dimensional} systems.
\newblock {\em Journal of Computational Physics}, 84(1):90 -- 113, 1989.

\bibitem{1990MaCom..54..545C}
B.~{Cockburn}, S.~{Hou}, and C.-W. {Shu}.
\newblock The {Runge-Kutta} local projection discontinuous {Galerkin} finite
  element method for conservation laws. {IV. The} multidimensional case.
\newblock {\em Mathematics of Computation}, 54:545--581, April 1990.

\bibitem{jiang1994cell}
Guang~Shan Jiang and Chi-Wang Shu.
\newblock On a cell entropy inequality for discontinuous {Galerkin} methods.
\newblock {\em Mathematics of Computation}, 62(206):531--538, 1994.

\bibitem{barth2001energy}
Timothy Barth, Pierre Charrier, and Nagi~N Mansour.
\newblock Energy stable flux formulas for the discontinuous {Galerkin}
  discretization of first order nonlinear conservation laws.
\newblock Technical Report 20010095444, NASA Technical Reports Server, 2001.

\bibitem{hou2007solutions}
Songming Hou and Xu-Dong Liu.
\newblock Solutions of multi-dimensional hyperbolic systems of conservation
  laws by square entropy condition satisfying discontinuous {Galerkin} method.
\newblock {\em Journal of Scientific Computing}, 31(1-2):127--151, 2007.

\bibitem{Godunov1959}
S.~K. Godunov.
\newblock A~difference method for numerical calculation of discontinuous
  solutions of the equations of hydrodynamics.
\newblock {\em Mat. Sb. (N.S.)}, 47(89):271--306, 1959.

\bibitem{COSTA2007970}
Bruno Costa and Wai~Sun Don.
\newblock Multi-domain hybrid {spectral-WENO} methods for hyperbolic
  conservation laws.
\newblock {\em Journal of Computational Physics}, 224(2):970 -- 991, 2007.

\bibitem{doi:10.1002/fld.2654}
A.~Huerta, E.~Casoni, and J.~Peraire.
\newblock A simple shock-capturing technique for high-order discontinuous
  {Galerkin} methods.
\newblock {\em International Journal for Numerical Methods in Fluids},
  69(10):1614--1632, 2012.

\bibitem{10.1007/978-3-319-05591-6_96}
Matthias Sonntag and Claus-Dieter Munz.
\newblock Shock capturing for discontinuous {Galerkin} methods using finite
  volume subcells.
\newblock In J{\"u}rgen Fuhrmann, Mario Ohlberger, and Christian Rohde,
  editors, {\em Finite Volumes for Complex Applications VII-Elliptic, Parabolic
  and Hyperbolic Problems}, pages 945--953, Cham, 2014. Springer International
  Publishing.

\bibitem{Dumbser2014a}
Michael Dumbser, Olindo Zanotti, Raphaël Loubère, and Steven Diot.
\newblock A posteriori subcell limiting of the discontinuous {Galerkin} finite
  element method for hyperbolic conservation laws.
\newblock {\em Journal of Computational Physics}, 278:47 -- 75, 2014.

\bibitem{BOSCHERI2017449}
Walter Boscheri and Michael Dumbser.
\newblock {Arbitrary-Lagrangian–Eulerian} discontinuous {Galerkin} schemes
  with a posteriori subcell finite volume limiting on moving unstructured
  meshes.
\newblock {\em Journal of Computational Physics}, 346:449 -- 479, 2017.

\bibitem{Zanotti:2015mia}
Olindo Zanotti, Francesco Fambri, and Michael Dumbser.
\newblock {Solving the relativistic magnetohydrodynamics equations with ADER
  discontinuous Galerkin methods, a posteriori subcell limiting and adaptive
  mesh refinement}.
\newblock {\em Mon. Not. Roy. Astron. Soc.}, 452(3):3010--3029, 2015.

\bibitem{Fambri:2018udk}
Francesco Fambri, Michael Dumbser, Sven Köppel, Luciano Rezzolla, and Olindo
  Zanotti.
\newblock {ADER discontinuous Galerkin schemes for general-relativistic ideal
  magnetohydrodynamics}.
\newblock {\em Mon. Not. Roy. Astron. Soc.}, 477(4):4543--4564, 2018.

\bibitem{NUNEZDELAROSA2018113}
Jonatan {Núñez-de la Rosa} and Claus-Dieter Munz.
\newblock Hybrid dg/fv schemes for magnetohydrodynamics and relativistic
  hydrodynamics.
\newblock {\em Computer Physics Communications}, 222:113--135, 2018.

\bibitem{Boyle:2019kee}
Michael Boyle et~al.
\newblock {The SXS Collaboration catalog of binary black hole simulations}.
\newblock {\em Class. Quant. Grav.}, 36(19):195006, 2019.

\bibitem{SpECwebsite}
\url{https://www.black-holes.org/SpEC.html}.

\bibitem{Scheel:2008rj}
Mark~A. Scheel, Michael Boyle, Tony Chu, Lawrence~E. Kidder, Keith~D. Matthews,
  and Harald~P. Pfeiffer.
\newblock {High-accuracy waveforms for binary black hole inspiral, merger, and
  ringdown}.
\newblock {\em Phys. Rev. D}, 79:024003, 2009.

\bibitem{Szilagyi:2009qz}
Bela Szilagyi, Lee Lindblom, and Mark~A. Scheel.
\newblock {Simulations of Binary Black Hole Mergers Using Spectral Methods}.
\newblock {\em Phys. Rev. D}, 80:124010, 2009.

\bibitem{Lovelace:2010ne}
Geoffrey Lovelace, Mark.~A. Scheel, and Bela Szilagyi.
\newblock {Simulating merging binary black holes with nearly extremal spins}.
\newblock {\em Phys. Rev. D}, 83:024010, 2011.

\bibitem{Buchman:2012dw}
Luisa~T. Buchman, Harald~P. Pfeiffer, Mark~A. Scheel, and Bela Szilagyi.
\newblock {Simulations of non-equal mass black hole binaries with spectral
  methods}.
\newblock {\em Phys. Rev. D}, 86:084033, 2012.

\bibitem{Hemberger:2012jz}
Daniel~A. Hemberger, Mark~A. Scheel, Lawrence~E. Kidder, Béla Szilágyi,
  Geoffrey Lovelace, Nicholas~W. Taylor, and Saul~A. Teukolsky.
\newblock {Dynamical excision boundaries in spectral evolutions of binary black
  hole spacetimes}.
\newblock {\em Class. Quant. Grav.}, 30:115001, 2013.

\bibitem{Scheel:2014ina}
Mark~A. Scheel, Matthew Giesler, Daniel~A. Hemberger, Geoffrey Lovelace, Kevin
  Kuper, Michael Boyle, B.~Szil\'agyi, and Lawrence~E. Kidder.
\newblock {Improved methods for simulating nearly extremal binary black holes}.
\newblock {\em Class. Quant. Grav.}, 32(10):105009, 2015.

\bibitem{Szilagyi:2014fna}
B\'ela Szil\'agyi.
\newblock {Key Elements of Robustness in Binary Black Hole Evolutions using
  Spectral Methods}.
\newblock {\em Int. J. Mod. Phys. D}, 23(7):1430014, 2014.

\bibitem{Bonazzola:1998ge}
S.~Bonazzola, E.~Gourgoulhon, and J.~A. Marck.
\newblock {Spectral methods in general relativistic astrophysics}.
\newblock {\em J. Comput. Appl. Math.}, 109:433, 1999.

\bibitem{Meringolo:2021yjh}
Claudio Meringolo and Sergio Servidio.
\newblock {Aliasing instabilities in the numerical evolution of the Einstein
  field equations}.
\newblock {\em Gen. Rel. Grav.}, 53(10):95, 2021.

\bibitem{Hilditch:2015aba}
David Hilditch, Andreas Weyhausen, and Bernd Br\"ugmann.
\newblock {Pseudospectral method for gravitational wave collapse}.
\newblock {\em Phys. Rev. D}, 93(6):063006, 2016.

\bibitem{Rashti:2021ihv}
Alireza Rashti, Francesco~Maria Fabbri, Bernd Br\"ugmann, Swami~Vivekanandji
  Chaurasia, Tim Dietrich, Maximiliano Ujevic, and Wolfgang Tichy.
\newblock {New pseudospectral code for the construction of initial data}.
\newblock {\em Phys. Rev. D}, 105(10):104027, 2022.

\bibitem{Meringolo:2020jsu}
Claudio Meringolo, Sergio Servidio, and Pierluigi Veltri.
\newblock {A spectral method algorithm for numerical simulations of
  gravitational fields}.
\newblock {\em Class. Quant. Grav.}, 38(7):075027, 2021.

\bibitem{Tichy:2009zr}
Wolfgang Tichy.
\newblock {Long term black hole evolution with the BSSN system by
  pseudo-spectral methods}.
\newblock {\em Phys. Rev. D}, 80:104034, 2009.

\bibitem{Kidder:2016hev}
Lawrence~E. Kidder et~al.
\newblock {SpECTRE: a task-based discontinuous Galerkin code for relativistic
  astrophysics}.
\newblock {\em J. Comput. Phys.}, 335:84--114, 2017.

\bibitem{deppe_nils_2021_5501002}
Nils Deppe, William Throwe, Lawrence~E. Kidder, Nils~L. Vu, Fran\c{c}ois
  H\'ebert, Jordan Moxon, Crist\'obal Armaza, Gabriel~S. Bonilla, Yoonsoo Kim,
  Prayush Kumar, Geoffrey Lovelace, Alexandra Macedo, Kyle~C. Nelli, Eamonn
  O'Shea, Harald~P. Pfeiffer, Mark~A. Scheel, Saul~A. Teukolsky, Nikolas~A.
  Wittek, et~al.
\newblock \texttt{SpECTRE v2022.04.04}.
\newblock
  \href{https://doi.org/10.5281/zenodo.6412468}{10.5281/zenodo.6412468}, 4
  2022.

\bibitem{Baumgarte:2010ndz}
Thomas~W. Baumgarte and Stuart~L. Shapiro.
\newblock {\em {Numerical Relativity: Solving Einstein's Equations on the
  Computer}}.
\newblock Cambridge University Press, 2010.

\bibitem{2013rehy.book.....R}
L.~{Rezzolla} and O.~{Zanotti}.
\newblock {\em {Relativistic Hydrodynamics}}.
\newblock Oxford University Press, September 2013.

\bibitem{2006ApJ...637..296A}
Luis Ant\'on, Olindo Zanotti, Juan~A. Miralles, Jos\'e~M. Mart\'i, Jos\'e~M.
  Ib\'a\~nez, Jos\'e~A. Font, and Jos\'e~A. Pons.
\newblock {Numerical 3+1 general relativistic magnetohydrodynamics: a local
  characteristic approach}.
\newblock {\em The Astrophysical Journal}, 637:296--312, January 2006.

\bibitem{Font:2008fka}
Jose~A. Font.
\newblock {Numerical hydrodynamics and magnetohydrodynamics in general
  relativity}.
\newblock {\em Living Rev. Rel.}, 11:7, 2008.

\bibitem{GravitationMTW}
Charles~W.\ Misner, Kip~S.\ Thorne, and John~Archibald Wheeler.
\newblock {\em Gravitation}.
\newblock Freeman, New York, New York, 1973.

\bibitem{2002JCoPh.175..645D}
A.~{Dedner}, F.~{Kemm}, D.~{Kr{\"o}ner}, C.-D. {Munz}, T.~{Schnitzer}, and
  M.~{Wesenberg}.
\newblock {Hyperbolic divergence cleaning for the {MHD} equations}.
\newblock {\em Journal of Computational Physics}, 175:645--673, January 2002.

\bibitem{Mosta:2013gwu}
Philipp M\"osta, Bruno~C. Mundim, Joshua~A. Faber, Roland Haas, Scott~C. Noble,
  Tanja Bode, Frank L\"offler, Christian~D. Ott, Christian Reisswig, and Erik
  Schnetter.
\newblock {GRHydro: A new open source general-relativistic magnetohydrodynamics
  code for the Einstein Toolkit}.
\newblock {\em Class. Quant. Grav.}, 31:015005, 2014.

\bibitem{Teukolsky:2015ega}
Saul~A. Teukolsky.
\newblock {Formulation of discontinuous {Galerkin} methods for relativistic
  astrophysics}.
\newblock {\em J. Comput. Phys.}, 312:333--356, 2016.

\bibitem{Hesthaven2008}
J.S. Hesthaven and T.~Warburton.
\newblock {\em Nodal Discontinuous Galerkin Methods: Algorithms, Analysis, and
  Applications}.
\newblock Springer-Verlag New York, New York, 2008.

\bibitem{Scheel:2006gg}
Mark~A. Scheel, Harald~P. Pfeiffer, Lee Lindblom, Lawrence~E. Kidder, Oliver
  Rinne, and Saul~A. Teukolsky.
\newblock {Solving {Einstein's} equations with dual coordinate frames}.
\newblock {\em Phys. Rev.}, D74:104006, 2006.

\bibitem{DUMBSER20091731}
Michael Dumbser, Manuel Castro, Carlos Parés, and Eleuterio~F. Toro.
\newblock {ADER} schemes on unstructured meshes for nonconservative hyperbolic
  systems: {Applications} to geophysical flows.
\newblock {\em Computers \& Fluids}, 38(9):1731 -- 1748, 2009.

\bibitem{DALMASO1995}
G~Dal~Maso, P.~G LeFloch, and F~Murat.
\newblock Definition and weak stability of nonconservative products.
\newblock {\em Journal de mathématiques pures et appliquées}, 1995.

\bibitem{MINOLI20111876}
Cesar A.~Acosta Minoli and David~A. Kopriva.
\newblock Discontinuous {Galerkin} spectral element approximations on moving
  meshes.
\newblock {\em Journal of Computational Physics}, 230(5):1876 -- 1902, 2011.

\bibitem{Shu1988439}
Chi-Wang Shu and Stanley Osher.
\newblock Efficient implementation of essentially non-oscillatory
  shock-capturing schemes.
\newblock {\em Journal of Computational Physics}, 77(2):439 -- 471, 1988.

\bibitem{doi:10.1137/19M1292692}
William Throwe and Saul Teukolsky.
\newblock A high-order, conservative integrator with local time-stepping.
\newblock {\em SIAM Journal on Scientific Computing}, 42(6):A3730--A3760, 2020.

\bibitem{mersman1965self}
William~A Mersman.
\newblock {\em Self-starting multistep methods for the numerical integration of
  ordinary differential equations}.
\newblock National Aeronautics and Space Administration, 1965.

\bibitem{cockburn2000development}
Bernardo Cockburn, George~E Karniadakis, and Chi-Wang Shu.
\newblock The development of discontinuous {Galerkin} methods.
\newblock In {\em Discontinuous Galerkin Methods}, pages 3--50. Springer, 2000.

\bibitem{cockburn2001runge}
Bernardo Cockburn and Chi-Wang Shu.
\newblock {Runge-Kutta} discontinuous {Galerkin} methods for
  convection-dominated problems.
\newblock {\em Journal of Scientific Computing}, 16(3):173--261, 2001.

\bibitem{KRIVODONOVA20131}
Lilia Krivodonova and Ruibin Qin.
\newblock An analysis of the spectrum of the discontinuous {Galerkin} method.
\newblock {\em Applied Numerical Mathematics}, 64:1 -- 18, 2013.

\bibitem{Deppe:2021DgLimiterComparison}
Nils Deppe et~al.
\newblock {Simulating magnetized neutron stars with discontinuous Galerkin
  methods}.
\newblock Submitted.

\bibitem{1998JCoPh.141..199C}
B.~{Cockburn} and C.-W. {Shu}.
\newblock {The {Runge-Kutta} discontinuous {Galerkin} method for conservation
  laws {V. Multidimensional} systems}.
\newblock {\em Journal of Computational Physics}, 141:199--224, April 1998.

\bibitem{Krivodonova2004}
L.~Krivodonova, J.~Xin, J.-F. Remacle, N.~Chevaugeon, and J.E. Flaherty.
\newblock Shock detection and limiting with discontinuous {Galerkin} methods
  for hyperbolic conservation laws.
\newblock {\em Applied Numerical Mathematics}, 48(3):323 -- 338, 2004.

\bibitem{Krivodonova:2007}
Lilia Krivodonova.
\newblock Limiters for high-order discontinuous {Galerkin} methods.
\newblock {\em Journal of Computational Physics}, 226(1):879 -- 896, 2007.

\bibitem{2013JCoPh.232..397Z}
X.~{Zhong} and C.-W. {Shu}.
\newblock A simple weighted essentially nonoscillatory limiter for
  {Runge-Kutta} discontinuous {Galerkin} methods.
\newblock {\em Journal of Computational Physics}, 232:397--415, January 2013.

\bibitem{2016CCoPh..19..944Z}
J.~{Zhu}, X.~{Zhong}, C.-W. {Shu}, and J.~{Qiu}.
\newblock {Runge-Kutta} discontinuous {Galerkin} method with a simple and
  compact {Hermite WENO} limiter.
\newblock {\em Communications in Computational Physics}, 19:944--969, April
  2016.

\bibitem{WANG2012653}
Cheng Wang, Xiangxiong Zhang, Chi-Wang Shu, and Jianguo Ning.
\newblock Robust high order discontinuous {Galerkin} schemes for
  two-dimensional gaseous detonations.
\newblock {\em Journal of Computational Physics}, 231(2):653 -- 665, 2012.

\bibitem{1989ddcm.proc..392M}
Yvon {Maday}, Cathy {Mavriplis}, and Anthony~T. {Patera}.
\newblock {Nonconforming mortar element methods - Application to spectral
  discretizations}.
\newblock In {\em Domain Decomposition Methods}, pages 392--418, January 1989.

\bibitem{KOPRIVA1996475}
David~A. Kopriva.
\newblock A conservative staggered-grid {Chebyshev} multidomain method for
  compressible flows. {II. A} semi-structured method.
\newblock {\em Journal of Computational Physics}, 128(2):475 -- 488, 1996.

\bibitem{Kopriva2002}
David~A. Kopriva, Stephen~L. Woodruff, and M.~Y. Hussaini.
\newblock Computation of electromagnetic scattering with a non-conforming
  discontinuous spectral element method.
\newblock {\em International Journal for Numerical Methods in Engineering},
  53(1):105--122, 2002.

\bibitem{Thanh2012}
Tan Bui-Thanh and Omar Ghattas.
\newblock Analysis of an $hp$-nonconforming discontinuous {Galerkin} spectral
  element method for wave propagation.
\newblock {\em SIAM Journal on Numerical Analysis}, 50:1801--1826, 01 2012.

\bibitem{PerssonTci}
Per-Olof Persson and Jaime Peraire.
\newblock Sub-cell shock capturing for discontinuous {Galerkin} methods.
\newblock In {\em 44th AIAA Aerospace Sciences Meeting and Exhibit}. {American
  Institute of Aeronautics and Astronautics, Inc.}, 2006.

\bibitem{Gottlieb1977}
David Gottlieb and Steven~A. Orszag.
\newblock {\em Numerical Analysis of Spectral Methods}.
\newblock Society for Industrial and Applied Mathematics, 1977.

\bibitem{DORMAND198019}
J.R. Dormand and P.J. Prince.
\newblock A family of embedded runge-kutta formulae.
\newblock {\em Journal of Computational and Applied Mathematics}, 6(1):19--26,
  1980.

\bibitem{Balsara2001}
Dinshaw Balsara.
\newblock Total variation diminishing scheme for relativistic
  magnetohydrodynamics.
\newblock {\em The Astrophysical Journal Supplement Series}, 132(1):83--101,
  January 2001.

\bibitem{Giacomazzo2006}
Bruno {Giacomazzo} and Luciano {Rezzolla}.
\newblock {The exact solution of the Riemann problem in relativistic
  magnetohydrodynamics}.
\newblock {\em Journal of Fluid Mechanics}, 562:223--259, September 2006.

\bibitem{2005A&A...436..503L}
T.~{Leismann}, L.~{Ant{\'o}n}, M.~A. {Aloy}, E.~{M{\"u}ller}, J.~M.
  {Mart{\'{\i}}}, J.~A. {Miralles}, and J.~M. {Ib{\'a}{\~n}ez}.
\newblock {Relativistic MHD simulations of extragalactic jets}.
\newblock {\em \aap}, 436:503--526, June 2005.

\bibitem{DelZanna:2007pk}
L.~Del~Zanna, O.~Zanotti, N.~Bucciantini, and P.~Londrillo.
\newblock {ECHO}: an {Eulerian} conservative high order scheme for general
  relativistic magnetohydrodynamics and magnetodynamics.
\newblock {\em Astron. Astrophys.}, 473:11--30, 2007.

\bibitem{1999JCoPh.149..270B}
Dinshaw~S. {Balsara} and Daniel~S. {Spicer}.
\newblock A staggered mesh algorithm using high order {Godunov} fluxes to
  ensure solenoidal magnetic fields in magnetohydrodynamic simulations.
\newblock {\em Journal of Computational Physics}, 149(2):270--292, March 1999.

\bibitem{2000JCoPh.161..605T}
G{\'a}bor {T{\'o}th}.
\newblock {The {\ensuremath{\nabla}}{\textperiodcentered} B=0} constraint in
  shock-capturing magnetohydrodynamics codes.
\newblock {\em Journal of Computational Physics}, 161(2):605--652, July 2000.

\bibitem{2010PhRvD..82h4031E}
Zachariah~B. {Etienne}, Yuk~Tung {Liu}, and Stuart~L. {Shapiro}.
\newblock {Relativistic magnetohydrodynamics in dynamical spacetimes: A new
  adaptive mesh refinement implementation}.
\newblock {\em \prd}, 82(8):084031, October 2010.

\bibitem{2003A&A...400..397D}
L.~{Del Zanna}, N.~{Bucciantini}, and P.~{Londrillo}.
\newblock {An efficient shock-capturing central-type scheme for
  multidimensional relativistic flows. II. Magnetohydrodynamics}.
\newblock {\em \aap}, 400:397--413, March 2003.

\bibitem{1991JCoPh..92..142D}
C.~Richard {DeVore}.
\newblock Flux-corrected transport techniques for multidimensional compressible
  magnetohydrodynamics.
\newblock {\em Journal of Computational Physics}, 92(1):142--160, January 1991.

\bibitem{2011ApJS..193....6B}
Kris {Beckwith} and James~M. {Stone}.
\newblock A second-order {Godunov} method for multi-dimensional relativistic
  magnetohydrodynamics.
\newblock {\em Astrophysical Journal, Supplement}, 193(1):6, March 2011.

\bibitem{2005JCoPh.205..509G}
Thomas~A. {Gardiner} and James~M. {Stone}.
\newblock {An unsplit Godunov method for ideal MHD via constrained transport}.
\newblock {\em Journal of Computational Physics}, 205(2):509--539, May 2005.

\bibitem{2008ApJS..178..137S}
James~M. {Stone}, Thomas~A. {Gardiner}, Peter {Teuben}, John~F. {Hawley}, and
  Jacob~B. {Simon}.
\newblock {Athena: A New Code for Astrophysical MHD}.
\newblock {\em Astrophysical Journal, Supplement}, 178(1):137--177, September
  2008.

\bibitem{Tolman:1939jz}
Richard~C. Tolman.
\newblock {Static solutions of Einstein's field equations for spheres of
  fluid}.
\newblock {\em Phys. Rev.}, 55:364--373, 1939.

\bibitem{Oppenheimer:1939ne}
J.~R. Oppenheimer and G.~M. Volkoff.
\newblock {On massive neutron cores}.
\newblock {\em Phys. Rev.}, 55:374--381, 1939.

\bibitem{Cipolletta:2019geh}
Federico Cipolletta, Jay~Vijay Kalinani, Bruno Giacomazzo, and Riccardo Ciolfi.
\newblock {Spritz: a new fully general-relativistic magnetohydrodynamic code}.
\newblock {\em Class. Quant. Grav.}, 37(13):135010, 2020.

\bibitem{2002PhRvD..65h4024F}
Jos{\'e}~A. {Font}, Tom {Goodale}, Sai {Iyer}, Mark {Miller}, Luciano
  {Rezzolla}, Edward {Seidel}, Nikolaos {Stergioulas}, Wai-Mo {Suen}, and
  Malcolm {Tobias}.
\newblock {Three-dimensional numerical general relativistic hydrodynamics. II.
  Long-term dynamics of single relativistic stars}.
\newblock {\em Phys. Rev. D}, 65(8):084024, April 2002.

\bibitem{1992ApJ...398..203C}
Gregory~B. {Cook}, Stuart~L. {Shapiro}, and Saul~A. {Teukolsky}.
\newblock Spin-up of a rapidly rotating star by angular momentum loss: Effects
  of general relativity.
\newblock {\em The Astrophysical Journal}, 398:203, October 1992.

\bibitem{1994ApJ...424..823C}
Gregory~B. {Cook}, Stuart~L. {Shapiro}, and Saul~A. {Teukolsky}.
\newblock Rapidly rotating neutron stars in general relativity: Realistic
  equations of state.
\newblock {\em The Astrophysical Journal}, 424:823, April 1994.

\bibitem{laxmikant_kale_2020_3972617}
Laxmikant Kale, Bilge Acun, Seonmyeong Bak, Aaron Becker, Milind Bhandarkar,
  Nitin Bhat, Abhinav Bhatele, Eric Bohm, Cyril Bordage, Robert Brunner, Ronak
  Buch, Sayantan Chakravorty, Kavitha Chandrasekar, Jaemin Choi, Michael
  Denardo, Jayant DeSouza, Matthias Diener, Harshit Dokania, Isaac Dooley,
  Wayne Fenton, Juan Galvez, Fillipo Gioachin, Abhishek Gupta, Gagan Gupta,
  Manish Gupta, Attila Gursoy, Vipul Harsh, Fang Hu, Chao Huang, Narain
  Jagathesan, Nikhil Jain, Pritish Jetley, Prateek Jindal, Raghavendra
  Kanakagiri, Greg Koenig, Sanjeev Krishnan, Sameer Kumar, David Kunzman,
  Michael Lang, Akhil Langer, Orion Lawlor, Chee~Wai Lee, Jonathan Lifflander,
  Karthik Mahesh, Celso Mendes, Harshitha Menon, Chao Mei, Esteban Meneses,
  Eric Mikida, Phil Miller, Ryan Mokos, Venkatasubrahmanian Narayanan, Xiang
  Ni, Kevin Nomura, Sameer Paranjpye, Parthasarathy Ramachandran, Balkrishna
  Ramkumar, Evan Ramos, Michael Robson, Neelam Saboo, Vikram Saletore, Osman
  Sarood, Karthik Senthil, Nimish Shah, Wennie Shu, Amitabh~B. Sinha, Yanhua
  Sun, Zehra Sura, Ehsan Totoni, Krishnan Varadarajan, Ramprasad Venkataraman,
  Jackie Wang, Lukasz Wesolowski, Sam White, Terry Wilmarth, Jeff Wright,
  Joshua Yelon, and Gengbin Zheng.
\newblock Uiuc-ppl/charm: Charm++ version 6.10.2, August 2020.

\bibitem{Hunter:2007}
J.~D. Hunter.
\newblock Matplotlib: A 2d graphics environment.
\newblock {\em Computing in Science \& Engineering}, 9(3):90--95, 2007.

\bibitem{thomas_a_caswell_2020_3948793}
Thomas~A Caswell, Michael Droettboom, Antony Lee, John Hunter, Eric Firing,
  Elliott~Sales de~Andrade, Tim Hoffmann, David Stansby, Jody Klymak, Nelle
  Varoquaux, Jens~Hedegaard Nielsen, Benjamin Root, Ryan May, Phil Elson,
  Darren Dale, Jae-Joon Lee, Jouni~K. Seppänen, Damon McDougall, Andrew Straw,
  Paul Hobson, Christoph Gohlke, Tony~S Yu, Eric Ma, Adrien~F. Vincent, Steven
  Silvester, Charlie Moad, Nikita Kniazev, hannah, Elan Ernest, and Paul
  Ivanov.
\newblock matplotlib/matplotlib: Rel: v3.3.0, July 2020.

\bibitem{tikz}
T.~Tantau.
\newblock The tikz and pgf packages.

\bibitem{paraview}
Utkarsh Ayachit.
\newblock {\em The ParaView Guide: A Parallel Visualization Application}.
\newblock Kitware, Inc., Clifton Park, NY, USA, 2015.

\bibitem{paraview2}
J.~Ahrens, Berk Geveci, and C.~Law.
\newblock Paraview: An end-user tool for large-data visualization.
\newblock In {\em The Visualization Handbook}, 2005.

\bibitem{sebastian2003multidomain}
Kurt Sebastian and Chi-Wang Shu.
\newblock Multidomain {WENO} finite difference method with interpolation at
  subdomain interfaces.
\newblock {\em Journal of Scientific Computing}, 19(1-3):405--438, 2003.

\bibitem{doi:10.1002/fld.1469}
W.~R. Wolf and J.~L.~F. Azevedo.
\newblock High-order {ENO} and {WENO} schemes for unstructured grids.
\newblock {\em International Journal for Numerical Methods in Fluids},
  55(10):917--943, 2007.

\bibitem{TSOUTSANIS2014254}
Panagiotis Tsoutsanis, Antonios~Foivos Antoniadis, and Dimitris Drikakis.
\newblock {WENO} schemes on arbitrary unstructured meshes for laminar,
  transitional and turbulent flows.
\newblock {\em Journal of Computational Physics}, 256:254 -- 276, 2014.

\bibitem{Tsoutsanis2019}
Panagiotis Tsoutsanis.
\newblock Stencil selection algorithms for {WENO} schemes on unstructured
  meshes.
\newblock {\em Journal of Computational Physics: X}, 4:100037, 08 2019.

\bibitem{FARMAKIS2020112921}
Pericles~S. Farmakis, Panagiotis Tsoutsanis, and Xesús Nogueira.
\newblock {WENO} schemes on unstructured meshes using a relaxed a posteriori
  {MOOD} limiting approach.
\newblock {\em Computer Methods in Applied Mechanics and Engineering},
  363:112921, 2020.

\bibitem{DUMBSER2007204}
Michael Dumbser, Martin Käser, Vladimir~A. Titarev, and Eleuterio~F. Toro.
\newblock Quadrature-free non-oscillatory finite volume schemes on unstructured
  meshes for nonlinear hyperbolic systems.
\newblock {\em Journal of Computational Physics}, 226(1):204 -- 243, 2007.

\bibitem{doi:10.2514/6.2019-2955}
Chunhua Sheng, Qiuying Zhao, Dongdong Zhong, and Ning Ge.
\newblock A strategy to implement high-order {WENO} schemes on unstructured
  grids.
\newblock In {\em AIAA Aviation 2019 Forum}. {American Institute of Aeronautics
  and Astronautics, Inc.}, 2019.

\bibitem{doi:10.2514/6.2017-0845}
Lucie Freret, Lucian Ivan, Hans~De Sterck, and Clinton~P. Groth.
\newblock A high-order finite-volume method with anisotropic {AMR} for ideal
  {MHD} flows.
\newblock In {\em 55th AIAA Aerospace Sciences Meeting}. {American Institute of
  Aeronautics and Astronautics, Inc.}, 2017.

\end{thebibliography}
\end{document}